\documentclass[aps,pre,reprint,showpacs]{revtex4-1}%

\DeclareMathAlphabet{\mathpzc}{OT1}{pzc}{m}{it}
\usepackage{amsmath}
\usepackage{amsfonts}
\usepackage{amssymb}
\usepackage{graphicx}
\usepackage{amsthm}
\usepackage{mathrsfs}
\usepackage[mathscr]{euscript}
\usepackage{dsfont}
\usepackage{txfonts}
\usepackage{lineno}
\usepackage{enumerate}

\begin{document}
\pagestyle{myheadings}

\title{Entropy and its relationship to allometry}
\author{Robert Shour}
\affiliation{Toronto, Ontario}
\date{\today}

\begin{abstract}
Researchers have found that the metabolisms of organisms appear to scale proportionally to a 3/4 power of their mass.  Mathematics in this article suggests that the capacity of isotropic energy distribution scales up by a 4/3 power as size, and therefore the degrees of freedom of its circulatory system, increases. Cellular metabolism must scale inversely by a 3/4 power, likely to prevent the 4/3 scaling up of the energy supply overheating the cells. The same 4/3 power scaling of  energy distribution may explain cosmological dark energy. 
\end{abstract}

\maketitle 

\noindent \textbf{Keywords}\: allometry --- cosmological parameters --- dark energy --- entropy --- The 4/3 Ratio of Degrees of Freedom Theorem --- The 4/3 Ratio of Lengths Theorem --- inflation

\tableofcontents

\section{Introduction and Background}

The universe is not only expanding (Hubble 1929), it is accelerating (Riess et al. 1998; Perlmutter et al. 1999). The unknown cause of the acceleration has been called dark energy (Turner 1998). This article proposes that the distance light travels in a radiation reference frame is 4/3 greater when measured in a spa
ce reference frame. The  4/3 stretching of radial length in a space reference frame relative to a radiation-space  reference frame (the 4 dimensional combination of radiation and space) is due, according to the mathematics developed in this article, to the $4/3 : 1$ ratio of  the scaling of radiation energy distribution relative to the scaling of a corresponding space.

The same  $4/3 : 1$ ratio of radiation scaling  plays a role in allometry. Allometry is the study of biological scaling. Studies suggest that metabolism scales by a 3/4 power of the size of an organism's mass. Metabolism scaling proportionate to a 3/4 power of an organism's mass can be explained by the organism's energy distribution system --- its circulatory system --- scaling up by a 4/3 power as it increases in size. 

Thus, the mathematics in this article suggests that laws of nature general in scope may explain both dark energy and metabolic scaling as well as other phenomena, such as the fractal dimension of Brownian motion. 

This article: 
\begin{itemize}
	\item Begins with an overview of ideas leading to a mathematical model applicable to allometry and dark energy.
	\item Describes development of the idea that  radiated energy distribution involves two reference frames, namely a 4 dimensional radiation-space reference frame and a 3 dimensional space reference frame. 
	\item Discusses how the idea of degrees of freedom of the base of a scale factor arises.
	\item Derives The $4/3$ Ratio of Degrees of Freedom Theorem (The $4/3$ RDFT).
	\item Shows how The $4/3$ RDFT might account for 3/4 metabolic scaling. 
	\item Shows how The 4/3 RDFT implies The $4/3$ Ratio of Lengths Theorem (The $4/3$ RLT). 
	\item Connects The $4/3$ RLT to  dark energy.
	\item Connects  ideas related to 3/4 metabolic scaling to dark energy.
	\item Gives other possible manifestations of the two $4/3$ theories and their mathematics. 
\end{itemize}

The ideas in this article arose indirectly out of a question  about IQs (which itself arose out of earlier questions): does constant improvement in society's abstractions and ideas explain the rate of increase in average IQs. In September 2005, I found that the average English lexical growth rate from 1657 to 1989 is close to a measured 3.3\% per decade rate of increase in average IQs.  I wondered whether society's lexical growth rate could be modeled using a scale factor. 

At the end of May 2007 I found that a network's mean path length $\mu$ measured in steps worked as the base of a scale factor $\mu^k$. The mean path length determined the number of degrees of freedom $\log_\mu(n)$ in a network relative to its mean path length. The mean path length as the  base of any scale factor common to all nodes in the network implied homogeneous and isotropic scaling of the network by the mean path length. 

In March 2008, I adapted  mean path length  scaling to 3/4 metabolic scaling. Both mean path length scaling and 3/4 metabolic scaling involve degrees of freedom. Expecting that the algebra would output 3/4 scaling, the mathematics instead implied that circulatory system capacity scaled by a 4/3 power at every level of the  circulatory system.  I found in April 2008 that Stefan's Law implied that similar 4/3 scaling applied to black body radiation. In September 2012, it appeared that  in cosmology 4/3 scaling described how radiation  energy density  compared to  matter energy density varied with distance. 

Applying mean path length  scaling to a variety of phenomena since 2007 helped test, explain, correct, verify, improve, refine and simplify the mathematical model. Part of this process is exhibited by the sequence of versions of articles I submitted to arXiv.org. 

The Overview  sketches the mathematical elements that lead to an explanation for 4/3 scaling, which give an explanation of 3/4 metabolic scaling and lead to a mathematical explanation for dark energy. The Overview has the advantage of simplicity and the disadvantage that it omits the development of the ideas leading to its focus on degrees of freedom. The Overview is intended to be a self-contained summary; the part of the article following it elaborates and discusses ideas in it. This Overview (in the version 17 article) is intended to improve on and refine the ideas contained in the Overview in arXiv version 16 of this article. An important difference is distinguishing between dimension and degrees of freedom in relation to a radiation cone volume increment. The Overview in this article uses slightly different notation than contained in the rest of the article. The Overview now refers to the 4/3 ratio relationships as laws as they appear to be laws of nature, but they can be described as theorems as well. The older notation in the article outside The Overview has for the most part not been updated, in order to reduce the amount of time required to revise version 16 of this article.

\section{The Overview}

Two new laws of nature are here proposed,
\begin{itemize}
	\item A radiation  quantum radial length  has 4/3 the degrees of freedom of its image in  empty space: The 4/3 Ratio of Degrees of Freedom Law (The 4/3 RDFL) and 
	\item  A quantum radial length measured in 3 dimensional space is 4/3 as long as a quantum radial length measured in 4 dimensional radiation-space:  The 4/3 Ratio of Lengths Law (The 4/3 RLL)
\end{itemize}

\noindent that may explain various phenomena. To derive them requires solving  the following  problem.

\subsection{The 4/3 dimension ($Dim$) problem} \label{Subsec-St-Dim-problem}

A radiation cone volume increment $V$ has volume $V=AL$ where $A$ is the average cross-sectional area of $V$, and $L$ is  radiation's  quantum length. $L$'s dimension  is 1/3 of $V$'s. If $v$ is $V$'s scale factor, and $a$ is $A$'s scale factor, then $vV/V=aAL/AL$ since $L$ is constant, so $v=a$; $Dim(V)=Dim(A)$. The dimension of $V$ when scaled appears to be 
\begin{equation} \label{Eq-St-of-4-3-Scaling-Problem}
\begin{split}
Dim(V)& =Dim(AL) \\
&=Dim(A)+Dim(L) \\
& = Dim(V) + (1/3)Dim(V) \\
& = (4/3) Dim (V).
\end{split}
\end{equation}
$Dim(V)=(4/3) Dim (V)$, from the left and right extremes of equation (\ref{Eq-St-of-4-3-Scaling-Problem}), is impossible. Can  equation (\ref{Eq-St-of-4-3-Scaling-Problem})'s  inconsistency be resolved?

Deriving The 4/3 Ratio of Lengths Law can explain the inconsistency as described below. 

\subsection{Clausius and Boltzmann: equations describing entropy}

The history of entropy begins with the analysis  by  Sadi Carnot  (1824) of an idealized steam engine. 
\'Emile Clapeyron's memoir (1834) added diagrams and equations. Rudolf Clausius,  using an absolute temperature devised by William Thomson (Lord Kelvin),  found (p. 111, 1854) that for an ideal heat engine, in effect 

\begin{equation} \label{Eq-Clausius-Ent-Eq-100}
\frac{dQ_1}{T_1} = dS_1= dS_2 = \frac{dQ_2}{T_2}, 
\end{equation}
where $S$ represents entropy, $T$ a temperature in absolute degrees Kelvin, and $Q$ an amount of heat. The subscripts 1 and 2 respectively denote the portions of an ideal heat cycle where a furnace and heat sink are in contact with a heat engine chamber. 

Max Planck (p. 119, Planck)  reconfigured  Ludwig Boltzmann's $H$ Theory (Ch. 1, Sec. 6, Boltzmann Gastheorie) using a combinatorial approach to obtain a logarithmic formula for entropy:

\begin{equation} \label{Eq-Planck-Boltzmann-log-Ent-Eq-100}
S= k \log W.
\end{equation}


\subsection{Entropy as degrees of freedom of a scale factor's base} \label{Sec-Ent-As-SF}

\subsubsection{Equating entropy equations} \label{Subsec-Eqatg-Ent-eqns-SF-as-unit}

Substitute total entropy $S$ and total energy $Q$ for equation (\ref{Eq-Clausius-Ent-Eq-100})'s  differences in entropy $dS$ and heat (or energy) $dQ$. Equate equations (\ref{Eq-Clausius-Ent-Eq-100}) and (\ref{Eq-Planck-Boltzmann-log-Ent-Eq-100})  as
\begin{equation} \label{Eq-Clausius-equals-Boltzmann-Eq-100}
S = \frac{Q}{T}= \log W
\end{equation}
for $T$ a measure of absolute temperature (not degrees Kelvin) proportional to an amount of energy in  a given thermodynamically isolated chamber and $Q$  a quantity of heat providing  uniform temperature or, equivalently, isotropically and homogeneously distributed energy in the chamber.  $W=x^k$ for some $x$ and $k$.

$Q/T$ in equation (\ref{Eq-Clausius-equals-Boltzmann-Eq-100}) would count the number of times, $\log W$, that $T$ is contained in $Q$ if $T$ were the base of $\log W$. Choose $T$ as the base of the logarithm $\log W$:
\begin{equation} \label{Eq-log-W-mult-energy-unit}
Q = \log_T(T^k) T.
\end{equation}
It follows that 
\begin{equation} \label{Eq-Entropy-as-E-div-by-SF}
\frac{\log_T(T^k) T} {T} = \log_T(T^k) =S.
\end{equation}
In equation (\ref{Eq-log-W-mult-energy-unit}) $\log_T(T^k)=k$  multiplies an amount of energy proportional to $T$.  Entropy $\log W$  is proportional to the energy \textit{capacity} of a given system if the base of the scale factor  satisfies equation (\ref{Eq-Entropy-as-E-div-by-SF}). Equation (\ref{Eq-log-W-mult-energy-unit}) shows that $k$ times $T$ spans total energy $Q$. If a system's output is proportional to its energy input, then $k$ in equation (\ref{Eq-log-W-mult-energy-unit})  also is proportional to \textit{output} capacity (assuming no energy losses) in output units proportional to $T$. Entropy is thus the exponent, number of scalings or degrees of freedom  of the base of a scale factor for a homogeneous distribution of energy. The scale factor's base is  the base of corresponding logarithm.

The logarithm's base in $\log W$  is immaterial in classical thermodynamics but for equation (\ref{Eq-log-W-mult-energy-unit}) must be chosen so that the equation holds.

In equations (\ref{Eq-log-W-mult-energy-unit}) and (\ref{Eq-Entropy-as-E-div-by-SF}) the base of  the system's intrinsic scale factor  $T$ is the same as the energy unit, which is multiplied by $\log W$.

\subsubsection{Molecular gas mean path length as a scale factor} \label{Sec-entropy-eg-gas-mol}

As an example of section \ref{Subsec-Eqatg-Ent-eqns-SF-as-unit} characterization of entropy, consider an ideal gas in a chamber.  With $\mu$ the mean path length between all pairs of colliding gas molecules, equation (\ref{Eq-Entropy-as-E-div-by-SF}) gives the entropy $S$ of the system, with $\mu$ as both the scale factor and the unit length. As a (constant) statistical average, $\mu$  is analogous to constant $T$ in equation  (\ref{Eq-Entropy-as-E-div-by-SF}). The average energy of a gas molecule is proportional to $\mu$; the calculation of entropy using $\mu$ as the base of the logarithm describes the scaling of the energy of the ensemble of gas molecules.

\subsection{Radiation in space} 

\subsubsection{Modeling radiation by a radiation cone}

Model homogeneous radiation distribution in space by a radiation cone  $G_k = V_1 + V_2 + \ldots +V_k$ with radial length $kL$ where $V_k$ is a  $G_k$ volume increment with  quantum radial length $L$. A \textit{radiation event} adds $L$ to   $G_k$'s radial length $kL$. Having regard for equation (\ref{Eq-log-W-mult-energy-unit}), $L$ as a scale factor scales $L$ as a radial quantum length. Let $A_k$ be the average cross-sectional area of $V_k$ so
\begin{equation}
V_k = A_k L.
\end{equation}

\subsubsection{Radiation's constant quantum radial length $L$} \label{Sec-Radiation-constant-L}

Assume that $G_k$ grows at a constant rate.  $L$ is proportional to a quantum of energy  and  a quantum of time.   

Paraphrasing Newton's statement about time in his \textit{Principia}, absolute, true, and mathematical (idealized) \textit{radiation}, in and of itself and of its own nature, without reference to anything external, flows uniformly. $L$'s constancy is also similar to Einstein's postulate of definite (constant) velocity, independent of the motion of the emitting body, with which light propagates in empty space (1905).

\subsubsection{Distinguishing radiation}

When in empty space a sphere is scaled up in size, each radius of the sphere increases by the same scale factor. If the radial length of a radiation cone grows at a constant rate of $L$ per radiation event, the radial length does not scale (or even change) from one cone volume increment to the next. This distinguishes a growing radiation cone from a growing space sphere.


\subsection{A volume with radiation scales differently than one without radiation} \label{Sec-Diff-Scaling-Rad-Sp}

\subsubsection{Modeling a volume in empty isotropic space by a sphere}

The universe is isotropic to better than one part in 100,000 (Fixsen, 1996). Idealize this observation for empty  space at all scales.  Let $\Theta_k$ be a spherical volume representative of isotropic empty  space. For scale factor $\theta_k >1$ and $\ell_k$ a radius of $\Theta_k$, 
\begin{equation}
\frac{4}{3} \pi (\ell_{k+1})^3  = \Theta_{k+1}=  \theta_k \Theta_k = \frac{4}{3} \pi [(\theta_k)^{1/3}\ell_k]^3.
\end{equation}
The scale factor for each radius $\ell_k$ of $\Theta_k$ leading to $\Theta_{k+1}$ is $(\theta_k)^{1/3}$: $(\theta_k)^{1/3} \ell_k = \ell_{k+1}$.

Scaling up $\Theta_k$ results in a $\Theta_{k+1}$ that is a larger volume than $\Theta_k$ but has not expanded the size of the space in which $\Theta_k$ and $\Theta_{k+1}$ reside; the size of empty eventless space has not changed. 

\subsubsection{What causes space to expand}

To empty space  add radiation that grows by $L$ from every radiation point source. That could expand space. Mathematics that follows is consistent with that inference. 

\subsubsection{Scaling  average cross-sectional area in static space}

Imagine a space sphere $\Theta_k$ and a radiation cone volume increment $V_k$ side by side in space and that each has in the same plane average cross-sectional area $A_k$. If the scale factor for $A_k$ is $a_k$, the scale factor for every radius of $A_k$ both for $\Theta_k$ and  $V_k$ is $(a_k)^{1/2}$.

The scale factor for each $\Theta_k$ radius on scaling $\theta \Theta_k = \Theta_{k+1}$ is $(\theta_k)^{1/3}$. For $A_{k+1}$ in  $\Theta_{k+1}$ the scale factor that applies to $A_k$ is  
\begin{equation}
a_k= [(\theta_k)^{1/3}]^2 =(\theta_k)^{2/3}.
\end{equation}

Similarly, absent a radiation event, for volume $V_k$
\begin{equation} \label{Eq-SF-v-relp-SF-a}
a_k= [(v_k)^{1/3}]^2 =(v_k)^{2/3}.
\end{equation}

\subsubsection{Scale factor relationship for area and volume after a radiation event} \label{Contrast-SF-Rad-vs-Space-v-a}

Compare $V_{k}$'s $a_k$ scale factor \textit{after} a radiation event. Scale $V_k$ so that $v_k V_k = V_{k+1} = A_{k+1} L$, increasing $G_k$'s radial length from $kL$ to $(k+1)L$.   Then
\begin{equation} \label{Eq-equality-v-k-and-a-k}
v_k = \frac{V_{k+1}}{V_{k}} = \frac{A_{k+1}L}{A_{k}L} = a_k.
\end{equation}
unlike equation (\ref{Eq-SF-v-relp-SF-a})'s $(v_k)^{2/3}= a_k$ \textit{absent} a radiation event. 

Radial length orthogonal to the plane of the cross-sectional areas scales differently for  $\Theta_k$ (space) and  $V_k$ (a radiation cone increment), that is,   $V_k$ scales differently when there is a radiation event.  

\subsubsection{$G_k$ in radiation-space and space reference frames}

A radiation cone $G_k$ has features of both radiation and space. $G_k$'s radial length grows by $L$ on every radiation event. At an instant a radiation cone volume increment  has a measurable volume $V_k=A_kL$. $G_k$ inhabits a union of radiation and space, radiation-space. Radiation-space is analogous to the term space-time, with radiation corresponding to time.  

Radiation-space is not space because space is empty and unchanging, while radiation-space contains radiation.  Radiation-space is not only radiation, because radiation has only one dimension by analogy to the dimensions of space. One dimensional radiation by itself does not have a space length, volume or area. A radiation event manifests itself in radiation-space as a quantum radiation length. Radiation-space and space scale differently in the direction orthogonal to the cross-sectional areas of a radiation cone and space sphere. On a radiation event, space's 3 dimensions are augmented by a single radiation dimension. Since radiation is absent from and independent of empty space, its reference frame  is a fourth dimension orthogonal to space's three dimensions. Radiation added to space is denoted as 4 dimensional radiation-space.

\subsection{An algebra of degrees of freedom: $Deg$ and $deg$}

\subsubsection{Degrees of freedom as a concept}

Planck wrote that $W$ in equation (\ref{Eq-Planck-Boltzmann-log-Ent-Eq-100}), is ``always an integer'', a thermodynamic probability (p. 120, Planck), a combinatorial expression  (p. 124, Planck) of the number of possible choices. In modern terminology $\log W$ represents degrees of freedom relative to the base of the logarithm. In his \textit{Mechanics}, Arnold Sommerfeld  (1952, p. 1) remarked that ``the concept of degrees of freedom is not too well known.'' It is of central importance to entropy.

\subsubsection{Reasons for a degrees of freedom notation}

By compressing the degrees of freedom concept using notation, less time is required to manipulate aspects of degrees of freedom; more energy is available to focus on  the 4/3 inconsistency problem. 

\subsubsection{Degrees of freedom notation}

Use $Deg$ as a function and $deg$  as units as part of an algebra of degrees of freedom. 

Let $\ell_k$ be an arbitrary radial length with  scale factor base $\beta >1$ defined \textit{in general} as 
\begin{equation} 
\ell_{k+1} / \ell_k= \beta.
\end{equation} 

Define $Deg$, $Deg^*$  and $Deg^{**}$ as follows: 
\begin{equation} \label{Eq-DegDefn-1-deg}
\begin{split}
Deg_\ell^*(\ell_k) &\equiv Deg_\ell(\ell_{k+1} / \ell_k) \\ 
&= Deg_\ell(\beta\ell_{k} / \ell_k) \\
& = \log_\beta(\beta)  \ deg_\ell \\
&= 1 \ deg_\ell. 
\end{split}
\end{equation} 
\begin{equation}  \label{Eq-DegDefn-kdeg}
Deg_\ell^{**}(\ell_k) \equiv  Deg_\ell( \ell_{k+1} / \ell_1) = Deg_\ell(\beta^k \ell_1 / \ell_1) = k \ deg_\ell.
\end{equation}
A one asterisk $Deg$ finds the number of scalings of  $\ell_{k+1}$ relative to $\ell_{k}$. A two asterisk $Deg$ finds the number of scalings of $\ell_{k+1}$ relative to $\ell_{1}$. 

The argument of $Deg$ must be a fraction to give the number of scalings that have been applied to the unit (proportional to an amount of energy) in the numerator. Scaling involves \textit{multiplying} the parameter by a scale factor; to reveal a scale factor form a fraction (divide). The numerator and the denominator in $Deg$'s argument must contain  parameters as factors that cancel each other leaving the scale factor.  The subscript for $Deg$ specifies to which parameter in space --- length, area or volume --- the scaling is applied. If the subscript is $L$ or $\ell$, the scale factor of interest is the one that applies to a length. 

$Deg$ is an attribute of energy or radiation. Scaling due to a radiation event adds energy. Scaling  empty space does not change its dimension or energy content, only its size relative to the whole of unchanging space. 

\subsubsection{$Deg^{**}$ as entropy of gas molecules}

Using an ensemble of gas molecules with mean path length $\mu$ as an example
\begin{equation}
Deg_\mu^{**} (\mu_k) = Deg_\mu (\mu^k \mu/ \mu) =k \deg_\mu
\end{equation}
gives the entropy of the gas molecules relative to $\mu$.  $Deg^{**}$, by measuring the number of scalings, measures the entropy of the energy of the system relative to a scale factor. If the scale factor is a quantum or average scale factor, we may think of the $Deg^{**}$ of a system as its \textit{intrinsic entropy}.  

\subsection{Dimension and $deg$ in space} \label{Sec-Dim-and-deg}

\subsubsection{$Dim$ applied to lengths, areas and volumes in space} \label{Sec-Dim-of-L-A-V}

Let $Dim$  be a dimension function for space  analogous to the $Deg$ function. By convention, dimension is determined with reference to length: a line is one dimensional, an area two dimensional, a volume three dimensional.   Make the relativity of dimension explicit by using a  length, area or volume subscript for $Dim$. 

Dimensions form equivalences classes in empty space, the classes of all lines or $[\ell_k]$, all areas $[A_k]$ and all volumes  $[\Theta_k]$.  For a scale factor $\beta$ for $\ell_k$, $m>k$, and discarding for simplicity's sake varieties of area and volume shapes,
\begin{equation} \label{Eq-Dim-ell-in-Deg-and-deg}
\begin{split} 
Dim_\ell [\ell_k] & =[Deg_\ell(\beta^m \ell_k / \ell_k)] \\
& =Deg_\ell(\beta \ell_k / \ell_k) \\ 
&= Deg_\ell^*( \ell_k) \\
&= 1 \ deg_\ell. \\
\end{split}
\end{equation}
\begin{equation} \label{Eq-Dim-A-in-Deg-and-deg}
\begin{split}
Dim_\ell  [A_k] &=[Deg_\ell((\beta^m \ell_k)^2 / (\ell_k)^2)] \\
&=Deg_\ell((\beta \ell_k)^2 / (\ell_k)^2) \\
&= Deg_\ell^*(A_k) \\
&= 2 \ deg_\ell.
\end{split}
\end{equation}
\begin{equation} \label{Eq-Dim-Theta-in-Deg-and-deg}
\begin{split}
Dim_\ell [\Theta_k] &=[Deg_\ell((\beta^m \ell_k)^3 / (\ell_k)^3)] \\
&=Deg_\ell((\beta \ell_k)^3 / (\ell_k)^3) \\
& = Deg_\ell^*(\Theta_k) \\
& = 3 \ deg_\ell.
\end{split}
\end{equation}

\noindent Equations (\ref{Eq-Dim-ell-in-Deg-and-deg}), (\ref{Eq-Dim-A-in-Deg-and-deg}) and (\ref{Eq-Dim-Theta-in-Deg-and-deg}) express  the convention that motion along a line has 1 degree of freedom, motion in an area has 2 degrees of freedom and motion in a volume has 3 degrees of freedom. The $deg$ terminology is used for $Dim$ both for  convenience and because $Deg$ is   a generalization of observations about the degrees of freedom inherent in one, two and three dimensions.   Under any number of scalings of a line, area or volume their $Dim$'s are invariant.

$Dim$ would treat 4 dimensions  analogously to the way it treats 3 dimensions.

\subsubsection{Subscripts of $Deg$, $deg$ and $Dim$}

Subscripts appended to $Deg$, $deg$ and $Dim$ avoid ambiguity. The value of $Deg^*(\ell_k)$ depends on what the subscript of $Deg$ is: $Deg_\ell^*(\ell_k) = 1 \ deg_\ell$ but $Deg_\Theta^*(\ell_k) = 1/3 \ deg_\Theta$. Ratios and  additions involving  $Deg$ and  $Dim$  require subscripts of the same dimension. When there is no ambiguity, the subscripts can be dispensed with. 

Subscripts $L$ and $V$ signify a radiation-space reference frame. Subscripts $\ell$ and $\Theta$ signify a space reference frame.  

\subsubsection{Subscripts of scale factors}

Subscripts can be required for scale factors used for $L$ and $\ell$ if their scale factors  change from one radiation cone volume increment to the next. As the radiation cone $G_k$ grows,  constant radial increment $L$ changes relative to increasing $V_k$; while $A_k$'s area increases with $k$, $L$'s length is constant.

\subsubsection{Length, area and volume $Dim$ relationship}

From section \ref{Sec-Dim-of-L-A-V}

\begin{equation} \label{Eq-ell-deg-in-A-and-Theta}
Dim_\ell [\ell_k] = 1 \ deg_\ell = \frac{1}{2}Dim_\ell [A_k]  = \frac{1}{3}Dim_\ell [\Theta].
\end{equation}

If in  equation (\ref{Eq-ell-deg-in-A-and-Theta})  we substitute the subscript $\Theta$, then $Dim_\Theta^*(\ell_k)= (1/3) \ Dim_\Theta(\Theta)$, so $Dim_\Theta^*(\ell_k) :Dim_\Theta(\Theta) = 1/3 : 1 \equiv 1: 3$, the same as for $Dim_\ell [\ell_k] : Dim_\ell [\Theta]$ in equation (\ref{Eq-ell-deg-in-A-and-Theta}). 

Describe space with three orthogonal axes, $X_1$, $X_2$ and $X_3$. Then for a volume in space $\Theta = x_1x_2x_3$,
\begin{equation}  \label{Eq-Orth-Dim-Add}
\begin{split}
Dim_\ell(\Theta) & = Dim_\ell(x_1x_2x_3) \\
& = Dim_\ell (x_1) + Dim_\ell(x_2) +Dim_\ell(x_3). 
\end{split}
\end{equation}
The dimension of a system is the sum of the dimensions of its orthogonal components. This observation extends to a system with 4 (or more) orthogonal components. 

\subsection{Reference frame hypothesis} \label{Sec-Main-Ref-Fr-Hyp}

\subsubsection{$Deg$ in radiation and  space, of spheres and cubes} \label{Sec-Deg-applied-L,A,V}

In radiation-space's reference frame, let a quantum length be $L$. In space's reference frame, let a quantum length be $\ell$,  an area $A_k$ and a  volume $\Theta_k$. The scale factor for $A_k$, in both radiation-space and space, is  $a_k$ and for $\Theta_k$, $\theta_k$. 

For empty space, think of $\Theta_k$ as a spherical volume which scales the same along its three orthogonal axes. Discarding the constants 4/3 and $\pi$ when calculating the $deg$ of the ratio of two volumes, one based on $L$ and the other based on $\ell$, gives the same $deg$ result as for cubes based on $L$ and  $\ell$: 
\begin{equation} \label{Eq-ratio-Spheres-Cubes-equal-Isotropic}
Deg_L\left( \frac{(4/3) \pi (kL)^3}{ (4/3) \pi (k \ell)^3}\right) = Deg_L \left(\frac{L^3}{\ell^3}\right). 
\end{equation}
To calculate $Deg_L$ on the right side of  equation (\ref{Eq-ratio-Spheres-Cubes-equal-Isotropic}) it is necessary to express  one of $\ell$ and $L$ in terms of the other. To accomplish that (among other reasons), form the following hypothesis.

\subsubsection{Hypothesis:}

\begin{equation} \label{Eq-4-3-L-ell-Hyp}
\ell = \phi L.  
\end{equation}

If $L$ is constant, then so is $\ell = \phi L$; in space $\ell$ is a quantum length orthogonal to the plane of the average cross-sectional area of $V_k$.  $L$ in a radiation-space reference frame has image $\ell$ in a space reference frame, a reference frame hypothesis that emerges from and is required by the mathematics.

\subsection{Radiation-space and space as distinct reference frames}

\subsubsection{For radiation-space, $Deg_V^*(V_k)=Deg_A^*(A_k)$ on a radiation event} \label{Sec-DegV=DegA}

Section \ref{Contrast-SF-Rad-vs-Space-v-a} showed that for a radiation cone, for a radiation event $v_k = a_k$. Using $Deg$ notation,
\begin{equation} \label{Eq-rad-sp-Deg-V-comp-A}
Deg_A^*(A_k) = Deg_V^*(A_k) = Deg_V^*(V_k).
\end{equation}
If in equation (\ref{Eq-equality-v-k-and-a-k}) we substitute $\ell $ for $L$, for a radiation event we get the  $Deg$ relationship between volume and area:
\begin{equation} \label{Eq-Deg-A-rel-V-same-rad-and-space}
Deg_\Theta^*(A_k) = Deg_\Theta^*(V_k).
\end{equation}

\subsubsection{Different scaling radially}

Unlike equation (\ref{Eq-rad-sp-Deg-V-comp-A}), in empty space a radial length orthogonal to the cross-sectional area scales by a 1/3 power of the volume's scale factor and $Deg_\Theta^*(A_k) = (2/3)Deg_\Theta^*(V_k)$. 

\subsubsection{For a radiation event, $Deg_L^*(L) = 0 \ deg_L$} \label{Sec-Deg-ell=0}

Since $L$ is constant, for a radiation event, with (nominal) scale factor $x=1$, 
\begin{equation} \label{Eq-Deg-L-is-0}
Deg_L^*(L) = Deg_\ell(xL/L) = \log_x(x) = \log_x(1) = 0
\end{equation}
and, since $\ell$ is constant if $L$ is,
\begin{equation}
Deg_\ell^*(\ell) = \log_x(1) = 0.
\end{equation}

On the other hand, in empty space $Dim_\ell(\ell) =1 \ deg_\ell$. 

It follows from equation (\ref{Eq-Deg-L-is-0}) that applying $L$ as a scale factor to $L$ as a quantum length  on a radiation event
\begin{equation} 
Deg_L^*(LL) = Deg_L(x^2LL/LL) = \log_x(x^2) = \log_x(1) = 0.
\end{equation}

\subsubsection{Orthogonality of radiation and space}

Space cannot grow absent radiation. A radiation cone in space must grow when its radial length increases incrementally by length $L$. $Deg^{**}$ as a function helps characterize radiation and radiation-space; $Dim$ as a function  helps characterize space. 

This suggests that  radiation-space and space form distinct reference frames. Compare $Deg$ and $Dim$ for the   radiation-space radiation cone which has volume in space but grows by radiation events.

\subsection{4/3 $Dim$ of  radiation-space compared to space, and The 4/3 RDFL} \label{Sec-4-3-Scaling-derived}

\subsubsection{4/3  $Dim$ of $V_k$ in radiation-space}

Suppose a radiation event extends $G_k$'s radial length by $L$. $V_k$'s scale factor is $v_k$, $A_k$'s scale factor is $a_k$ and, by Section \ref{Subsec-Eqatg-Ent-eqns-SF-as-unit},  $L$'s scale factor is $L$.  Then
\begin{equation}
 V_{k+1}= v_kV_k = (a_kL) V_k  = (a_kA_k) (L L).
\end{equation}

Evaluate $Deg_V^*(V_k) = Deg_V(V_{k+1}/ V_k)$ on the radiation event. Since $Deg_V(L^2)=0$,
\begin{equation} \label{Eq-Deg-Vk-is-4-3}
\begin{split}
Deg_V^*(V_k) & = Deg_V\left(\frac{a_kA_kLL}{A_kL}\right) \\
& = Deg_A^*(A_k)  +Deg_V^*(L) \\
& = Deg_V^*(A_k) +0 \\
& = Deg_V^*(A_k)
\end{split}
\end{equation}
as in section \ref{Sec-DegV=DegA}.

Dimensionally in radiation-space, on a radiation event,
\begin{equation} \label{Eq-Dim-Vk-is-4-3}
\begin{split}
Dim_V( [V_{k+1} ] ) & = Dim_V( (a_kL) V_k ) \\
& = Dim_V( [a_kA_kLL] ) \\
&= Dim_V( [a_kA_kL ] )+ Dim_V([L] ) \\ 
& = Dim_V([V_k])+ (1/3) Dim_V([V_k]) \\
& =(4/3) Dim_V([V_k]).
\end{split}
\end{equation}

In equation (\ref{Eq-Dim-Vk-is-4-3})
\begin{itemize}
\item Absent a radiation event, $Dim([V_{k+1}])=Dim([V_k])$ and $Dim([A_{k+1}])=Dim([A_k])$. 
\item On a radiation event $Dim([L]) \neq Dim([LL])= Dim([L]) + Dim([L])$, since $L$ is both a length  and length $L$'s scale factor. Hence $Dim_V([a_kA_kL])+ Dim_V([L]) =Dim_V([a_kA])+ Dim_V([L])+ Dim_V([L])= (2/3 + 1/3 + 1/3) \ deg_V$.
\item A radiation event (scaling $L$ by $L$) adds a fourth (orthogonal) dimension to the three dimensions $Dim_\ell(V_k)$ of space. 
\end{itemize}

Subscript  $V$ denotes the radiation-space reference frame and  subscript $\Theta$ denotes the space reference frame. Compare $V_k$'s dimension on a radiation event to its dimension in space without one, described as
\begin{equation} \label{Eq-4-3-Dim-Ratio-V-RS-to-S}
\frac{Dim_V( [V_{k}])}{Dim_\Theta( [V_{k} ])} = \frac{(4/3) \ deg_V}{1 \ deg_\Theta},
\end{equation}
or, described in another way,
\begin{equation} \label{Eq-4-3-Ratio-Using-Single-Dim}
Dim \left( \frac {[V_{k}]_V } { [V_{k} ]_\Theta } \right)= \frac{(4/3) \ deg_V}{1 \ deg_\Theta}.
\end{equation}
Because of the relationship of $Dim$ and $Deg^*$, equation (\ref{Eq-4-3-Dim-Ratio-V-RS-to-S}) implies that in a space reference frame
\begin{equation} \label{Eq-4-3-Deg-Ratio-V-RS-to-S}
\frac{Deg_V^*( V_{k})}{Deg_\Theta^*(V_{k})} = \frac{(4/3) \ deg_V}{1 \ deg_\Theta}.
\end{equation}
From a dimensional point of view, $deg_V$ is equivalent to $deg_\Theta$.

\subsubsection{The 4/3 RDFL}

Equation (\ref{Eq-4-3-Dim-Ratio-V-RS-to-S})  implies that for a single radiation event, for the dimensional equivalence classes $[V_k]$, $[L]$  and $[\ell]$, with subscripts $L$ denoting the radiation-space reference frame and $\ell$ denoting the space reference frame,
\begin{equation} \label{Eq-1-3-Power-of-V-1-RS-to-S}
\begin{split}
\frac{Dim_V([V_{k}]^{1/3})}{Dim_\Theta( [V_{k}]^{1/3})} & = \frac{Dim_V([L])}{Dim_\Theta([ \ell] )}  \\
& = \frac{Dim_L([L])}{Dim_\ell([ \ell] )}  \\
& = \frac{(1/3)(4/3) \ deg_V}{(1/3)1 \ deg_\Theta} \\
& = \frac{(4/3) \ deg_L}{1 \ deg_\ell}.
\end{split}
\end{equation}
Changing the subscript in the second expression in equation (\ref{Eq-1-3-Power-of-V-1-RS-to-S}) from a volume to a length in both numerator and denominator does not change the ratio.  The fourth expression in equation (\ref{Eq-1-3-Power-of-V-1-RS-to-S}) is equivalent to observing that 
\begin{equation} 
(v^{4/3})^{1/3} = (v^{1/3})^{4/3} 
\end{equation}
and that $v^{1/3}$ is the scale factor for length $L$ in a space reference frame, that is, from a dimensional point of view. From a dimensional point of view, $deg_L = deg_\ell$; each of  $L$ and $\ell$ is a measure of length, $L$ in radiation-space, and $\ell$  in space.

From section \ref{Sec-Dim-of-L-A-V}, for dimensional equivalence classes $Dim=Deg^*$. Thus in equation (\ref{Eq-1-3-Power-of-V-1-RS-to-S}) the equality of the third  and fifth (last) expressions implies that
\begin{equation} \label{Eq-The-4-3-RLL}
\frac{Dim_L([L])}{Dim_\ell([ \ell] )}= \frac{(4/3) \ deg_L}{1 \ deg_\ell},
\end{equation} 
a quantum radial length  has 4/3 the degrees of freedom of its image in isotropic empty space, which is The 4/3 Ratio of Degrees of Freedom Law.

Having regard for the equivalence of entropy $S$ and $Deg^*$ on a single radiation event, for $S$ entropy
\begin{equation} \label{Eq-The-4-3-RLL-in -S-terms}
\frac{Dim_L([L])}{Dim_\ell([ \ell] )}= \frac{(4/3) \ deg_L}{1 \ deg_L}=\frac{S_L}{S_\ell},
\end{equation} 
which implies
\begin{equation} \label{Eq-4-3-relp-L-ell-in-S}
S_L = (4/3) S_\ell.
\end{equation}

\subsubsection{Back to the 4/3 dimension problem}

Section \ref{Subsec-St-Dim-problem} described the inconsistency $Dim(V) =(4/3)Dim(V)$. Reasons for the  inconsistency in  equation (\ref{Subsec-St-Dim-problem}) are that
\begin{itemize}
	\item $Dim(A) \neq Dim (V)$; area has 2 dimensions and a volume has 3 dimensions. Treating $Dim(A)$ as equal to $Dim(V)$ and adding a 1/3 dimension leads to an inconsistency.  
	\item  Only on a radiation event is $Deg_V^*(A) = Deg_V^* (V)$; for $Deg$ and not for $Dim$.
	\item On a radiation event, $Dim (V) \neq Dim(AL)$ but rather $Dim ([V]) = Dim([aA][LL])$; the dimension of radiation-space is 4 compared to the 3 dimensions of space.
\end{itemize}

\subsubsection{On deriving The 4/3 RDFL}

The 4/3 RDFL describes the $deg$ ratio of a radiation quantum length $L$ to a space quantum length $\ell$. To arrive at The 4/3 RDFL involves first finding the $deg$ ratio of a volume $V_k$ in a radiation-space reference frame compared to a space reference frame. Why must we proceed via a radiation-space volume to derive a law about quantum lengths?

The 4/3 ratio of $deg$ is a consequence of radiation-space having 4 dimensions compared to space's 3 dimensions. To reveal the  4 dimensions  of radiation-space requires a volume with features of both radiation and space, namely a volume in radiation-space undergoing a radiation event. To reveal the law requires  distinguishing between $Deg$ and $Dim$ to resolve the inconsistency described in section \ref{Subsec-St-Dim-problem}, facilitated by defining radiation cone volume increments and subscripted $Deg$, $deg$, and $Dim$.

\subsubsection{On two reference frames and 3/4 metabolic scaling}

From biological measurements, metabolism is observed to increase by a 3/4 power of an organism's mass. The circulatory system distributes energy at an approximately constant rate, analogous to a constant rate of radiation events. Since the rate at which metabolism scales up is less than the rate at which mass scales up, that suggests that an organism's radiation distribution system  (analogous to radiation-space) is distinct from the system (analogous to space) that receives the energy because they scale distinctly.

\subsubsection{4/3 scaling ratio of radial lengths} \label{Sec-4-3-scaling-due-Radial-length-Diff}

Radiation-space is modeled by a radiation cone, space by a sphere. Radial lengths orthogonal to the cross-sectional area scale differently. The scaling relationship between a radius and a cross-sectional area is the same for radiation-space and space reference frames. That suggests the 4/3 ratio of scaling arises because $\ell = \phi L$.

\subsection{The 4/3 Ratio of Lengths Law (The 4/3 RLL)}

\subsection{Deriving The 4/3 RLL from The 4/3 RDFL}

Section \ref{Sec-Main-Ref-Fr-Hyp} hypothesizes that $\ell= \phi L$. Applying equation (\ref{Eq-Clausius-equals-Boltzmann-Eq-100}), for $S$ entropy, and using energy $E$ instead of $Q$
\begin{equation} \label{Eq-L-entropy-relp100}
S_L = \frac{E}{L}
\end{equation}
and
\begin{equation} \label{Eq-ell-entropy-relp100}
S_\ell = \frac{E}{\ell}
\end{equation}
since $L$ and $\ell$ are quantum lengths. $E$ is the same in equations (\ref{Eq-L-entropy-relp100}) and (\ref{Eq-ell-entropy-relp100}) since $\ell$ is $L$'s image in space's reference frame.  From equations (\ref{Eq-L-entropy-relp100}) and (\ref{Eq-ell-entropy-relp100}), since from equation (\ref{Eq-4-3-relp-L-ell-in-S}) $S_l = (4/3)S_\ell$, 
\begin{equation} \label{Eq-L-S-ell-S-200}
S_\ell \ell =  S_\ell \phi L = S_L L = (4/3) S_\ell L. 
\end{equation}
From  equation (\ref{Eq-L-S-ell-S-200}) $S_\ell \ell = (4/3) S_\ell L $ implies $\ell = (4/3) L$. That is The 4/3 Ratio of Lengths Law. 

\subsubsection{$V_k$ in radiation-space and space reference frames}

The 4/3 ratio of $Deg$ for $V_k$ in its radiation-space reference frame compared to it in a space reference frame is entirely due to $\ell = 4/3 L$. As noted in section \ref{Sec-4-3-scaling-due-Radial-length-Diff}, the different scaling of radiation-space and space occurs only in the radial direction. The scaling of $V_k$ arises because $\ell=(4/3)L$. This can be shown as follows.

Denote the volume $V_k$ in a radiation-space reference frame as $(V_k)_V$. Denote the volume $V_k$ in a space reference frame as $(V_k)_\Theta$.
In a radiation-space reference frame
\begin{equation} \label{Eq-Vk-vol-in-Rad-RF}
(V_k)_V = A_kL.
\end{equation}

In a space reference frame 
\begin{equation} \label{Eq-Vk-vol-in-Space-RF}
(V_{k})_\Theta = A_k \ell= A_k (4/3)L = (4/3)(A_k L) = (4/3)(V_k)_V.
\end{equation}
Equation (\ref{Eq-Vk-vol-in-Space-RF}) shows that $V_k$'s radial length $\ell$ in a space reference frame is 4/3 longer than its radial length in a radiation-space reference frame, and so  $V_k$'s  volume seems 4/3 larger in the space reference frame compared to its volume in radiation-space. 

From a dimensional point of view, the scaling of the cross-sectional area does not play a role; in a space reference frame for the cross-sectional areas of a cone and of a sphere the scale factor of a radius is $a^{1/2}$. The different scaling of radial lengths for radiation-space compared to space probably connects to special relativity, in which only motion in the radial direction causes a length to contract. 

\subsubsection{Space distance stretches 4/3 relative to radiation distance at all scales}

Space stretches relative to the radiation reference frame. Moreover, the stretching is per scaling event, so that the stretching is scale invariant. That is,
\begin{equation}
k \ell = k [(4/3) L] = (4/3) [kL].
\end{equation}

Space therefore appears stretched by 4/3 relative to a corresponding light distance; light appears to go 3/4 as far as one expects compared to stretched space.  Homogeneously scaled radiation-space and its corresponding isotropic empty space provide two distinct reference frames related via The 4/3 RLL.

\subsection{Proposed applications of the foregoing theory}

The 4/3 RDFL and The 4/3 RLL imply each other --- are logically equivalent. The two 4/3 laws and  their related concepts may help explain various phenomena, as sketched below.

\subsection{Natural logarithm theorems}

\subsection{The natural logarithm as an intrinsic scale factor}

The foregoing implies that the natural logarithm is a base for energy scale factors. Ways to show that include: 

\subsubsection{Scale factor equals unit}

From section \ref{Subsec-Eqatg-Ent-eqns-SF-as-unit},  $T$ as unit changes by $T$ as scale factor: 
\begin{equation}
\frac{dT}{dt}=T,
\end{equation}
exactly as the exponential function $e^x$ is its own derivative relative to time $t$. 

\subsubsection{Network proof}

Section \ref{Subsec-Eqatg-Ent-eqns-SF-as-unit} also shows that entropy $S=k$ multiplies the capacity of an individual unit. The contribution to the multiplication of capacity per radiation node of a system with $n=\mu^k$ nodes, that is analogous to radiation-space, as a proportion of $S$ is
\begin{equation}\label{Eq SM-Ec 300.20}
\begin{split}
	 \frac{d S}{dn}&=\frac{ d \left[\log_{\mu}\mu^{k}\right]} {d(\mu^{k})}\\
&=\frac{1}{\ln\left(\mu\right){\mu^{k}}}.
\end{split}
\end{equation}
If an equal number of nodes (or sub-volumes) in the receiving system (corresponding to space) receive the radiated energy, the per node reception of their increase in capacity as a proportion of $S=k$ is $1 / n = 1 / \mu^k$. The receiving network receives the same amount of energy (assuming no energy losses) as the transmitting network, so 
\begin{equation}\label{Eq SM-Ec 400.20}
\frac{1}{\mu^{k}} =\frac{1}{\ln\left(\mu\right){\mu^{k}}}\Rightarrow \ln(\mu)=1\Rightarrow \mu = e.
\end{equation}

\subsubsection{Mean path length spans a network}

A mean path length $\mu$ spans a network of $n$ nodes. Also, $\log_\mu(n)=k$ times $\mu$ spans the network, a result arising from $\mu$'s dual role as unit and base of the network's scale factor. Per network radiation event
\begin{equation} \label{Eq-NatLogThm-using-ds-dS}
	\frac{d\mu}{d S} = S \times \mu
\end{equation}
\noindent which implies $\mu=e^S$, which for $S=1$ implies $\mu=e$.

\subsection{Degrees of freedom}

\subsubsection{Ergodic hypothesis}

To derive the $H$ \textit{Theorem}, Boltzmann assumed that every combinatorial state (phase) was equally likely and would occur eventually --- the ergodic hypothesis. ``This argument is wrong, because this infinite long time must be much longer than $O(e^N)$, while the usual observation time is $O(1)$'' (Shang-Keng Ma, p. 442). If the network is scaled by its mean path length $\mu$ in time proportional to $\log_e(O(e^N))=O(1)$, the ergodic hypothesis can be justified. Section \ref{Subsec-Eqatg-Ent-eqns-SF-as-unit}  shows that the base of the scale factor is $\mu$ and that the number of scalings is $\log_\mu(\mu^k)$, a linear relationship. The number of degrees of freedom relative to $\mu$ is linear. Every combinatorial state (phase) is equally available. 

\subsubsection{Many worlds hypothesis}

Hugh Everett  (1957) discussed a `many worlds' interpretation of quantum mechanics. Available degrees of freedom can replace `many worlds'.

\subsubsection{Black hole entropy}

$Deg(V)= Deg(A)$ on a radiation event is  similar to the  observation that entropy of the volume of a black hole is proportional to the entropy of 1/4 of its surface area (Bekenstein 1972,  Hawking 1975); the surface area is $ 4 \pi r^2$, and $\pi r^2$ can represent the average cross-sectional area of a radiation cone.

\subsection{Phenomena with 4/3 fractal dimension}

\subsubsection{Stefan's Law}

Stefan's Law (also known as the Stefan-Boltzmann Law) was  derived by Boltzmann (1884). Using Planck's notation (p. 62, Planck 1914), a step in the proof is 
\begin{equation}\label{Eq PlanckOnStefansLaw100}
\left( \frac{\partial S}{\partial V} \right)_T = \frac{4u}{3T}.
\end{equation}
$\partial S$ represents change in entropy, $\partial V$ represents change in volume, subscript $T$  that the equation applies at a given temperature, and $u/T$ the number of degrees of freedom in volume density of radiation $u$ (p. 61, Planck 1914) relative to a degree of absolute temperature.

Rewrite  equation (\ref{Eq PlanckOnStefansLaw100}) as
\begin{equation}\label{Eq PlanckOnStefansLaw200}
\partial S = \left(\frac{4}{3}\right) \frac{\partial Vu}{T};
\end{equation}
$\partial Vu$ is the total energy in the chamber, like $Q$ in equation (\ref{Eq-Clausius-equals-Boltzmann-Eq-100}). Using the ideas in section \ref{Subsec-Eqatg-Ent-eqns-SF-as-unit},  equation (\ref{Eq PlanckOnStefansLaw200}) says that in a cavity, when a radiation volume (analogous to the radiation cone increment $V_k$ above) grows due to a radiation event, the change in its entropy --- degrees of freedom --- relative to the change in volume scales by $4/3$ of  energy  relative to  scale factor $T$. Which is The 4/3 RDFL.

\subsubsection{Brownian motion}

The mean path length $\mu$ for random Brownian motion of particles on the surface of water is analogous to the constant quantum length $L$ of radiation-space. The water surface is analogous to the space reference frame. Based on The 4/3 RDFL, the fractal dimension of Brownian motion relative to its medium should be 4/3. In 2001, Lawler, Schramm and Werner (Lawler 2001, 2004) found that the fractal dimension of Brownian motion is 4/3 based on the difficult mathematics of  stochastic L\"owner evolution (Kager 2004).

\subsubsection{3/4 metabolic scaling}

Max Kleiber (1932) concluded that metabolism $Y =M^{3/4}$, where $M$ is organism mass. West, Brown and Enquist (1997 WBE) proposed a fractal explanation of the exponent 3/4 based on scaling but 1997 WBE has an internal inconsistency (Kozlowski 2004, 2005). 

The 4/3 RDFL can  explain 3/4 metabolic scaling. It must be applied twice. 

First, a scaled tubular circulatory system is analogous to  radiation space. The cross-sectional areas of same-sized tubes at a level in the circulatory system are  analogous to the average cross-sectional area of a $V_k$. Tube volumes  scale like $V_k$. The rate of blood flow is (approximately) constant through all tube levels, analogous to constant radiation distributed through successive levels of $V_k$. The same amount of blood flows through the cumulative cross-sectional area of a given level in the circulatory system. The organism's mass, analogous to space, is proportional to size. Accordingly an organism's circulatory system together with the fluid pumping through it has 4/3 the dimension of its mass.

Second, as a result of intracellular Brownian motion and the operation of a Brownian ratchet, intracellular distribution of energy (analogous to radiation-space) has  4/3 more $deg$ than  intracellular fluid (analogous to space). Thus the energy supply capacity of the circulatory system for a larger organism increases by a 4/3 power at the circulatory system level, and that increase is maintained at the cellular level: more energy supplied to a cell by a larger circulatory system increases $deg$ at the cellular level also by a 4/3 power. 

For circulatory system $Circ$, $Deg_V^*(Circ_{k}) = (4/3) \ deg$. If for metabolism $r$, $Deg_\Theta^*(r_{k}) = (3/4) \ deg$ then
\begin{equation} \label{Eq-3/4MetScaling100.100}
\begin{split}
Deg_V^*(Circ_{k}) \times  Deg_\Theta^*(r_k) & = (4/3) \ deg \times (3/4) \ deg \\
& = 1 \ deg^2.
	\end{split}
\end{equation}
Since the 4/3 $Deg$ relationship is based on dimension, subscripts $V$ and $\Theta$ in equation (\ref{Eq-3/4MetScaling100.100}) can be treated as  equivalent.

For organisms the product of the $Deg$ of energy supply capacity times the  $Deg$ of the rate of metabolism is invariant for all $k$, that is, for all sizes of organism.  If the biochemistry of cells is uniform or invariant for differently sized animals, then the temperature of the cells \textit{in organisms with similar biochemistries} should lie in a narrow range.

If cellular metabolism did not scale down  when an animal's size scaled up, the extra 1/3 power (4/3 compared to 1) of energy supplied by the circulatory system compared to that of  cells receiving energy would  overheat the organism's cells. Overheated cells would impair intracellular operation; the organism could not survive. Metabolism scales by a 3/4 power of size to prevent cells from overheating.

\subsubsection{MIMO}

In a multiple input multiple output (MIM0) antenna system such as a cell phone network, for the number of antennas at each node $M>1$, if `channel matrices are non-degenerate then the precise degrees of freedom $\eta^*_X = \frac{4}{3}M$' (Jafar and Shamai 2008), a result that appears to be analogous to the degrees of freedom ratio for radiation-space compared to space. 

\subsubsection{Cosmological distance factors for radiation and matter}

Cosmology reasons (p. 64, Ryden 2003; p. 17, Wang 2010)  that for distance scale factor $a$ the energy density of matter 
\begin{equation} \label{Eq-Cosmo-Rho-m-a3-100.100}
\rho_M \propto \frac{1}{a^3}
\end{equation}
\noindent and the energy density of radiation,
\begin{equation} \label{Eq-Cosmo-Rho-r-a4-100.100}
\rho_\gamma \propto \frac{1}{a^4}.
\end{equation}
Treat the matter scale factor $a^3$ as pertaining to space. Then the ratio of degrees of freedom for radiation (in radiation-space) compared to matter (space) is $4:3$ or $4/3: 1$, consistent with The 4/3 RDFL. 

\subsection{Phenomena exhibiting the 4/3 lengths ratio}

\subsubsection{Mean path length ratio for gas molecules}

Clausius in an 1858 paper (p. 140 in Brush) remarks about molecules in a gas:
\begin{quotation}
\noindent The mean lengths of path for the two cases (1) where the remaining molecules move with the same velocity as the one watched, and (2) where they are at rest, bear the proportion of $\frac{3}{4}$ to $1$. It would not be difficult to prove the correctness of this relation; it is, however, unnecessary for us to devote our time to it.
\end{quotation}
He gave a geometrical proof in 1860 (p. 434). 

An ensemble of $n$ gas molecules all moving with the same average velocity correspond to a radiation space with quantum radial length $L$. Stationary gas molecules correspond to space divided into $n$ sub-volumes with centers separated by quantum length $\ell$. Clausius used trigonometry and calculus to reach his conclusion, a result of The 4/3 RLL.
 
\subsubsection{Expansion of cosmological space}

The universe is not only expanding (Hubble 1929), it is accelerating (Riess et al. 1998; Perlmutter et al. 1999). The unknown cause of the acceleration has been called dark energy (Turner 1998).

Suppose light from a standard candle $SC$ with intrinsic brightness $B$ travels from $SC_1$  to Earth distance $d_1$ in the radiation-space reference frame. Suppose $SC_2$ with the same intrinsic brightness $B$ travels in  the radiation-space reference frame $2d_1$ to Earth. In the space reference frame, The 4/3 RLL predicts that $SC_2$ is 
\begin{equation}
\frac{B}{(4/3)d_1} = \frac{(3/4)B}{d_1}, 
\end{equation}
\noindent 3/4 as bright, or 25\% fainter, relative to $SC_1$, than it would be in the radiation-space reference frame. 1998 observations found that Type Ia supernovae appeared ``about 25\% fainter, that is, farther away than expected'' (Dark Energy Survey, 2013; p. 259, Cheng 2010) consistent with The 4/3 RLL.

\subsubsection{Energy density of dark energy}

Predict the ratio of  energy densities in radiation-space compared to space:  $\rho_\gamma: \rho_M$, having regard to the observations in section \ref{Sec-Deg-applied-L,A,V}  and $\ell=(4/3) L$, for energy $E$:

 \begin{equation} \label{Eq-ratioEnDensities100.100}
\frac{E /L^3}{E /\ell^3} =  \frac{E /L^3}{E /[(4/3)L]^3} = \frac{64}{27} :1 \approx \frac{0.7033}{0.2967} : 1.
\end{equation}
Treat  energy density of matter $\rho_M$ as equivalent to  energy density of space. $\rho_\Lambda$ denotes the energy density of the vacuum (or the cosmological constant).

Cosmology assumes that $\rho_\Lambda +\rho_\gamma +\rho_M =\rho_c$, the critical density. Dividing through by $\rho_c$, with $\Omega_X = \rho_X / \rho_c$, $\Omega_\Lambda +\Omega_\gamma +\Omega_M =1$. Since the measured value   $8.4 \times 10^{-5}$ of $\rho_\gamma$   is very small compared to $\rho_\Lambda$ and $\rho_M$, $\Omega_\Lambda +\Omega_M \approx 1$. Equation (\ref{Eq-ratioEnDensities100.100}) predicts
 \begin{equation} \label{Eq-ratioEnDensities200.200-1}
\frac{\Omega_\Lambda}{\Omega_M }  \approx \frac{0.7033}{0.2967} : 1.
\end{equation}

The Seven Year Wilkinson Microwave Anisotropy Probe (WMAP) (Jarosik et al. 2010), measured (p. 2)  dark energy density
\begin{equation} 
\Omega_{\Lambda}=0.728^{+0.015}_{-0.016},
\end{equation}
with a $68\%$ confidence limit (also Komatsu et al. 2011). 

The Planck satellite in March 2013  (p. 11) measured dark energy density
\begin{equation}  
\Omega_{\Lambda}=0.686 \pm{0.020},
\end{equation}
 with a $68\%$ confidence limit. 

The average of the WMAP and Planck measurements for $\Omega_\Lambda$, 0.707, is about one half per cent different than the 0.7033 in equation (\ref{Eq-ratioEnDensities100.100}) predicted by The 4/3 RLT. The WMAP and Planck measurements differ more with each other than they do with the value for $\Omega_\Lambda$ predicted by The 4/3 RLT. The 4/3 RLL may help explain dark energy.

\subsubsection{Hubble time and distance}

Let  $t$ be  time, $R(t)$ the radius of the universe, and $H(t)$ Hubble's constant. 
The 4/3 RLL implies
\begin{equation} \label{EqHubbleReason100.300-2}
	\frac{\dot{R}(t)}{R(t)} = \frac{4/3}{(4/3)t} = \frac{1}{t} =H(t).
\end{equation}

Classic cosmological solutions (p. 37, Liddle 2003) suggest that the elapsed age of the universe $t_0$ for the current matter dominated universe is 
\begin{equation} \label{EqAgeUni100.200}
t_0  =2/3 \times 1/H_0 
\end{equation}

\noindent (p. 37, Coles 2002; p. 40, Weinberg 2008). Equation (\ref{EqAgeUni100.200}) implies an age for the universe of 
\begin{equation} \label{EqAgeUni100.300}
t_0 = (6.5 - 10) \times 10^9 h^{-1} \; \; \mathrm{years}
\end{equation}
\noindent  (p. 84, Coles 2002), $h\equiv H_0/100 \mathrm{km} \mathrm{s}^{-1} \mathrm{Mpc}^{-1}$ where $h=0.67$ (p. 11, Planck Collaboration 2013), leading to an underestimate of the age of the universe compared to the current estimate of about 13.7 billion years (p. 84, Coles 2002). Equation (\ref{EqHubbleReason100.300-2}) avoids the factor 2/3 in equation (\ref{EqAgeUni100.200}) that leads to the underestimate.

\subsubsection{Mean path lengths in networks} \label{Subsec-measured-network-MPL's}

The Natural Logarithm Theorems predict that a network analogous to an ensemble of moving gas molecules or radiation-space will have a mean path length between a pair of nodes (the average number of collisions or steps to connect pairs of molecules), equal to $e \approx 2.71828$ steps. The 4/3 RLL predicts that a network  receiving information analogous to a stationary background or space will have a mean path length $(4/3)e \approx 3.624$ steps.

The mean path length for English words has been measured as 2.67 (Ferrer i Cancho 2001), for neurons in \textit{C. elegans} as 2.65 (Watts \& Strogatz 1998) and for the neurons of the human brain as 2.49 (Achard et al. 2006), all close to the value of $e$. The 4/3 RLL applies if these are isotropically transmitting networks. 

The mean path length for a network of 225,226 actors (Watts \& Strogatz 1998) was measured as 3.65, less than one per cent different than 3.624 predicted above for a network receiving information.

\subsection{Degrees of freedom as capacity multiplier}

\subsubsection{Network rate theorem}

Use a network's mean path length $\mu$ in equation (\ref{Eq-log-W-mult-energy-unit}) so:
\begin{equation} \label{Eq-log-Network-mu-unit}
Q = \log_\mu(\mu^k) \mu.
\end{equation}
$Q$ supplies the energy for the network's output. Assume there is no loss of energy on transmission of energy to the output network. For time $t$ compare network output rates $Q/t$ to nodal or individual rates $\mu/t$ by dividing equation (\ref{Eq-log-Network-mu-unit}) through by $t$:
\begin{equation} \label{Eq-log-Network-rate-per-mu}
\frac{Q}{t} = \log_\mu(\mu^k)\frac{\mu}{t}= \frac{\log_\mu(\mu^k)}{t} \mu.
\end{equation}
$Q/t$ gives the network's rate of output for some parameter proportional to $Q$. 

$C$, the clustering coefficient, is the average proportion of nodes one step away from given nodes that connect to those given nodes. Since for an actual network not all nodes are connected to the same degree, the network rate equation must reflect that. Network entropy $S=C\log_\mu(\mu^k)$. Modify equation (\ref{Eq-log-Network-rate-per-mu}) as follows: 
\begin{equation} \label{Eq-log-Network-rate-per-mu-w-C}
\frac{Q}{t} = C\log_\mu(\mu^k)\frac{\mu}{t}= \frac{C\log_\mu(\mu^k)}{t} \mu.
\end{equation}

If data is available for two of the three rates described in equation (\ref{Eq-log-Network-rate-per-mu}), that enables calculation of the third rate. For example, if $Q/t$ and $C \log_\mu(\mu^k) / t$ are known, it is possible to calculate $\mu /t$.

\subsubsection{Problem solving rates}

A society's problem solving capacity depends on the number of problem solvers and on the store of already solved problems, such as theorems, problem-solving techniques and other knowledge. From equation (\ref{Eq-log-Network-mu-unit}), a networked population of size $n$ multiplies individual capacity by its network entropy, $C\log(n)$, with its mean path length as the base of its scale factor and logarithmic function. The same observation applies to the network of solved problem resources. Section \ref{Subsec-measured-network-MPL's} shows that measured mean path lengths for networks are consistent with what theory predicts. Assume measured values are good estimates of $C$.

\renewcommand{\arraystretch}{1.1}
\begin{table}  
\begin{center}
\footnotesize
	\begin{tabular}{|c|c|c|c|c|c|c|}\hline
	\textbf{Network} & Nodes  & Number of nodes & $\mu$ & \textit{C} & $S$ & Notes\\ \hline
1657 English & words  & 200,000 & 2.67 & 0.437 & 5.431 & {\footnotesize\ 3, 4}\\ \hline
1989 English & words & 616,500 & 2.67 & 0.437 & 5.932 & {\footnotesize\ 2, 3}\\ \hline
1657 population & people & 5,281,347 &  3.65 & 0.79 & 9.445 & {\footnotesize\ 1, 6}\\ \hline
1989 population & people & 350,000,000 &  3.65 & 0.79 & 12.0 & {\footnotesize\ 1, 5}\\ \hline
	\end{tabular}
	\caption{Calculations of network entropy, $S$} \label{Table 1}
	\end{center}

\small
\begin{center}
\noindent \textbf{Notes to Table} \ref{Table 1} \\
	\end{center}
\footnotesize
\begin{flushleft}
1.\ $\mu$ and $C$ for people based on an actors network study, Watts and Strogatz (1998).\\
2.\ The number of words: OED. \\
3.\ $\mu$ and $C$: Ferrer, 2001 based on about 3/4 of the million words appearing in the British National Corpus. \\
4.\ The number of words: EMEDD.\\
5.\ The number is estimated for 1989 by adding  censuses: 1990  USA, 248.7 million; 1991 Canada 27,296,859; 1991 England 50,748,000; 1991 Australia, 16,850,540. These total 343,595,000 people.\\
6.\ The number of people in England: Table 7.8, following p. 207, for the year 1656, in Wrigley, 1989. 
\end{flushleft}
\end{table}
\normalsize

The rate of English lexical growth can be used as a proxy for society's collective rate of problem solving. All of society is involved in solving problems of how to form and choose phonemes, words, grammatical rules and hierarchical language structures and how to increase the efficiency of a language. The average English lexical growth rate from 1657 to 1989 is 3.39\% per decade. From 1,750 BCE to  1992  lighting efficiency increased about 345,800 times (Nordhaus), an average of 3.41\% per decade. The rate of increase of average IQs is about 3.00 to 3.63 IQ points per decade  (p. 113  and Table 1 at p. 180, Flynn 2007), consistent with these rates. 

Use Table \ref{Table 1} to calculate $C\log_\mu(n)$ for $n$ the population of English speakers and for $n$ the number of English words at 1657 and 1989. For 1657 to 1989 calculate the average English lexicon entropy $S_{lex}$,  and average English population entropy $S_{pop}$:
\begin{equation} 
S_{lex}=C_{lex}[\log_{\mu{_lex}}([\mu_{lex}]^k)]_{av}
\end{equation}
and
\begin{equation} 
S_{pop}=C_{pop}[\log_{\mu-pop}([\mu_{pop}]^k)]_{av}.
\end{equation}
The average rate of individual problem solving $\mu_{ps} /t$ is calculated as follows: 
\begin{equation} 
3.41\% \ \textrm{per decade} = S_{lex} S_{pop} \frac{\mu_{ps}}{t}.
\end{equation}
This gives $\mu_{ps}/t$ of 5.56\% per thousand years, an estimate of the innate individual problem solving capacity (problem solving rate) of  the physiologically average human brain from 1657 to 1989. The average rate of collective problem solving  is 61.28 times the rate of average individual problem solving.

From another perspective, $S_{pop} \mu_{ps}$ is the problem solving capacity multiplier for society's collective brain, which has access to society's entire network of ideas which has entropy $S_{lex}$.  

\subsubsection{Glottochronology and problem solving rates}

Glottochronology, conceived by Morris Swadesh, estimates the date of origin of a mother language based on the rate of divergence between two daughter languages. He estimated that cognates --- words that sound similar in daughter languages with a common mother language --- diverge by an average of about 14\% per thousand years per thousand years (p. 280, Swadesh) for Indo-European.  In 2003, 37 years after Swadesh's estimate,  a computer-assisted study estimated  ancestral Indo-European language began 8,700 years before, about 6,700 BCE (Gray 2003). Modernizing Swadesh's estimate  $7037/8700 \times 14\% = 11.32\%$ per 1000 years, half of which is 5.66\%, very close to the 5.56\% estimated using entropy. The two results corroborate each other. Each daughter language diverges from the historical mother language at the same average rate; if the entropies of the lexicon and population for the daughter populations are approximately equal only the  rate of average individual problem solving plays a role in the lexical divergence of daughter languages from the common mother language.

\subsection{Two reference frames}

\subsubsection{Aspects of two reference frames}

Aspects of the two reference frames that might apply to the phenomena mentioned in this section are
\begin{itemize}
	\item The two reference frames exist at all scales, since $k \ell = (4/3) k L$ for all $k$.
	\item 4/3 stretching  separates parts of space, but no  stretching occurs in radiation-space. 
\end{itemize}

\subsubsection{Space's scale factor}

If theory limits itself to one reference frame, then for space to expand in an accelerating way the expansion scale factor $a$ must be a function of time $t$; it is  $a(t)$. With two reference frames theory, space  stretches per radiation event and cumulatively by $t(4/3)$; stretching is scale invariant. 

\subsection{Radiation-space and space-time}

Radiation occurs in a one dimensional reference frame at a constant rate, The radiation reference frame is orthogonal to three dimensional space. Minkowski in 1908 declared in a talk in Cologne that space itself and time itself would henceforth only exist as a union of the two. Physics calls it space-time. Radiation-space is perhaps a better name.

\subsubsection{Radiation, time and reference frames}

Radiation events occur at a constant rate in the radiation-space reference frame. If time is proportional to the distance a quantum of light energy travels in both radiation-space and space, there can only be an absolute time in  radiation-space. Space  stretching per radiation event destroys the one to one relationship between distance $L$ and time $t$ that exists in the radiation-reference frame. If time is absolute in the  radiation-space reference frame, then no matter what the inertial or accelerating reference frame in space is,   light speed in radiation-space will always be the same. That inference is consistent with the assumptions of special relativity.

\subsubsection{Radiation and time's arrow}

Thomas Gold speculated that ``The large scale motion of the universe thus appears to be responsible for time's arrow'' which he thought might be deduced ``from an observation of the small scale effects only'' (Gold 1962). A quantum radiation event, a small scale effect,  expands space. The effect is scale invariant and so applies to ``large scale motion of the universe''. Time's arrow exists because light radiation occurs in one direction in radiation-space. 

\subsubsection{Quantum entanglement}

Perhaps particles  too far apart in space are not too far apart in radiation-space.

\subsubsection{The two slit experiment}

Perhaps  particles go through one of two slits they illuminate the space reference frame background that expands (and is not stationary) with each radiation event. The two slit experiment may reveal cosmological expansion of the background reference frame at the quantum level.

Compare the  stationary earth-bound reference frame of the Ptolemaic solar system around which the sun rotated. Aristarchus and Copernicus supposed that instead the earth reference frame rotated on its axis and revolved around the stationary sun. Reversing denotation of the stationary reference frame simplified the model. 

\subsubsection{Cosmological horizon problem}

The cosmological horizon problem asks how parts of the universe too far apart for light to traverse seem observably similar. This seems to be the cosmological analog of the problem of quantum entanglement. At all scales, too far apart in space  is not too far apart  in radiation-space, perhaps. 

\subsubsection{Cosmological inflation}

Radiation events occur at a steady rate in radiation-space. What is not steady, though, is the proportional increase in radius of the universe in the space reference frame. One million new radiation events that occur in a universe only one million radiation events old doubles the size of the universe. One million radiation events that occur in a universe $10^{20}$ radiation events old does not increase the rent in nearly the same proportion. Near the beginning of time space would appear to be inflating, even though radiation events occur at the same constant rate. Cosmological inflation may be inconsistent with constant light speed, which causes the space reference frame to expand. 

\subsubsection{Bondi's $k$ calculus}

Hermann Bondi explained special relativity using a scaling $k$ calculus (Bondi 1980) which is consistent with the scaling ideas sketched above.  

\subsubsection{Push or pull}

Radiation appears to cause radiation-space to expand. Once space is larger, one may ask if that requires radiation to continue to fill the expanded space created. 

\subsubsection{Emergence at all scales}

The 4/3 dimensions of radiation-space relative to space increase the degrees of freedom of space. Degrees of freedom multiply output capacity. Thus, in a space-like system, degrees of freedom increase, and the increase of degrees of freedom   increases output capacity (rate). . With the increasing energy resources (and degrees of freedom) supplied, and the effect of degrees of freedom on a system's energy capacity, these two effects combine to create conditions for newly emergent phenomena at all scales. 

Entropy is maximal when all possibilities are equal, an implication of Jensen's inequality (Jensen 1905) and Jaynes's maximum entropy principle (1957). 

\subsubsection{Economies of scale}

The scaling principles that permit larger organisms to reduce cellular metabolic rates likely also permit business entities to operate more efficiently as they get bigger. 

\subsubsection{The nature of energy}

When a radiation-space reference frame  growing by $L$ changes by 1 $deg$, a space reference frame growing by $\ell$ changes by 1 $deg$. The energy for $L$ and $\ell$ is the same. The $deg$ of space is 4/3 the $deg$ of radiation-space, when measured in terms of $L$. It is the 4/3 ratio which causes space to expand and which physicists have named dark energy. If the 4/3 ratio applies at all scales from quantum to cosmological, infer (possibly) that what we call energy is the 4/3 ratio of the degrees of freedom of radiation-space compared to space.









\subsection{Conclusions}

The 4/3 ratios of scaling and of lengths require two natural reference frames, related by $\ell=(4/3)L$.  Natural phenomena that appear to conform to the 4/3 ratios support the existence of those 4/3 ratios as general laws of nature. Some phenomena consistent with the 4/3 ratios have been identified in this article. There are likely many others. The greatest impediment in the way of accepting the existence of two reference frames, radiation-space and space, and their real world effects is likely an unwillingness to depart from  ideas  taken for granted.

\section{Background of ideas leading to two reference frames}

The Overview makes observations about degrees of freedom of space compared to radiation-space. 

Identification of assumptions logically necessary for a mathematical model can help isolate the model's necessary logical conditions. In the development of the ideas leading to The 4/3 RLT, two related approaches have been used. One approach is to suppose the existence of two systems, one an energy distribution system $S$ ($S$ for Supply) and the other an energy receiving system $\mathpzc{R}$ ($\mathpzc{R}$ for Receipt ). A second (related) approach is to observe that there are two reference frames, one for the distribution of radiated units, and the other a passive or stationary system that receives the radiated units. In the reference frame approach, the same amount of energy results in a  characteristic length $L$ in the radiation-space reference frame and a characteristic length $\ell$ in the space reference frame, with $L= (3/4)\ell$. 

This section sketches the development of these two points of view and their related concepts. The reference frame point of view has advantages of simplicity. Each point of view highlights different issues.

\subsection{Increasing average IQs}

The role of scaling in the ideas leading to a mathematical model of dark energy began with a seemingly unrelated question: why do average IQs increase? 

A history of mathematics shows that its ideas improve. Likewise, studies in historical linguistics show that language becomes more energy efficient over time. Words and phrases are truncated and condensed, short forms and contractions are adopted. Newly coined words stand for new concepts and allude, in an information-compressing way, to whole eras, events, and technologies.  

The linguist Otto Jespersen observed (p. 324, Jespersen 1922) that language develops with regard to energetics: speakers of a language seek the most efficient means for transmitting and receiving information. Zipf  (1949) explored Jespersen's observation about language efficiency at some length. 

The development of software data compression suggests a modern  analogy to language. Just as software engineers improve the compression of information transmission, so too society collectively improves the compression of information transmission by improving the energy efficiency of spoken and written language. 

These observations suggest an explanation for increasing average IQs.   Since about 1970 average IQs have been observed to increase at about the rate of 3.3\% per decade. I hypothesized in 2005 that this was due to language increasing the compression of information represented by words, phrases, and so on, in all the hierarchies of a language's structure. That would enable human beings as problem solvers to deploy --- to wield --- more compressed, and therefore more efficient, language tools  and concepts in solving problems. 

To test this hypothesis required comparing the rate of increasing average IQs to some collective rate involving words. A proxy for word inventiveness is the size of a lexicon. I found in August 2005 that the size of the English lexicon since about 1657 had increased by an average of about 3.3\% per decade. Thus there was some basis for supposing that improving average IQs might be attributable to improvements in the efficiency of language. 

One way to test the hypothesized connection between increasing average IQs and increasing information compression in language is to find other collective problem solving rates. Over the course of the next 4 years or so, I found various collective problem solving rates close to 3.3\% per decade, such as, in terms of its labor cost, the rate of increase in lighting efficiency (p. 33, Cowles Paper, Nordhaus 1997). As mentioned in the Overview, from 1,750 BCE to  1992  lighting efficiency increased about 345,800 times (Nordhaus), an average of 3.41\% per decade. 

Another way is, if possible, to compare 3.3\% per decade to other measures of lexical growth. Glottochronology is such a measure. 

Glottochronology  was devised by Morris Swadesh (Swadesh 1972) to measure the rate at which two daughter languages diverged from a common mother tongue. By studying the historical written usage of words, he could estimate the rate at which daughter languages diverged from their mother tongue. Around 1966 he calculated that the rate of divergence of Indo-European languages from each other was an average of about 14\% per thousand years. More recent work dating Indo-European  (Gray \& Atkinson 2003) gives a basis to revise Swadesh's divergence rate to about 11.2\% per thousand years, which is equivalent to each language diverging from its ancestral language at the rate of about 5.6\% per thousand years. 

A third approach is to attempt to fashion a formula that measures the rate of lexical growth and test whether it is consistent with observation. Since glottochronology and lexicons both exhibit a rate of change,  some formula might connect them.

The  5.6\%   per thousand years rate of change of each daughter language differs from the 3.3\% per decade average rate of English lexical growth. Might the two rates be related by a formula for lexical growth? Is the 5.6\%  per thousand years divergence rate a kind of `fossil rate' embedded in the 3.3\% per decade average rate of English lexical growth? Glottochronology helps date the age of the mother tongue. Might a `fossil rate' be used to date the beginning of language? One can estimate what the rate of growth in the size of a lexicon might be by supposing that language grew from one word  to the 616,500 words of the Oxford English Dictionary in 1989, and calculating what the rate would be depending on the start date.  These questions arose beginning in September 2005.

\subsection{Lexical scaling, a logarithmic formula}

Naive reasoning suggests that a formula for lexical growth is logarithmic: parents in each generation transmit information to their children. If information was solely acquired in this way, the accumulation of information would be based on the number of generations, $\log_2(2^k)=k$, where the log is calculated using base 2. The base of the logarithm, and of the  scale factor, would be 2. But information is acquired both from contemporary peers and others. What, if the formula is logarithmic, is the appropriate base of the logarithm and of scale factors?  The quest for a number for the base of the logarithm began about September 2005. 

At the end of May 2007 calculations suggested that the mean path length $\mu$ of a social network worked as the base of the logarithm. For calculations for real networks, $\log_\mu(n)$ is multiplied by the network's clustering coefficient, $C$ to give a formula $C\log_\mu(n)$.  $C$ measures the average connectedness of nodes to adjacent nodes (Watts \& Strogatz 1998). If all adjacent nodes connect, then  $C=1$. Since in a social network on average nodes do not connect to all adjacent nodes, $0<C<1$. The appearance of the formula $C\log_\mu(n)$ resembles the appearance of the log formula for entropy. That suggests that the growth of a lexical network is connected to principles of thermodynamics and statistical mechanics.

\subsection{The Network Rate Theorem (The NRT)}

The mean path length of a social network (of actors) was measured in 1998 as  3.65 (Watts and Strogatz 1998). The value of $C$ was also measured then. 

The  formula $C \log_\mu(n) \equiv \eta$ leads to a rate theorem, derived in The Overview, here called The Network Rate Theorem,
\begin{equation} \label{Eq-eta-fr-LexGr-Clogn}
r_n=C\log_\mu(n) r_1,
\end{equation} 
that calculates the multiplicative effect of networking on the rate of individual problem solving capacity without regard for the additional effect provided by language. From another perspective, it allows one to calculate collective capacity when individual capacity is known, and vice versa. In equation (\ref{Eq-eta-fr-LexGr-Clogn}), $r_n$ is the collective rate of problem solving, $C$ is the network's clustering coefficient, $\log_\mu(n)$ is the log of the number $n$ of nodes in the network to base $\mu$, $\mu$ is the network's mean path length, and $r_1$ is the average problem solving capacity of an individual (a node) in the network. 

The problem solving capacity of a human society $r_{(P-S)}$ which uses a language for communication is 
\begin{equation} \label{Eq-Coll-Prob-Solvg-Incl-Lang}
r_{(P-S)} = C_{Soc}\log_{\mu_{Soc}}(n_{Soc}) r_1 \times C_{Lex}\log_{\mu_{Lex}}(n_{Lex}).
\end{equation}

In equation (\ref{Eq-Coll-Prob-Solvg-Incl-Lang}) the degrees of freedom of social interaction $C_{Soc}\log_{\mu_{Soc}}(n_{Soc}) = \eta_{Soc}$ (the Greek letter eta standing for entropy) is multiplied by the degrees of problem solving freedom afforded by words (concepts), $C_{Lex}\log_{\mu_{Lex}}(n_{Lex}) = \eta_{Lex}$, giving the total problem solving degrees of freedom available to the society. Both $\eta_{Soc}$ and $\eta_{Lex}$ can be calculated using data on network sizes, and on measured values of $\mu$ and $C$, for populations and for lexicons. Then for $r_{(P-S)}$ and $r_1$, if one is known the other can be calculated. 

Since there are measurements for $C$ and $\mu$ for social networks and for the English language, it is possible to calculate the individual rate of problem solving based on a collective problem solving rate of about 3.41\% per decade. The calculated rate for $r_1$ is almost exactly 5.6\% per thousand years. With hindsight, this follows from Swadesh's approach. If as a result of collective problem solving each of two languages grows by an average 5.6\% per thousand years, then their rate of divergence from each other should be twice that,  an average 11.2 \% per thousand years.

Applying 5.6\% per thousand years to the 616,500 words of the OED gives an estimate of the start date for language of about 150,000 years ago. Experts in language have expressed doubt about glottochronology's reliability in estimating the era for Indo-European, about 8,700 years ago, because spoken language leaves no physical artifacts. So an estimate that goes much further back in time to  the start of language seems even more doubtful. On the other hand, a formula that could validly estimate the beginning of spoken language  suggests it is a powerful mathematical characterization of network change. 

Having regard for reservations in June 2007 about the validity of the formula in equation (\ref{Eq-eta-fr-LexGr-Clogn}), a convincing  physical model leading to the formula would increase its plausibility. 

\subsection{A physical model for The NRT} \label{Sec-Phys-basis-for-NRT}

The average amount of information per mean path length $\mu$ is proportional to $\mu$, to the mean energy $\mathpzc{E}(\mu)$ used to create or transmit that information, and to the mean time $t(\mu)$ it takes to travel the mean path length. The proportionality of average information, length, energy, and time explains why $\mu$ works well as a scale factor. Scaling $n$ by $\mu$ is equivalent to scaling a network's energy $\mathpzc{E}(\Theta)$  per time unit by $\mathpzc{E}(\mu)$.

A network of size $n$ with mean path length $\mu$ can be spanned, for some $k$, by $\log_\mu(n) =k$ lengths  each   $\mu$ steps long. This implies that $n= \mu^k$. Since $\mu$ is the mean path length, the mean \textit{connection distance} between nodes is $\mu$ steps. If the rate per $\mu$, or per $\mu^k$ for $k=1$ is known, then the rate per single degree of freedom of $\mu^1$ nodes or units of energy, length, or time is known. It takes $k$ steps (or degrees of freedom) of length $\mu$ to span the network, $k$ units of time where each unit of time spans a single $\mu$, and $k$ units of energy where each unit of energy spans a single $\mu$. In other words, if each generation is $\mu$ times bigger than the preceding one (starting with a generation of size $\mu$), there are $\log_\mu(\mu^k)=k$ such generations. 

\subsection{$\eta$ and nested clusters}

The function $\log_mu(n)$ can be thought of as the entropy $\eta$ of a system, and the function $C\log_mu(n)$ can be thought of as a network's entropy $\eta$.

Attempting to fashion a model for equation (\ref{Eq-eta-fr-LexGr-Clogn}) and for $\eta$ raises various problems, including: 

\begin{itemize}
	\item The counting problem: if each of the $n$ nodes in a network has $\mu+\mu^2 + \dots +\mu^k$ sources of information where $\mu^k=n$ (that is, $\mu^k$ sources in  generations 1 to $k$) there would be more information sources than nodes. 
	\item The $n-1$ problem: how can one out of $n$ nodes receive a multiplicative capacity benefit with $n$ as the argument in $\log_\mu(n)$; that seems to require a node transmitting information to itself.  
	\item 	A spanning problem: it seems that $\log_\mu(n) \times \mu$ is required to span the network, consistent with Jensen's inequality, but a mean path length $\mu$ is sufficient to span the network by itself. It cannot be that both $\log_\mu(n) \times \mu$ and $\mu$ are the most efficient ways to span the network.
\end{itemize}

These problems are resolved if scaling occurs in a nested hierarchy. For example if there is a single cluster of $\mu^k$ nodes, one scaling of the single cluster by $\mu$ results in $\mu$ clusters each with $\mu^{k-1}$ nodes, and so on. So if there are $\mu^k$  energy units, that cluster of energy divide into $\mu$ clusters each with $1/ \mu$ as much energy.  If there are $k$ scalings, then each node can be considered to be simultaneously in $k$ differently sized clusters, and within all $k$ generations. The counting problem is resolved because clusters of size $\mu^k$ reside in a next larger size cluster of size $\mu^{k+1}$. The $n-1$  problem is resolved because a node resides in a cluster; the network benefit the node receives is a result of a node being in $k$ levels of clusters or having $k$ degrees of freedom. The spanning problem is resolved because $\log_\mu(n)=k$ measures degrees of freedom, not a cumulative distance. 

If nested scaling validly models $\log_\mu(n)$, there should be  physical systems that manifest this nested hierarchical structure. Each of the $k$ different clusters in which a node can appear in one of the node's degrees of freedom. Nested scaling of a network is one way to visualize its degrees of freedom. 

To better understand the model, it is helpful to identify its key components and therefore to consider a possible  set of postulates. In 2008, those initial postulates were:

\begin{itemize}
	\item A network \textbf{N} exists. Its nodes require energy, are connectable and can transmit and receive benefits. 
	\item Every node in \textbf{N} can respond to its environment and will minimize its use of energy for the acquisition of each unit of benefit received from the environment or from another node in \textbf{N}, and will maximize the benefits it receives for each unit of energy it expends. 
\end{itemize}

These two postulates assume that the network  isotropically distributes information, which is proportional to energy. 

Another reason to look for postulates from which the features of the model can be derived is as an aid in identifying  material underlying physical conditions. That can help understand what makes the system work, and permit the observer to discard data irrelevant to the material conditions. 

\subsection{Jensen's inequality} \label{Sec-Jensens-Ineq}

Shannon (1949) observed in his information theory monograph that entropy would be maximal if the probability $p_i$ were equal for every scaling. The same applies for $\log(n)$; it is maximal for a network if the base of the logarithm is an average, in the case of a network, the mean path length. If every node has the same scale factor, then the entropy is maximal; energy distribution is isotropic. Hence, Shannon's observation is related to Jensen's inequalities about convex and concave curves (Jensen 1905). Jaynes (1957) called Shannon's observation  the maximum entropy principle. 

The maximum entropy principle is related to the evolution of organisms: an organism that is more efficient through its maximization of entropy has an energy efficiency advantage compared an organism that does not maximize entropy.

\subsection{The Network Rate Theorem (The NRT)}

For $d(\Theta_k)$ representing a cumulation of $\ell_1$ lengths,
 
\begin{equation} \label{Eq-Dist-Theta-NRT-deriv.100}
\begin{split}
d(\Theta_k) & = \ell_1 + \beta\ell_1 + \ldots + \beta^{k-1}\ell_1 \\ &
= k\ell_1  \\
&= log_\beta(\beta^k) \ell_1  \\
&= Deg^{**}(\Theta_k) \ell_1. 
\end{split}
\end{equation}
This is also discussed in connection with equation  (\ref{Eq-Dist-Theta-100.100}) below. 

The most efficient spanning of a system $\Theta_k$ in units equal to the mean path length (or its proportional mean amount of energy or time) is expressed by equation (\ref{Eq-Dist-Theta-NRT-deriv.100}). If we use a unit $X$ larger than the mean path length $\ell= \mu$, then $kX$ will be longer than $k \ell$. If we use a unit shorter than $\mu$, the number of scalings, relative to that unit, will increase. Jensen's inequality (Jensen 1905) tells us that the most efficient way to span the distance  $d(\Theta_k)$ with given units is $k \times \mu$, that is, to traverse the networks in units equal to the mean (or average) path length. 

For a continuous convex function $\phi$ (a different use of The Greek letter $\phi$ than in $\ell = \phi L$), Jensen's inequality \cite{Jensen1905} is
\begin{equation}\label{EqJensen100.100}
\phi \left( \frac{\sum p_i x}{\sum p_i} \right) \leq \frac{\sum p_i \phi(x)}{\sum p_i}, 
\end{equation}

\noindent where $\sum p = 1$. If the $p_i$ are equal, $\phi's$ value on the left side of equation (\ref{EqJensen100.100}) equals $\phi's$ value on the right side and so is maximal. If the function $\phi = \log$, then the formula in Jensen's inequality is the formula for entropy. Stipulating that the $p_i$ are equal is equivalent to requiring homogeneous scaling (such as scaling by a constant  by $\Theta_k$'s mean path length) and to maximizing entropy. 

This aspect of Jensen's inequality leads to the apparently paradoxical, but resolvable, observation mentioned above in section \ref{Sec-Phys-basis-for-NRT} in connection with traversing a network. Jensen's inequality and equation (\ref{Eq-Dist-Theta-NRT-deriv.100}) imply that the most efficient way to traverse $d(\Theta_k)$ is by $k$ lengths (or iterations) of $\mu$. But $\mu$ as the mean path length traverses (on average) the entire length of $d(\Theta_k)$ and is much shorter than $d(\Theta_k)$ in equation (\ref{Eq-Dist-Theta-NRT-deriv.100}). These can't be both true. The resolution is to think of $k \mu$ not as $k$ lengths, but as $k$ degrees of freedom of $\mu$. Each node has $Deg^{**}(\Theta_k)$ degrees of freedom. 

In the early stages of The NRT the situation was characterized not by using degrees of freedom but by uniformly nested and scaled clusters; uniform scaling meant each node was scaled by the same scale factor. $Deg$ was characterized as entropy. Degrees of freedom is in some ways a better characterization of scaling than entropy.

If real world phenomena can be modelled by degrees of freedom manifested as nested scaling, one would expect there to be many instances of hierarchical nested uniform scaling. Hierarchical nested scaling occurs in a wide variety of phenomena, as has long been recognized (for example (Simon 1962). Degrees of freedom, or nestedness, may apply to Everett's 1957 many worlds hypothesis (Everett 1957) and to the problems raised by the ergodic hypothesis (p. 442, Ma 2000). 

Now for an example applying $C\log_\mu(n)$. 

Suppose a society has $n$ problem solvers with an average individual problem solving rate of $r_1$ per brain. Then the collective problem solving capacity of the society's brains is $r_n = C \log_\mu(n) r_1 $.   $ C \log_\mu(n)$ gives the multiplier effect of networking for problem solving brain capacity. The mean path length and clustering coefficient have been measured for various networks of known size so the effect of networking can be calculated by inserting $C$, $\mu$ and $n$ into the formula in equation (\ref{Eq-eta-fr-LexGr-Clogn}). If the collective rate $r_n$ is known, then the individual rate $r_1$ can be calculated. 

The problem solving capacity of an individual brain is increased by language. Neurons or synapses network in an individual brain. Individual brains can network via language and other signs. The significance of language appears to be that it multiplies a society's and an individual's  problem solving degrees of freedom. If a society has a collective problem solving capacity or rate of $r_n$ without language, then we multiply $r_n$  by the degrees of freedom afforded by ideas or, as representative of ideas, by the degrees of freedom of the lexicon. In this view, the enormous enhancement of human problem solving capacity is achieved mostly as a result of collective effort;  language collectively invented by societies multiply problem solving degrees of freedom attributable to neurons. Written language enables a larger portion of previously solved problems to be contemporaneously disseminated and to be preserved for posterity, and increases posterity's problem solving capacities.

One of the most striking results of using The NRT is the connection between glottochronology and the average rate of increase in IQs, mentioned above. This connection encouraged persistence in efforts to validate the formula $C \log_\mu(n)$ from June of 2007 to August of 2009, when an explanation of the connection came into view.

Not only does nature manage a $4/3$ stretching of space using an algebra of degrees of freedom, using the same principles it creates the basis for a way to mathematically model emergence. 

In this way, The NRT can be derived based on the ideas of the algebra of degrees of freedom connected to The 4/3 RDFT.  

\subsection{A natural logarithm theorem}

The dual role of the mean path length $\mu$ in a network as a scale factor and as the units scaled in a cluster of size $\mu^k$ implies that a cluster of size $\mu$ to a power  changes at a rate proportional to $\mu$.  In a network the cumulative length  $k\mu$ is determined by $\log_\mu(\mu^k) =k$; $k\mu$  spans the network's $k$ generations. 

That suggests that for a network, $\mu$ is proportional to the natural logarithm. Since the mean path length $\mu$ was the base of the logarithmic formula that multiplied the capacity of an individual network node, that suggests $\mu$ is a scale factor for a cluster of size $\mu^k$ for $k$, including $k=1$. This raises a new problem: is there a way to show the role of the natural logarithm in a network without assuming the mean path length is the base of a logarithmic formula representing the benefit of networking?

This natural logarithm theorem, that for mean path $\mu$, $d \mu / dt = \mu$, supports the validity of the ideas relating scaling, entropy, scale factors and capacity; it is an unanticipated and plausible inference based on the mathematics. The emergence of a natural logarithm theorem suggests the mathematics is on the right path. Trying to determine how scaling relates to increasing average IQs leads to a theorem connecting the natural logarithm to a mathematical model. 

\subsection{Allometry and 3/4 metabolic scaling} \label{Sec-Allom-3-4-Met-Sc}

\subsubsection{Metabolic scaling as a test of the lexical scaling model}

In September 2005 I began to look for the base of a logarithm that would work in a logarithmic formula describing the multiplicative effect of networking on lexical growth. In June 2007, a formula, $C\log_\mu(n)$, where the base $\mu$ of the log was the network's mean path length, seemed to work. By the end of February 2008, a nested hierarchy of scaled energy clusters seemed to provide a physical model of the $\log_\mu(n)$ part of the formula (Shour 2008, lexical growth). There was good agreement between the model and the rate of lexical growth. The rate of lexical growth appeared to have some special correspondence to half the rate of lexical divergence of daughter languages under the suppositions of glottochronology. The lexical growth rate corresponded to the rate of increase in average IQs and also seemed to work in other settings. 

The mathematical model for $C\log_\mu(n)$ seems to assist in answering otherwise difficult problems. That success is reason to be skeptical of its validity or its novelty, or both; the formula for the entropy of a network was not mentioned in the literature and one would suppose that it was unlikely that it had been overlooked.  In light of that unlikelihood, I looked for other scaling phenomena to test the validity of the formula. 

If the mathematical model of lexical scaling could explain other phenomena, that would increase confidence in the mathematical model. Moreover, the role of the mean path length as a scale factor was still unclear then (and for quite some time thereafter). The observation of mean path length scaling in connection with other phenomena might assist in understanding mean path length scaling. With that in mind, seeking other scaling phenomena to which mean path length scaling ideas might apply seemed to be an appropriate next step.

The likely connection (via the logarithmic formula) between thermodynamics and energy scaling applicable to lexical growth suggested that biological scaling might supply ways to test  the mathematical model. In that case, a good place to start learning about biological scaling is  Whitfield (2006).  Whitfield recounts scientific investigation of the relationship between thermodynamics and biology. In particular, he discusses allometry and metabolic scaling. 

\subsubsection{Metabolic scaling and the surface law}

The word allometry was coined by Huxley \& Teissier (1936). Allometry is the study of quantitative relationships of organisms (Gayon 2000). 

Historically, the surface law for metabolism was proposed as a solution for $b$ in the formula
\begin{equation} \label{Eq-3-4-met-scaling-formula}
Y = a M^b
\end{equation}
where $Y$ represents the whole organism's metabolism, $a$ is a constant for a species, and $M$ is the organism's mass (Whitfield 2006). The surface law $b = 2/3$ was probably first proposed in 1838 by Sarrus and Rameaux (Hulbert 2014) and first formulated in 1839 by Robiquet and Tillaye;  $b = 2/3$ was proposed because an organism's surface area dispersing body heat grows by a power of 2, while its mass that supplies the heat grows by a power of 3. 

Max Rubner (1883) published statistics consistent with $b = 2/3$  (and see Rubner 1902, translated in 1982 into English). Max Kleiber (1932) analyzed the data and arguments in favor of the surface law, and concluded that a weight-power law should supplant the surface law, and that researcher's measurements favored $b =3/4$. 

\subsubsection{WBE 1997}

West, Brown and Enquist gave a geometrical explanation (1997) (hereafter WBE, sometimes in reference to the article and sometimes in reference to its authors) for $3/4$ metabolic scaling. They choose an organism's circulatory system as a demonstration of their mathematical model of transport of materials to all parts of an organism through a linear branched network. WBE notes that, at that time in relation to equation (\ref{Eq-3-4-met-scaling-formula}) and other relationships with 4 in the exponent's denominator, `No general theory explains the origin of these laws.' Some (White 2003, 2005) still claim metabolic scaling is $2/3$. The debate on this and other issues (Etienne 2006;  Price 2007) continues. 

A synopsis of WBE's 1997 derivation of 3/4 metabolic scaling is relevant here for these reasons: (1) To set the stage for showing how WBE's scaling ideas were modified by scaling ideas used to model lexical growth (Shour 2008, on lexical growth); (2) To distinguish between WBE's derivation and this article`s derivation of 3/4 metabolic scaling; (3) To show why the 4/3 energy scaling related to metabolism  supports characterizing The 4/3 RDFT as a general law which provides a possible mathematical model of dark energy. 

WBE  makes three assumptions to derive the power $b$ in equation (\ref{Eq-3-4-met-scaling-formula}):
\begin{enumerate}
	\item Fractal-like branching supplies the organism.
	\item Capillaries for differently sized animals are size invariant. 
	\item Energy used for blood distribution is minimized. 
\end{enumerate}

WBE's second assumption is necessary for WBE's mathematical derivation of 3/4 scaling.  The validity of WBE's second assumption  has been doubted (Dawson 1998); this assumption is unnecessary if 3/4 metabolic scaling is derived based on The 4/3 RDFT.

WBE's third assumption is equivalent to Jaynes's maximum entropy principle in this way: for a given amount of energy, maximize the output per unit energy. This is in turn related to Jensen's inequality: entropy is maximized when the probabilities for different possible paths are equal, that is, when there is uniform scaling. If fractal-like branching is based on \textit{uniform} scaling (in contrast to non-uniform scaling), the organism's energy distribution system --- its circulatory system --- will have maximum entropy or degrees of freedom. For a given organism, the third assumption therefore implies the circulatory system consists of scaled nested tube clusters or, put another way, a uniformly scaled hierarchical system. 

WBE's second assumption implies capillary volume $V_c$, length $l_c$, and radius $r_c$ are independent of body size. 

In WBE terminology and reasoning, there are $N$ branchings from the aorta's volume $V_0$ at level 0 to the capillaries. Capillaries each have volume $V_c$. $B$ is the metabolism of the whole organism, and $B \propto M^a$, for mass $M$ and some $a$, or $B= B_0M^a$ for some appropriate proportionality factor $B_0$. 

The blood flowing through each $k^{th}$ level  of the circulatory system has the same volume. WBE reasons that an organism's blood volume and mass is proportional to the number of capillaries, so that the total number of capillaries 
\begin{equation} \label{Eq-Nc-eq-M-power-a}
N_c \propto M^a
\end{equation}
 for  $a< 1$. The validity of this inference may be doubted since it leads to an inconsistency noted by Kozlowski and Konarzewski (Kozlowski 2004). If instead the number of capillaries $N_c \propto M$ then the scaling down of metabolism might occur some way other than a scaled down number of capillaries. Equation (\ref{Eq-Nc-eq-M-power-a}) figures prominently in the WBE derivation of 3/4 scaling, as described below.

In the terminology of The Overview earlier in this article, the aorta (or perhaps better, the heart) would be considered $V_1$ instead of $V_0$ as in WBE, since level 0 better describes the external supply of energy, namely food; the subdivision of energy as it moves away from the distribution source is consistent with a radiation model of energy distribution. $V_1$ contains the first cluster of energy  to be scaled. 

For a circulatory system, $N_k$ is level $k$'s number of tubes, $r_k$ is level $k$'s tube radius and $l_k$ is level $k$'s tube length. $N_0=1$ in WBE's terminology, where $N_0$ represents the value associated with the aorta. Each tube at the $k^{th}$ level branches (is scaled into) $n_k$ tubes. WBE's scale factors are as follows:
\begin{equation} \label{Eq-def-N-k-SF}
\frac{N_{k+1}}{N_k} \equiv n_k,
\end{equation}
\begin{equation} \label{Eq-def-radius-k-SF}
\frac{r_{k+1}}{r_k} \equiv \beta_k,
\end{equation}
\begin{equation} \label{Eq-def-length-k-SF}
\frac{l_{k+1}}{l_k} \equiv \gamma_k.
\end{equation}

At a $k^{th}$ branching into smaller tubes, tube radius \textit{decreases} by scale factor $\beta_k$, tube length \textit{decreases} by scale factor $\gamma_k$, and the number of tubes increases by scale factor $n_k$. 

Compare radiation cone volume increments to circulatory system tubes. Radiation cone volume increments  \textit{increase} in size for a given amount of energy as the radiated energy moves away from the source. For radiation cone volume increments it is not the \textit{number} of tubes that increases but the \textit{volume}. This difference does not affect the algebra relating to   scale factors.

WBE states that it can be shown that for all $k$ the scale factors (or  the bases of scale factors as they would be denoted by the terminology in this article's Overview) are constant:
\begin{equation} \label{Eq-def-N-SF}
\frac{N_{k+1}}{N_k} \equiv n,
\end{equation}
\begin{equation} \label{Eq-def-radius-SF}
\frac{r_{k+1}}{r_k} \equiv \beta,
\end{equation}
\begin{equation} \label{Eq-def-length-SF}
\frac{l_{k+1}}{l_k} \equiv \gamma.
\end{equation}

WBE reasoning is roughly as follows. 

$n_k=n$ because the circulatory system is \textit{assumed} to be a self-similar fractal. This is an approximation and an idealization. The optimality of uniform scaling can be shown by Jensen's inequality which is discussed in section \ref{Sec-Jensens-Ineq}. The fractal nature of energy distribution is a consequence of the 4/3 fractal dimension of radiation-space compared to space; WBE's assumption that the circulatory system is a self-similar fractal accords with The 4/3 RDFT. 

$\gamma_k=\gamma$ because for the fluid in the circulatory system to reach all parts of the organism, the fluid must be distributed by \textit{space-filling} fractal structures. The `service volume' reached by a capillary is assumed to be a sphere with radius equal to $l_c/2$. WBE assumes that this observation about service volumes applies at all $k$ levels, an assumption which follows from their assumption of fractality. 

For $N$ large and with $r_k<<l_c$, the sum of all spherical service volumes  reaches all parts of the organism. 

On this reasoning, the total service volume equals the organism's total volume, which is proportional to the organism's total mass. Since WBE assumes that capillaries for all animals are size invariant, the number of capillaries would be proportional to the service volume, $N_c \propto M$, inconsistent with equation (\ref{Eq-Nc-eq-M-power-a}). The derivation of 4/3 scaling in this article assumes that $N_c \propto M$, which is to say, the circulatory system volume grows in proportion to $M$.

For WBE's analysis, one must overlook overlapping spheres or gaps between spheres. The sum of the service volume spheres at the $k^{th}$ level is then:
\begin{equation} 
(4/3) \pi (l_c/2)^2 N_k,
\end{equation}
where $N_k$ is the number of spheres at the $k^{th}$ level.

Then
\begin{equation}
(\gamma_k)^3 \equiv \frac{(l_{k+1})^3}{(l_k)^3} \approx \frac{N_{k}}{N_{k+1}} = \frac{1}{n},
\end{equation}
showing that 
\begin{equation}
\gamma_k = n^{-1/3}.
\end{equation}
(More precisely, $\gamma_k \approx n^{-1/3}$.) In this way, if we accept that $N_{k}/N_{k+1}$ scales in a uniform way, we find that the scale factor $\gamma$ does not depend on $k$.

Similar reasoning applies to obtain $\beta_k= \beta$. Since the blood flowing through the circulatory system is incompressible, it must be that the cross-sectional areas $A_k=\pi (r_k)^2$  of every $k^{th}$ level is the same. In particular, the $k+1$ tubes that are smaller than a $k$ level tube of which they are branches have a radius such that
\begin{equation} \label{Eq-WBE-r-n-relp}
r_{k} = nr_{k+1}.
\end{equation}
Dividing both sides of equation (\ref{Eq-WBE-r-n-relp}) by $r_k$ gives
\begin{equation}
\beta_k \equiv \frac{r_{k+1}}{r_k} = n^{-1/2},
\end{equation}
and since $n$ is not dependent on $k$ neither is $\beta$.

To show that $n$, $\gamma$, $\beta$ are constants in an energy distribution system, we could instead follow the ideas in the Overview above, and compare the exponents of the scale factors via $Deg$. 

WBE notes that $N_c=n^N$: on their assumptions, the number of capillaries $N_c$ along one branch from the aorta scales by $n$ at each level $k$ for $N$ levels. This calculation shows why the aorta is assigned level $0$ by WBE; the $N$ branches occur after the $0^{th}$ level. This implies that
\begin{equation} \label{Eq-Nc-in-terms-of-n}
\ln(N_c)= \ln(n^N) = N \ln(n).
\end{equation}

Using WBE's point at equation (\ref{Eq-Nc-eq-M-power-a}) that $N_c \propto M^a$,
\begin{equation} \label{Eq-Nc-in-terms-of-M}
\ln(N_c) = \ln(M^a / M_0) = a\ln(M/M_0)
\end{equation}
where $M_0$ is used to permit the formation of an equality based on the proportionality $N_c \propto M^a$. 

Equating $\ln(N_c)$ in equations (\ref{Eq-Nc-in-terms-of-n}) and (\ref{Eq-Nc-in-terms-of-M}) gives the equality
\begin{equation}
N \ln(n)=a\ln(M/M_0),
\end{equation}
that is,
\begin{equation} \label{Eq-N-in-terms-of-M-and-ln-n}
N =\frac{a\ln(M/M_0)}{\ln(n)},
\end{equation}
which is the result in equation (3) of WBE's paper.

Using the scale factors defined in WBE,
\begin{equation} \label{Eq-showing-r-in-geo-series-in-r}
\begin{split}
\frac{\pi(r_k)^2l_kn^k}{\pi(r_{k+1})^2l_{k+1}n^{k+1}} &=\frac{\pi(r_k)^2l_kn^k}{\pi(\beta r_k)^2 (\gamma l_k)n n^k} \\
&= \frac{1}{n \gamma \beta^2}.
\end{split}
\end{equation}
The  terms on the left side of the first line of equation (\ref{Eq-showing-r-in-geo-series-in-r}) are shown on the right side with their scale factors. The right side is then simplified to yield the second line in equation (\ref{Eq-showing-r-in-geo-series-in-r}). The factor $(n \gamma \beta^2)^{-1}$ is used as the $r$ term appearing in a geometric series $S_n=1+r +r^2 + \ldots +s^n$ to arrive at the second line in equation (\ref{Eq-WBE-paper-4}), also using the fact that each $k^{th}$ level has volume $n^N V_c$.

WBE calculates the total blood volume $V_b$ in the circulatory system as
\begin{equation} \label{Eq-WBE-paper-4}
\begin{split}
V_b=\sum^N_{k=0}{N_kV_k} & = \sum^N_{k=0} \pi(r_k)^2l_kn^k \\
&= \frac{(n\gamma\beta^2)^{-(N+1)}-1}{(n\gamma\beta^2)^{-1} -1}n^NV_c.
\end{split}
\end{equation}

In the first line of equation (\ref{Eq-WBE-paper-4}), there are $N_k$ tubes of volume $V_k$ at each $k^{th}$ level in the idealized circulatory system, leading to the expression $N_kV_k$, and $n^k$ tubes of volume $\pi(r_k)^2l_k$ leading to the expression $\pi(r_k)^2l_kn^k$.

WBE states that $n \gamma \beta^2 <1$ and $N>>1$. This observation is used to justify treating the exponent  $-(N+1)$ of the first expression in the numerator in the second line of equation (\ref{Eq-WBE-paper-4}) as approximately $-N$.

WBE finds, based on equation (\ref{Eq-WBE-paper-4}) that 
\begin{equation} \label{Eq-WBE-papr-Vb-follg-Eq4}
\begin{split}
V_b &= \frac{V_0}{1 - n \gamma (\beta)^2} \\
& = \frac{V_c (\gamma \beta^2)^{-N}}{1 - n \gamma \beta^2}.
\end{split}
\end{equation}

The first line in equation (\ref{Eq-WBE-papr-Vb-follg-Eq4}) is found by substituting $V_0$ for $n^NV_c = $ in the second line of equation (\ref{Eq-WBE-paper-4}).  $n^NV_c = V_0$ since every level has the same amount of fluid. The second line in equation (\ref{Eq-WBE-papr-Vb-follg-Eq4}) is because $(n\gamma\beta^2)^{-(N+1)}-1 \approx (n\gamma\beta^2)^{-N}-1$. Then the $n^N$ term in $(n\gamma\beta^2)^{-N}$ cancels the $n^N$ term in $n^N V_c$ in equation (\ref{Eq-WBE-paper-4}). 

With these steps, and assuming that capillaries are invariant for all organisms, WBE arrives at
\begin{equation} \label{Eq-gamma-beta2-propto-M}
(\gamma \beta^2)^{-N} \propto M.
\end{equation}
This conclusion critically relies on $V_c$ in the second line of equation (\ref{Eq-WBE-papr-Vb-follg-Eq4}) performing the role of a constant. Without the assumption of the invariance of capillaries,  equation (\ref{Eq-gamma-beta2-propto-M}) would not follow from equation (\ref{Eq-WBE-papr-Vb-follg-Eq4}). This also relies on WBE's assumption that the capillaries irrigate the whole organism which has mass $M$.

Using $M_0$ as in equation (\ref{Eq-Nc-in-terms-of-M}), and taking the log of each side of equation (\ref{Eq-gamma-beta2-propto-M}) gives
\begin{equation} \label{Eq-N-ln-n-in-terms-of-ngammabeta2}
-N \ln(n \gamma \beta^2) = \ln(M/M_0),
\end{equation}
and so
\begin{equation} \label{Eq-N-in-terms-of-ngammabeta2}
N= - \frac{\ln(M/M_0)}{\ln(n \gamma \beta^2)}.
\end{equation}

The right sides of equations (\ref{Eq-N-in-terms-of-M-and-ln-n}) and (\ref{Eq-N-in-terms-of-ngammabeta2}) both equal $N$. Equating them, and substituting for $\gamma$ and $\beta$ their values in terms of $n$
\begin{equation} \label{Eq-fr-WBE-arrivingat-3-4}
\begin{split}
a & = -\frac{\ln(n)}{\ln(\gamma \beta^2)} \\
& =  - \frac{\ln(n)}{\ln(n^{-1/3} (n^{-1/2})^2)} \\
& = - \frac{\ln(n)}{\ln(n^{-4/3})} \\
&= - \frac{\ln(n)}{(-4/3) \ln(n)}
\end{split}
\end{equation}
and so $a=3/4$. The last two steps in equation (\ref{Eq-fr-WBE-arrivingat-3-4}) are omitted in WBE. Those last two steps involve only algebraic manipulation. In the mathematical model of dark energy the last two steps in equation (\ref{Eq-fr-WBE-arrivingat-3-4}) are significant. The last two steps are not only algebraical steps; they hint at the 4/3 geometrical scaling of an energy distribution system.

That concludes a summary of the portions of WBE 1997 relevant to this paper.

\subsubsection{Kozlowski and Konarzewski on WBE 1997}

About WBE 1997 Kozlowski and Konarzewski note (Kozlowski 2004): ``The assumption that the final branch is size-invariant causes the number of levels to be a function of body size (or vice versa): more levels are required to fill a larger volume with the same density of final vessels.'' They argue  (Kozlowski 2004, 2005) that  the assumption of size invariant terminal supplying vessels  leads to an inconsistency: ``for large animals the volume of blood vessels would exceed body volume'' because ``the number of capillaries is, according to their model, proportional to body size'' (p. 284, Kozlowski 2004), contradicting the model's assumption that the number of vessels equals $M^a$.  

Kozlowski and Konarzewski note  (Kozlowski 2004) that it cannot be both that $N_c \propto M$ and $N_c \propto M^a$, unless $a=1$.

WBE respond (Brown 2005): ``the service volumes of tissue supplied by each capillary are free to vary \ldots''.  This is  unlikely if the biochemistry and bio-mechanics of energy distribution and tissue are similar for different animals as may be supposed (p. 2249, Gillooly 2001).

In the derivation of 3/4 metabolic scaling set out later in this article, the number of levels $k$  is proportional to mass $M$. The idea that the number of levels $k$ is proportional to mass $M$ is consistent with the supposition that an organism's circulatory system is uniformly scaled (or branching) and hierarchical. It is also consistent with assigning an important role to the number of degrees of freedom $k$ of an organism's circulatory system, and with the mathematical model of scaling applicable to lexical scaling described earlier in this article.

\subsubsection{Adapting lexical scaling to metabolic scaling}

Metabolic scaling offers a possible way to test the plausibility of the mathematical model of lexical scaling. WBE 1997 terminology suggests analogies between WBE's model of 3/4 metabolic scaling and the mean path length scaling of lexical growth. In 2008, listing similarities and differences between WBE 1997 and mean path length scaling was a starting point. 

Mean path length scaling and  WBE's mathematics used to explain 3/4 metabolic scaling possibly  have these    similarities:
\begin{enumerate}
	\item Both involve networks, one a circulatory system, and the other networked people and networked words.
	\item Capillaries are assumed size invariant. Network nodes are fundamental units.
	\item Energy to distribute resources is minimized.
	\item The number of circulatory system branch generations scales by $n$. The network's number of cluster generations scales by the average energy required to connect two nodes.
	\item Circulatory system fluid volume $V_k$ in the $k^{th}$ branch generation of the circulatory system is the same for all $k$. Total network capacity is $k$ times the capacity of a single cluster generation for a network for a network with $k$ scaled generations.
\end{enumerate}

Mean path length scaling and  WBE's mathematics used to explain 3/4 metabolic scaling  possibly have these differences:
\begin{enumerate}
	\item Metabolism $Y$ in equation (\ref{Eq-3-4-met-scaling-formula}) measures the rate of energy use by all the organism's cells, not the average energy cost, proportional to the mean path length $\mu$ between nodes.
	\item The circulatory system is a sub-network of the organism. Lexical scaling is about the whole network.
	\item Circulatory system fluid is distributed in one direction. Social and lexical networks have a bidirectional exchange of information.
	\item The circulatory system is a network of pipes physically different in scale from one generation to the next. In  social  and lexical  networks, the nodes in different cluster generations are the same nodes.
	\item Metabolic scaling remains the same   through 21 orders of magnitude. Network entropy for lexical scaling increases as the number of nodes  increases.
\end{enumerate}

Suppose the same mathematical model can account for lexical and 3/4 metabolic scaling. If so, then there must be a way to resolve the apparent differences listed above. In April 2008 the differences resolved due to two observations. 

First, in networked lexical scaling, directly or indirectly, each node can receive information from any other node, and can transmit information to any other node. For an organism, a cell in the circulatory system can benefit a cell in, for example, the lungs, or receive a benefit from a cell in the lungs, if not directly through networked cells, then at least indirectly from  a system within the organism. In an organism, directly or indirectly, each cell can receive benefits from any other cell, and can transmit benefits to any other cell. 

Second, an idealized social or lexical network has the same $n$ nodes in every cluster generation, while a circulatory system levels have tubes of different sizes. In lexical growth scaling though energy scaling leads to a flattened hierarchy (nested scaling); a physically observable hierarchy may manifest itself in networks of cells in organisms. To analogize a network to a circulatory system, the circulatory system must have only $n$ nodes in all generations or levels. This is so if each node corresponds to a single scaling energy cluster per time unit from heart to capillary. The perceived `different' tube sizes of a circulatory system compare to distinct energy cluster generations in a network. In other words, treat the circulatory system as having an energy cluster (at the heart level) that scales at each subsequent level or generation per scaling, just as with lexical scaling. 

Infer that an organism's tissue is arranged to maximize entropy, that is, to maximize the efficiency of energy distribution. 

Following WBE 1997, each capillary irrigates a corresponding volume in an organism.  Therefore, the rate of energy use by the organism's mass equals the rate of energy supply from the  organism's circulatory system $Circ$. Assume that the organism's energy use is proportional to both its mass and its volume. All this suggests that the volume $V_{Circ}$ of the circulatory system is in proportion to the volume of the organism and also in proportion to its mass $M$. One volume unit of the organism receiving energy distributed by the circulatory system contains one mass unit than requires one energy unit per time unit. What $Y \propto M^a $ says is that circulatory system  only needs to use $(V_{Circ})^a \propto M^a$ energy units to irrigate all $M$ mass units, for $0<a<1$.

If per unit time, $(V_{Circ})^a$ energy units are sufficient to supply energy to $M$ mass units, then $[(V_{Circ})^a]^{1/a} =V_{Circ}  \propto M^{1/a}$ is sufficient to supply energy to a mass of $M^{1/a}$ mass units. This amounts to observing that $V_{Circ}$ has too much capacity to distribute energy unless that capacity is scaled down by $a<1$. Put another way: $Y \propto M^a $ is not really a statement about how metabolism scales; it is a consequence of how the capacity of the circulatory system to deliver energy scales up with size.  

It follows  that if, observationally, $a=3/4$,  then the capacity of the circulatory system to distribute energy is scaling with size by a 4/3 power.  

Instead of the short inference in equation (\ref{Eq-fr-WBE-arrivingat-3-4}) about $1/a$, I used the following observation about scaling. If $\beta r_1 = r_{2}$, then $\beta r_2= \beta \beta r_1 = \beta^2 r_1$, and in general $r_{k+1} =\beta r_k = \beta^k r_1$. For $r_{k+1}$ there are $k=\log_\beta(\beta^k)$ scalings. If each scaling adds to a cumulative length with units each $r_1$ long, then we can think of $k$ as the entropy of a cumulative radius $kr$  or as $k$ degrees of freedom relative to the scale factor $\beta$, as in the terminology adopted in The Overview above. 

In 2008, the observation that  $r_{k+1} =\beta r_k = \beta^k r_1$ seemed a mathematically advantageous way of characterizing scaling, because it simplifies calculation. By 2012 or so, it began to seem that regarding the exponent of a scale factor as its degrees of freedom was of even more consequence. 

In 2008  these Three Assumptions,
\begin{enumerate}
	\item $\beta r_k = \beta^k r_1$,
	\item the capacity of a system equals the number of scalings (degrees of freedom) times the capacity of single generation, and
	\item $(V_{Circ})^a \propto M^a$ energy units are sufficient to irrigate $M$ mass units,
\end{enumerate}
permitted a shorter, simpler derivation of 3/4 metabolic scaling. Most of the steps in WBE represented in the present article by equations (\ref{Eq-Nc-in-terms-of-n}) to (\ref{Eq-fr-WBE-arrivingat-3-4}) can then be omitted, and the assumption that all animals have the same size capillaries can be dispensed with. 

Suppose a $k^{th}$ generation `tube' has radius $r_k$, length $l_k$, and that for the number $N_k$ of $k^{th}$ generation circulatory system tubes $N_{k+1}=nN_k$.

As in WBE 1997, let 
\begin{itemize}
	\item $r_{k+1}/r_k \equiv \beta$,
	\item $l_{k+1} /l_k \equiv \gamma$. 
\end{itemize}
 In general for the $(k+1)^{st}$ generation of tubes, a circulatory system tube has cross-sectional area $\pi \beta^{2k}(r_k)^2$ and the volume  it irrigates is $(4/3)\pi \gamma^{3k}(l_k/2)^3$. As in WBE 1997, $\beta = n^{-1/2}$ and $\gamma= n^{-1/3}$. 

Suppose that the energy content of blood is proportional to its volume. Then the supply volume of a $(k+1)^{st}$ level of a circulatory system expressed in terms of tube volume, $\pi (r_k)^2 l_k$, is the same as $V_k$. A representative volume of a tube at the $(k+1)^{st}$ level of the circulatory system is 
\begin{equation} \label{Eq-beta-gamma-Vol-1}
\pi \beta^{2k} (r_1)^2 \gamma^k l_1
\end{equation}
or 
\begin{equation} \label{Eq-n-SF-Vol-1}
n^k V_1.
\end{equation} 
Now equate the expressions in equations (\ref{Eq-beta-gamma-Vol-1}) and (\ref{Eq-n-SF-Vol-1}) for volume after $k$ scalings, as in equation (\ref{Eq3-1st-Ax-allom-paper}). Scale factors $\beta$ and $\gamma$ are on the left side and the scale factor $n$ is on the right side.

With $k$ scalings there are $k+1$ tube levels. Using the Three Assumptions just above, for the cumulative volume of $k$ levels of the circulatory system for one scaling tube (that is, totaling the volumes along all generations following a single path)
\begin{equation} \label{Eq3-1st-Ax-allom-paper}
(k+1) \pi \beta^{2k} (r_1)^2 \gamma^k l_1 = ((k+1) n^k V_1)^{1/a}.
\end{equation}

Since equation (\ref{Eq3-1st-Ax-allom-paper}) holds for all sizes of mass, suppose its fractality results in it  also holding at all tube levels. This simplifies the mathematics. Then equation (\ref{Eq3-1st-Ax-allom-paper}) would hold in particular for $k = 1$. With $k=1$ equation (\ref{Eq3-1st-Ax-allom-paper}) becomes
\begin{equation} \label{Eq precede-eq4-1st-Ax-allom-paper}
2 \pi \beta^{2} (r_1)^2 \gamma l_1 = (2 n V_1)^{1/a}
\end{equation}
which is equivalent to
\begin{equation} \label{Eq eq4-1st-Ax-allom-paper}
\frac{\beta^{2} \gamma}{n^{1/a}} = \frac{(2V_1)^{1/a}}{2 \pi (r_1)^2  l_1}.
\end{equation}
For a given animal, the right side of equation (\ref{Eq eq4-1st-Ax-allom-paper}) is a number because all the components of the numerator and denominator are numbers (2, $\pi$) or have numeric values ($V_1$, $r_1$, $\ell_1$, 1/a). We are interested in the scaling relationship pertaining to only the left side of equation (\ref{Eq eq4-1st-Ax-allom-paper})  which are not affected by constants that multiply a scale factor, so for scaling purposes treat the right side of equation (\ref{Eq eq4-1st-Ax-allom-paper}) as 1. We then exploit the scaling relationships among $\beta^2$, $\gamma$ and $n$ on the left side of equation (\ref{Eq eq4-1st-Ax-allom-paper}) to solve for $a$. Multiply the top and bottom of the left side of equation (\ref{Eq eq4-1st-Ax-allom-paper}) by $n^{-1/a}$. Then we have
\begin{equation} \label{Eq eq4-1st-Ax-allom-paper-infrnce}
\beta^{2} \gamma = n^{-1/a},
\end{equation}
and so $(n^{-1/2})^2 n^{-1/3} =n^{-1/a}$ which gives $-4/3 = -1/a$, so $a = 3/4$.

It is possible to abbreviate derivation of the scaling of supply or circulatory system tubes still further, with $\beta = n^{-1/2}$ and $\gamma=n^{-1/3}$. This derivation is of particular interest because of its brevity. Its brevity facilitates compact\ derivations and proofs.  After $k$ scalings (that is, at the $(k+1)^{st}$ level), collect powers on the left side to give the right side in the following equation:
\begin{equation} \label{Eq-4-3-Supplyentropy-in-allomtry100}
\pi n^{-k} (r_1)^2 n^{-k/3} l_1 =  n^{k(-4/3)} \pi r_1 l_1.
\end{equation}

The corresponding  spherical recipient  volume irrigated by a circulatory system tube is equivalent to 
\begin{equation} \label{Eq-4-3-Receiptentropy-in-allomtry100}
(4/3) \pi \gamma^{-3k} (l_1/2)^3  = \theta^{-1k}\Theta_1,
\end{equation}
for a sphere $\Theta_1$ which has $\theta$ as the base of its scale factor. Spheres scale by unity relative to other spheres since they are symmetric to their origin. In this perspective, circulatory system tubes scale per generation by a 4/3 power (the 4/3 power of $n^{-k(4/3)}$) while the corresponding recipient spherical volume scales by  a power of 1 (the 1 power of $-1k$) as on the right side of equation (\ref{Eq-4-3-Receiptentropy-in-allomtry100}). 

Thinking of the exponents of the scale factors as representing entropy (or degrees of freedom), the exponent of circulatory system supply is 4/3 and the exponent of the corresponding receipt is 1. This applies for every $k^{th}$ generation. It is not necessary to assume that capillaries are the same size for all animals. Since the energy distribution system scales by a 4/3 power and the energy receipt system scales by a 1 power, this suggests treating them as distinct systems; they scale differently. Designate the components of the distribution system differently than the components of the receiving system. 

\subsubsection{Two systems}

There are two systems in an organism, one distributing energy --- the subsystem that is the circulatory system --- and the other --- the entire mass of the organism --- receiving that energy. There are also two systems in  an information or  social network. 

An information network has a system that distributes information and a system that receives it. But the members of society who distribute information are the same people who, at different times, also receive it. In the case of an organism's circulatory system, nature requires us to consider there to be two systems scaling differently. The analogous requirement for an information network is obscured because the two systems have, at different times, identical membership.

The exponent 4/3 of the base of the scale factor for the circulatory system found in April 2008 was initially perplexing. The exponent sought was 3/4, not 4/3. Despite redoing the calculation many times, whether the capillaries were generation 0 or generation N, the answer was 4/3. The provenance of 4/3 resolved upon realizing that the metabolic rate scales down by a 3/4 power to offset the capacity of the circulatory system to distribute energy scaling up by a 4/3 power.  

The entropy of a distribution network increases the network's energy capacity. The entropy (the exponent of the base of the scale factor for energy distribution) of the circulatory system's capacity to distribute energy scales by 4/3 with size. This observation is consistent with the idea that the entropy of a network using  a base equal to its mean path length measures the increased collective capacity of the network relative to the capacity of a single node; that is, that capacity is proportional to degrees of freedom relative to the appropriate base of a scale factor. This observation is as significant as  the idea that the mean path length scales a network, which initially and for some time seemed to be the important principle. 

The initial attempt to model metabolic scaling contained   technical and algebraic mistakes. It left unanswered the question of why distribution by the capillaries to a spherical volume as supposed in WBE 1997,  as opposed to some other shape of volume, worked. 

\subsection{Stefan-Boltzmann Law}

In April 2008 the article on metabolic scaling was submitted to arXiv. After it appeared, that same month, while reading Allen and Maxwell's A Text-book of Heat (1939), page 742 described Stefan's law. Stefan's law is so called because Stefan first conjectured it. It is also sometimes called the Stefan-Boltzmann law, because Boltzmann later gave a mathematical explanation of it. 

Stefan's Law is, in Allen and Maxwell's notation,
\begin{equation}
E \propto T^4
\end{equation}
where $E$ is energy density and $T$ is temperature in degrees Kelvin. This applies to an enclosed space with perfectly reflecting walls, what is sometimes described as the setting for black body radiation. 

In Allen and Maxwell's presentation of Boltzmann's proof, in an intermediate step they have this inference:
\begin{equation} \label{Eq-p743-AllenMaxwell-deltaS}
\frac{\partial S}{\partial v} = \frac{4}{3} \frac{E}{T}.
\end{equation}

In equation (\ref{Eq-p743-AllenMaxwell-deltaS}), $S$ on the left side is entropy, $\partial S$ is the change in entropy, and entropy's change is per volume $\partial v$. The relationship in equation (\ref{Eq-p743-AllenMaxwell-deltaS})  is the same as in the case of metabolic scaling, where 4/3 scaling is relative to the \textit{volume} of a circulatory system tube. Since $S$ can be represented by a logarithmic function, so too in the case of a circulatory system tube scaling, the scaling is represented by the exponent of a scale factor. 

These considerations suggest that equation (\ref{Eq-p743-AllenMaxwell-deltaS}) represents the same kind of 4/3 power scaling derived in connection with 3/4 metabolic scaling and that the scaling method used for 3/4 metabolic scaling can be adapted to radiation.

Why a spherical volume based on a one half a circulatory system tube's length worked is a puzzle arising out of the derivation of 3/4 metabolic scaling in WBE 1997. The intermediate step in equation (\ref{Eq-p743-AllenMaxwell-deltaS}) in the derivation of Stefan's Law provides a clue connecting metabolic scaling and Stefan's Law. 

The sphere receiving fluid from capillaries connects to the isotropy of black body radiation and of the cosmic background radiation, which is  isotropic to one part in 100,000 (Fixsen 1996). The sphere scales in the same way for all axes from its center, and so models the  isotropic distribution of energy. This suggests that the 4/3 power scaling of energy distribution connects to the  4/3 power scaling of a circulatory system, black body radiation, and cosmic background radiation; one mathematical model for life, energy  and light.  The isotropic distribution of energy by capillaries in organisms models the isotropic distribution of energy in the universe; 3/4 metabolic scaling provides a clue to the mystery of dark energy.

$E \propto T^4$ may describe the 3 degrees of freedom for energy within a static volume relative to the 4 degrees of freedom of radiating energy represented by $T^4$.

\subsection{Adapting circulatory system scaling to radiation scaling}

\subsubsection{Analogies}

Radiation from a point source can be modeled as energy distribution via radiation cones. A scaled circulatory system is analogous to a radiation cone. A radiation cone scales uniformly as it grows larger; it is in the nature of a cone. Mark fixed lengths radially along the radiation cone axis, with each length defining a radiation cone increment. Analogize the radiation cone radial length to a circulatory system tube length. Analogize the average radius of a radiation cone increment to the fixed radius of a circulatory system tube. These analogies permit us to adapt derivation of 4/3 power scaling for an organism's circulatory system to derive scaling of light or energy radiated distribution from a point. 

The scale factor for the average radius of radiation cone increments from one generation to the next can be set as $\beta$ as with circulatory system tubes. The radial length of a radiation cone increment would be constant, not scaling up or down, if the rate or speed of radiation is constant as it would be for light. 

The space receiving  isotropically distributed radiation can be described by a spherical volume, with a radius equal to one half the length of the radiation cone increment. As with the circulatory system, we can use $\gamma$ as the scale factor for the radial length; here $\gamma$ helps keep track of the effect of scaling, since the radial length itself is constant.  (This article uses $s$ or $s_k$ to scale radiation length $L$.) Now apply the same method used for the circulatory system to the radiation system to derive 4/3 power scaling. 

$L$'s scale factor $s_k$ changes relative to $V_k$'s scale factor $v_k$, considered from a $Deg$ perspective; as $V_k$ increases in volume, $V_k$'s height $L$ becomes relatively smaller, and $s_k$ becomes smaller relative to $v_k$. In The Overview the changing of $s_k$ relative to $v_k$ plays no role, since the extra 1/3 $deg$ arises as an aspect of the dimensional relationship of $L$ to $V_k$. The extra 1/3 dimension of radiation-space arises per radiation event; the change from one volume increment to the next affects the $Deg$ relationship of $v$ and $s$ but not the $Dim$ relationship of $L$ and $V_k$.

\subsubsection{Differences and similarities}

An organism's circulatory system distributing energy has its tubes \textit{scale down} in size. As radiation is distributed from a point source, the volume of the radiation cone increments, each with the same radial length, \textit{scale up} in size. At any instant, summing the volumes from larger to smaller as with the circulatory system, or from smaller to larger as with radiation, does not affect the scaling relationship between length and volume (except perhaps for a plus or minus sign in exponents) which is per generation. 

Theory suggests denoting the start of circulatory system energy distribution at the heart, since that is where energy distribution for the circulatory system begins. We might instead think of capillaries scaling up in size until circulatory system tubes reach the heart. Since the number of branchings for a circulatory system is finite, the adding direction does not affect the mathematics. As an energy distribution system, though, the energy is being subdivided as it courses through circulatory system; as such it is better to consider scaling of circulatory system tubes as beginning at the heart. 

The radiation case mathematically suggests, perhaps requires, that the counting of levels must begin at the first radiation cone increment next to the point source. Assigning numbers to radiation cone increment levels is mathematically consistent if each additional radiation cone increment farther from the point source increases the level number. Counting does not entail shifting an entire array farther from the beginning to relabel level  1 as level 2 and start with a new number 1 closer to the point source. If our usual notion of counting reflects an underlying physical reality to the structure of the universe (a not unlikely inference), then counting must begin at the heart for circulatory systems and at the first radiation cone increment (nearest the point source) for radiation.

Mathematical consistency therefore suggests denoting the heart and the first radiation cone increment closest to the point source as level 1. An argument can be made to treat the heart as level 1 for an organism's circulatory system, since the circulatory system can be considered to be a set of nested energy clusters manifested as a branched linear system. 

If distributed energy is being scaled for each generation or at each level, and if each level has been scaled, then the source of energy must be external to the energy distribution system. The sources of food and oxygen that are transported by circulatory system are external to the organism. This observation suggests designating the external source of energy as level $0$. This would support the heart as level 1 for an organism's circulatory system.

The amount of energy per circulatory system level is constant. As the energy is scaled into smaller tubes, the energy per fluid volume remains the same. The amount of energy per radiation cone volume increment is constant. The \textit{energy density} per volume of a radiation cone volume increment scales down. 

For the circulatory system, tube size scales down and energy density is constant. For radiation, the volume of the radiation cone increments increases as the distance from the source increases, and energy density scales down. From this it appears that for radiation, energy density scaling down and the constant energy content of each radiation cone volume increment are respectively analogous to, for an organism's circulatory system, the scaling of the number of tubes and the constancy of energy density per unit of fluid volume. For the circulatory system, energy density per fluid volume is constant and the number of tubes scales per level; for radiation, energy density scales per level and the energy content per radiation cone volume increment is constant.    

The circulatory system has a finite number of branchings. Radiation constantly adds radiation cone increments. 

\subsubsection{Radiation scaling as a general law}

Suppose scaling by a 4/3 power is a valid attribute of radiation. Radiation in the universe preceded life. One can infer then that the 4/3 scaling of the capacity of the circulatory system of organisms is modeled on, or represents nature adapting a physical law governing isotropic energy distribution to living organisms. 

From a different perspective, one might observe that the same physical principles that govern the distribution of energy in the universe and that apply to black body radiation (and so apply at a quantum level) also apply to the distribution of energy in our own bodies. Suppose the distribution of energy in the universe via light is maximally efficient, perhaps perfectly efficient. Then subsystems within the universe, including organisms, might emulate the universe's maximally efficient energy distribution as they evolve over time. 

If the analogy of energy distribution by radiation to energy distribution by the circulatory system
is valid, then energy scaling applies at all scales from the scale of light up to the scale of a  circulatory systems, and presumably at all scales in between. Generalizing, if energy scaling applies to energy radiation, since radiation is a fundamental mechanism for energy distribution and since every array of energy is built up from energy clusters eventually comprised of light energy, then energy scaling applies at all scales. If energy scaling applies at all scales, then the lexicon as a system of networked concepts is a particular instance of energy scaling. These observations suggest that the scaling ideas here outlined have universal applicability at all scales. 

One might ask: did the way the universe emerged dictate our mathematics, or does mathematics constrain the universe's choices about how to maximize the efficiency of energy distribution? It is, perhaps, more likely that the mathematics we have developed reflects and conforms to the manner of the universe's emergence. If so, then the mathematical ideas and structures that have emerged and evolved due to our contemplation of the universe, including its earth-bound and starry features,  indirectly model the emergence of the universe. 

\subsubsection{The problem of the extra 1/3 power} \label{Sec-leading-to-extra-1-3-problem}

A mistaken way of examining  4/3 scaling for radiation helps illustrate the problem of the extra 1/3 power in the 4/3 power of energy distribution scaling that is mentioned in The Overview.

Suppose a radiation cone volume increment $V_k$ scales by scale factor $v$ such that
\begin{equation} \label{Eq-v-SF for V-extra1-3problem}
vV_k=V_{k+1},
\end{equation} 
and the average radius $r_k$ of a radiation cone  increment $V_k$ scales by $\beta$ such that
\begin{equation} \label{Eq-beta-SF for r-extra1-3problem}
\beta r_k = r_{k+1}.
\end{equation}
The relationship between scale factors $\beta$ and $v$ can be shown as
\begin{equation} \label{Eq-beta-SF relp to vSF-extra1-3problem}
\begin{split}
V_{k+1} / V_k & = v  \\
& =\pi (\beta r_k)^2 sL_k / \pi (r_k)^2 L_k \\
& =\beta^2
\end{split}
\end{equation}
so that $\beta = v^{1/2}.$ 

Suppose the radial length $L$ of a radiation cone increment has scale factor $s$ such that  
\begin{equation} \label{Eq-v-relp-to-s-extra1-3problem}
vV_k = (sL_k)^3 = s^3(L_k)^3
\end{equation}
giving
\begin{equation} \label{Eq-v-1-3Powr-relp-to-s-extra1-3problem}
s=v^{1/3}.
\end{equation}

Using the preceding equations (\ref{Eq-v-SF for V-extra1-3problem}) to (\ref{Eq-v-1-3Powr-relp-to-s-extra1-3problem}), and in particular  $v^{1/2} = \beta$ and $v^{1/3} = s$, describe the scaling that occurs from $V_k$ to $V_{k+1}$ as follows:
\begin{equation}  \label{Eq-V-has-2Dff-SFs-extra1-3problem}
\begin{split}
V_{k+1} & = vV_k  \\
& = \pi (\beta r_k)^2 sL_k   \\
& = \pi \beta ^2 (r_k)^2 sL_k  \\
&= \pi (v^{1/2})^2 (r_k)^2 (v^{1/3}) L_k \\
&= \pi v (r_k)^2 (v^{1/3}) L_k \\
&= v^{4/3} \pi (r_k)^2 L_k \\
&= v^{4/3} V_k.
\end{split}
\end{equation}
We used the substitutions $v^{1/2} = \beta$ and $v^{1/3} = s$ in the fourth line of equation (\ref{Eq-V-has-2Dff-SFs-extra1-3problem}).

Observations and problems arising from equation (\ref{Eq-V-has-2Dff-SFs-extra1-3problem}) include:
\begin{itemize}
	\item The extra $1/3$ in the exponent of the scale factor $v^{4/3} $ for the radiation cone increment volume $V_k$, in the last line of equation (\ref{Eq-V-has-2Dff-SFs-extra1-3problem}) is the same as the exponent for $v$ when used to represent the scale factor  $s= v^{1/3}$ for radiation's radial length $L_k$. This observation suggests that the extra $1/3$ power that leads to $v^{1/3}$ can be explained by the radial trajectory of radiated energy, by the  $s= v^{1/3}$ relationship.
	\item In equation (\ref{Eq-V-has-2Dff-SFs-extra1-3problem}) it impossible that $V_{k+1} = vV_k$ in the first line and at the same time $V_{k+1} =v^{4/3} V_k$  in the last line. In the attempt to  resolve this problem  the $Deg$ notation was developed as outlined in The Overview (and previous versions of The Overview).  Having observed 4/3 power scaling of radiation cone increment volumes, this impossibility requires identification, explanation, and resolution. The explanation set out in The Overview in version 16 of this article was based on observing that for radiation $Deg^*( L_k) = 0$ and  $Deg^*(A_k) = Deg^*(V_k)$. That explanation however did not eliminate the inconsistency between $V_{k+1} = vV_k$ and $V_{k+1} =v^{4/3} V_k$.  To explain the inconsistency it is necessary to distinguish between $Deg$ and $Dim$.
	\item $Deg$ counts scalings; more scalings increases $deg$. $Dim$ compares dimensions; scaling a length, area or volume does not change its $Dim$; scaled  lengths, areas and  volumes form equivalence classes of lengths, areas and  volumes respectively. 	
	\item With respect to the scale factor $\beta$ of the average radius $r_k$, first we find that $\beta = v^{1/2}$ in equation (\ref{Eq-beta-SF relp to vSF-extra1-3problem}). Then in equation (\ref{Eq-V-has-2Dff-SFs-extra1-3problem}), and in the derivations relating to 3/4 metabolic scaling, we square $\beta = v^{1/2}$, ending with $\beta^2 = v$ in equation (\ref{Eq-V-has-2Dff-SFs-extra1-3problem}). $\beta$ is used to scale the radius or average radius, but only $\beta^2$ plays a role in the derivation. Infer that we can dispense with the scaling of the average radius, and need only have regard for the scaling of the average cross-sectional area $A$ of a radiation cone increment. Scaling the average radius is an unnecessary additional step. This partly explains why only the scaling of the average cross-sectional area $A$ of a radiation cone increment plays a role in The Overview, and why it is unnecessary to use scaling of the average radius. Another reason may be that it only matters how much radiation is passing through the cross-sectional area $A$. This observation likely relates to the special role of the radial direction in special relativity. 
	\item The foregoing observations and problems relating to equation (\ref{Eq-V-has-2Dff-SFs-extra1-3problem}) require that a source system be recognized distinct from a receipt system. The problem of the two different scalings, $vV_k$ and  $v^{4/3}$ may be resolved by recognizing the existence of two reference frames. The Overview finds that $Deg^{**}$  helps characterize the entropy of energy. The  $s= v^{1/3}$ relationship about $Dim$ in three dimensional space. 
	\item Even without using the difference between $Dim$ and $Deg$ to explain where the extra 1/3 $deg$ comes from, it was possible in version 16 of this article to arrive at the ratio $4/3 : 1$ for the ratio of degrees of freedom of a radiation cone volume increment to that of a sphere in space. (That was possible  in version 16 by in effect examining equation (\ref{Eq-V-has-2Dff-SFs-extra1-3problem}) first in a radiation-space and then in a space reference frame.) Mathematical consistency points to the ratio  $4/3 : 1$ even without the advantage of differentiating between $Dim$ and $Deg$; the mathematical model is robust even in the face of a shortcoming in the derivation of the ratio b  $4/3 : 1$. 
\end{itemize}

The result of the foregoing list of observations is that if we can resolve the problems they raise--- which is what The Overview attempts to do --- we may be able to explain dark energy. 

\subsubsection{Two systems, two reference frames}

Relative to 3 dimensional volume, an energy distribution system scales by a 4/3 power, and the space into which the energy is distributed scales by the power 1. This suggests characterizing them as distinct systems. To distinguish volumes in the two systems, designate a radiation volume as $V$ (for volume) and a space volume as $\Theta$ (the shape of the Greek letter $\Theta$, almost circular, connotes isotropy). Since they scale differently, denote their scale factors differently, using corresponding small letters for their scale factors.  

The problematic equation (\ref{Eq-V-has-2Dff-SFs-extra1-3problem}) has a radiation cone  volume increment  scale by a 4/3 power in the last line, whereas a spherical isotropic static space scales by power 1. This, by reason of the $4/3 : 1$ ratio, suggests that radiation causes radiation-space to expand, enlarges space, and perhaps creates space. 

This exploration of ideas would be assisted by finding postulates on which to base the mathematics. 

One approach is to suppose that there are two systems in the universe. One, $S$, can be considered the Source or Supply of energy or radiation The other, $\mathpzc{R}$, can be considered the Receipt of the energy or radiation. $S$ grows at a constant rate, at least in its own reference frame. $\mathpzc{R}$ is static. 

In 2011, my article on Isotropy, entropy and energy scaling used a descriptive notation $Deg$ to describe (as opposed to, analyze) the relationship between $S$ and $\mathpzc{R}$:  $Deg(S)= (4/3) Deg(\mathpzc{R})$. This way of expressing the relationship between $S$ and $\mathpzc{R}$ has the advantage of economy but has the disadvantage of not revealing the fractal nature of the relationship; the $4/3 : 1$ ratio occurs at every scale. Another problem with $Deg(S)= (4/3) Deg(\mathpzc{R})$ is that it is   wrong, as discussed in section \ref{Sec-why-SR-R-NOT-S-and-R}; the correct description would be $Deg(S+\mathpzc{R})= (4/3) Deg(\mathpzc{R})$. In 2012, I decided to try using $Deg$ notation for analysis of how degrees of freedom results in an extra 1/3 power for the scaling of radiation cone volume increments. 

An advantage of $Deg(S+\mathpzc{R})= (4/3) Deg(\mathpzc{R})$ apart from economy is that this form of expression leads to questions about cosmogenesis: what starts things going? how does $S$ create $\mathpzc{R}$? what are the fundamental attributes of each? Focusing on degrees of freedom via the notation $Deg$, $Deg^*$,  $Deg^{**}$ and $deg$ avoids these cosmogenesis questions; perhaps that is a superior approach.

Notwithstanding the shortcomings of using the $S$ and $\mathpzc{R}$ terminology, a different approach to the same problem --- increasing the degrees of problem-solving freedom --- may improve the odds of learning something about it. Because of that possibility, I have included reference to  $S$ and $\mathpzc{R}$ in what follows. 

The Overview on the other hand mostly adopts the degrees of freedom point of view. That approach has many advantages for presentation of these ideas. In The Overview, the two systems are treated as two reference frames. The two reference frames, radiation and space, are (dimensionally) related: $Deg^*(V_k) = (4/3) Deg^*(\Theta_k)$.

\section{Equivalence of The NRT and The 4/3 RDFT; The NRT as a special case}

The 4/3 RDFT and the related mathematics of scaling helps to explain 3/4 metabolic scaling and in so doing also provides evidence of the validity of The 4/3 RDFT. Adapting mean path length scaling, that was used to model  lexical growth and The NRT, to  metabolic scaling led to The 4/3 RDFT. Now proceed in the reverse direction, using The 4/3 RDFT to find mean path length scaling.

If mean path length scaling can imply The 4/3 RDFT, and if The 4/3 RDFT can imply mean path length scaling, then each implies the other; they are equivalent or are based on  the same principles. If so, then evidence in support of The NRT is also evidence in favor of The  4/3 RDFT. 

The principle common to The NRT and The 4/3 RDFT is that the degrees of freedom of the scale factor of a homogeneously scaled system multiplies the capacity of the system. The 4/3 RDFT connects to capacity by showing that a Receipt receiving radiation from a source having 4/3 the Receipt's degrees of freedom must grow; the Receipt cannot receive 4/3 of the Receipt's intrinsic degrees of freedom without growing. This suggests that the degrees of freedom affects  Receipt's capacity to accept energy, or for an information system, to accept information. In fact, there is a linear relationship between a system's degrees of freedom and its system-wide capacity. 

Consider the mathematics pertaining to degrees of freedom in the context of a network $\mathpzc{R}$ with $n$ nodes, mean path length $\mu$, an average individual rate  of $r_1$ of information transmission between nodes, and a collective rate of transmission $r_n$. $\mathpzc{R}$'s collective rate $r_n$ is related to $r_1$ as follows:
\begin{equation}
r_n = Deg_\mu^{**}(n) r_1.
\end{equation}
What is $Deg_\mu^{**}(n)$? Since the mean path length of $\mathpzc{R}$ is $\mu$, then
\begin{equation}
Deg_\mu^{**}(n) = \log_\mu(n),
\end{equation}
since that is the number of degrees of freedom available to $\mathpzc{R}$ to form paths with a measuring stick $\mu$ units long. 

In 2007 the mean path length as the scale factor for lexical scaling seemed to be the governing principle. But the mean path length is the scale factor only in the special case that occurs at a particular instant in the life of a network. A general principle is that the degrees of freedom of the scale factor is linearly related to a system's capacity or cumulative energy. Another general principle is that entropy counts the degrees of freedom of a system relative to a quantum of energy.

Since The NRT describes an emergent phenomenon, the possibility of actually deducing The NRT based on the ideas used to derive The 4/3 RDFT is probably close to nil. Supposing a $\log$ formula might exist that relates society's collective knowledge to average individual knowledge and finding that the mean path length $\mu$ works as a scale factor led to The NRT. Mean path length scaling reveals the role of degrees of freedom in calculating a collective network capacity and therefore in  calculating a system's capacity generally.

\section{An algebra of degrees of freedom}\label{secDegOfF100.100}

\subsection{Postulate of two systems versus reference frames} \label{subsec Postulates}

\subsubsection{Systems and reference frames assumptions}

To arrive at a mathematical model of radiated energy distribution, one approach is that set out in the Overview: regard radiative distribution and stationary space, related by two 4/3 theorems, as occurring in distinct but related reference frames. Another approach is to postulate two systems, one that distributes energy homogeneously and isotropically, and another that receives that energy. The approach in the Overview is simpler, requires fewer assumptions and avoids problems that arise in trying to formulate postulates for two systems. 

Postulating two systems provides a different point of view and raise some different issues, including some related to the origins of the universe. I currently prefer the approach in The Overview, mainly because it seems to simplify the task of presentation. It is possible that if the postulational approach can be improved, it might be a more effective way of presenting these ideas than the two reference frames approach. 

Bearing that in mind,  postulates that describe two systems, one radiating energy, the other receiving it, follow. 

\begin{enumerate}
	\item \label{Postulate: S Exists100.100} A system $S$ --- a Supply of radiation --- exists. Its initial point source, a singularity, is designated \{0\}. Whether as a matter of logic \{0\} should or should not belong to $S$ seems to make no difference to the mathematics in this article. But, logically, once \{0\} has `scaled', it no longer exists. Once \{0\} has been scaled, any subsequent scaling must relate to a generation later than the \{0\} generation. So assume \{0\} is not part of the distribution system. Radiation itself has no volume in $S$.
	
The most important feature of $S$ is its radial distribution from every point of $S$. We require that feature because from the discussion to this point, uniform scaling is necessary to the mathematical model, and uniform scaling is a feature of homogeneity and isotropy. For our purposes, any system that distributes energy in that way is considered a radiation system. Borrowing from the reference frame approach, we observe that measurements of scalings within $S$ occur in a \textit{radiation-space reference frame} $S+\mathpzc{R}$ with $\mathpzc{R}$ as  described in the next postulate.
	
In view of the application of the mathematics to various systems, including Brownian motion and the circulatory systems of organisms, it seems the this postulate might be summarized as a postulate of homogeneous radial distribution. 

	\item \label{Postulate: R Exists100.100} A Receipt or Space $\mathpzc{R}$ exists. It is three dimensional and isotropic along all axes for a spherical volume. An unconstrained particle in $\mathpzc{R}$ has three independent degrees of freedom.  $\mathpzc{R}$ itself has no radiation. Based on the mathematics that follows, it seems that $S$ somehow creates $\mathpzc{R}$, but that supposition plays no express role in the inferences that follow. No radiation originates in $\mathpzc{R}$.
	
An important aspect of $\mathpzc{R}$ or space is that it is a stationary environment.  Measurements within that system occur in a space reference frame. A paradigm instance of a Receipt would be the stationary fluid in which randomly moving particles are suspended,  attributes of Brownian motion.
	\item \label{Postulate: S Transmits,R Receives100.100} The energy of each radiation pulse that originates in $S+\mathpzc{R}$  is entirely received without loss of energy by a corresponding part of $\mathpzc{R}$. This is an idealization and simplification that models light transmission. Every  iteration or pulse of radiation homogeneously radiates with a constant length $L_k=L$ into a corresponding \textit{spatial volume} $\Theta_k$ in $\mathpzc{R}$. 
	
This last assumption appears to be equivalent to the law of conservation of energy: an amount of energy measured in a volume in the radiation-space reference frame is the same as  in a corresponding volume in the \textit{space reference frame}. 
\end{enumerate}

It follows that a radiation volume has \textit{both} radiation and volume and so is within, and has features of, both $S+\mathpzc{R}$. The $S$ feature  of $S+\mathpzc{R}$ results in radial distribution and motion. The $\mathpzc{R}$ feature  of $S+\mathpzc{R}$ is that, at an instant (that is, dimensionally), $Deg^*(L_k)= (1/3)Deg^*(V_k)$.

\subsection{The role of $S+\mathpzc{R}$} \label{Sec-why-SR-R-NOT-S-and-R}

The volumes in a hierarchical energy distribution system, such as that represented by a circulatory system, scale by a 4/3 power, compared to the 1 power of the  corresponding energy Receipt space volumes. This different scaling suggests that there are two distinct systems involved in energy distribution. Since volumes in a radiation-space system $S+\mathpzc{R}$ scale by 4/3 and corresponding volumes in the  Receipt $\mathpzc{R}$ scale by 1, it might seem that it is appropriate to describe the relationship as $Deg(S) = (4/3) Deg(\mathpzc{R})$. This section discusses why  $Deg(S+\mathpzc{R}) = (4/3) Deg(\mathpzc{R})$ is a superior  characterization.  The reasons are based on dimension, observations, logic, and the mathematical derivation of The 4/3 RDFT.

Using the idea of dimension, apply the postulate about $S$. For $S$, there is distribution from every point. If we add the requirement of homogeneous scaling, then for a characteristic radiation length $L$, we have $sL_k=L_{k+1}$, and after $k$ scalings that starts with $L_1$ there is a cumulative length of $(k+1)L$. Radiation is a one dimensional system; it resembles a line. Space by postulate has 3 dimensions. Together radiation and space $S+\mathpzc{R}$ have 4 dimensions. 

Observations support the notion that the radiation system $S$ manifests itself or is perceived as part of radiation-space  $S+\mathpzc{R}$. A circulatory system, for example,  is part of an organism. The sun and stars as sources of radiation appear in  radiation-space. 

The derivation of The 4/3 RDFT compares how radiation cone volume increments scale compared to how space volumes scale. Volume is a feature of space, not of radiation. A radiation  cone volume increment is a hybrid thing: it distributes radiation, but is also has volume, so it must be in both  $S$ and $\mathpzc{R}$, here denoted $S+\mathpzc{R}$.

Thus $Deg(S+\mathpzc{R}) = 4$ whereas $Deg(\mathpzc{R}) = 3$, and it follows that 
\begin{equation} \label{Eq-4-4-Dim-SR-R}
Deg(S+\mathpzc{R}) = (4/3) Deg(\mathpzc{R}). 
\end{equation}

The following observations apply to equation (\ref{Eq-4-4-Dim-SR-R}). 

The 4 dimensions of $S+\mathpzc{R}$ resemble the 4 dimensions of space-time. If $S+\mathpzc{R}$ can be identified with space-time, then we may infer that the rate at which radiation occurs, or the scaling of the characteristic radial length (the progress of radiation events in radiation-space), is proportional to time. Since the cumulative effect of scaling the characteristic radial length can be expressed as a logarithm $\log_s(s^k)$, we may infer that, cumulative time in $S+\mathpzc{R}$ (with the proviso that it is unclear that time can be defined for $S$) is proportional to cumulative radiation distance in $S+\mathpzc{R}$, which is the entropy of the energy distribution. From this one might suppose that entropy and the expansion of the universe are in proportion at least in the $S+\mathpzc{R}$ reference frame, and that radiation is responsible for the expansion of the universe. 

Equation (\ref{Eq-4-4-Dim-SR-R}) also describes the difficulty observers face in inferring the existence of the distinct systems $S$ and $\mathpzc{R}$. We do not perceive separate systems $S$ and $\mathpzc{R}$; we perceive one system, $S+\mathpzc{R}$. To reconcile our observations of two systems that we perceive as one requires work arounds such as special relativity (with inertial reference frames relative to each other), and makes deciphering dark energy difficult.  

We may therefore characterize $S$ as $(S+\mathpzc{R}) - \mathpzc{R}$, and characterize $\mathpzc{R}$ as $(S+\mathpzc{R}) - S$. 

The different scaling of an organism's circulatory system compared to its mass provides a clue to the existence of separate systems $S$ and $\mathpzc{R}$. It is difficult, if not impossible, to imagine how one might guess that a clue to a cosmological problem might be found in the scaling of organisms (or gas molecular motion or Brownian motion). The likelier path of discovery is to observe a  cosmological relationship and subsequently connect it to dark energy.

If the foregoing observations have any value at all, that would make considering the 4/3 theorems from the point of view of two systems $S$ and $\mathpzc{R}$ potentially helpful.

\subsection{$S$ and $\mathpzc{R}$ and making inferences about them}

The salient mathematical feature of  $S+\mathpzc{R}$ is homogeneous radiation into (stationary) $\mathpzc{R}$ at a constant length $L$ per radiation iteration or pulse. The salient mathematical features of $\mathpzc{R}$ are, at all scales, its three degrees of freedom (dimensions) relative to a length, its passive receipt of radiation and its isotropy. Passive receipt is intended to mean that the recipient system is unmoving relative to the homogeneously scaling transmitting system. The salient attributes connecting $S$ and $\mathpzc{R}$  are that radiation volumes reside in $S+\mathpzc{R}$ and that each transmission pulse (denoted as a radiation event in The Overview) of a radiation volume in $S+\mathpzc{R}$  has a corresponding  recipient volume in the Receipt $\mathpzc{R}$. The connection between $S+\mathpzc{R}$ and $\mathpzc{R}$ might be better described as $\ell=(4/3)L$, for $\ell$ the characteristic (or quantum) length for  $\mathpzc{R}$  and $L$ the characteristic  (or quantum) length for $S+\mathpzc{R}$. 

No postulates assign any other physical attributes to $S$ and $\mathpzc{R}$. The physical nature of $S$ and $\mathpzc{R}$ are not otherwise supposed. How physical counterparts of these abstractions $S$ and  $\mathpzc{R}$ (if there are such) came into existence, why they have the analogs of these attributes, and why \{0\} begins to radiate are outside the domain of this article.

The advantage of the postulates is simplicity: logical inferences arising from the postulates are spared the clutter of extraneous assumptions. (The Overview is even simpler.) The mathematical reasoning leading to The 4/3 Ratio of Degrees of Freedom Theorem (The 4/3 RDFT) that arises out of these postulates ignores everything else. Observing the universe leads to mathematical abstractions and relationships used to describe the universe. By idealizing the abstractions as far as possible, and so divorcing them from reality, the possibility of identifying the underlying  principle --- The 4/3 RDFT --- is, ironically, increased. 

Modeling radiation-space and space using scaling does not require that the model correspond to what physically occurs in reality. All we may infer is that by using scaling we are able to construct models that help us to infer outcomes based on logic that correspond to observable outcomes. In other words, we cannot be sure that scaling as a means to model reality describes physical events, only that it can model them. 

A general mathematical principle is supported by particular instances of it. Conversely, a mathematical treatment of a particular instance, generalized, proves the general case. This aspect of generalization and particularization will be utilized repeatedly in what follows. 

A mathematical model is constrained by the assumption that the universe is self-consistent. An observation consistent with a model of a feature of the universe supports it. If a model leads to inconsistency with accurate observation or physical law, then the model's inferences or assumptions must be wrong. (On the other hand, sometimes theory can create doubt about observations or previously accepted theories.) If mutually inconsistent implications are compatible with the model, then the model itself can not be right; the universe is not self-contradictory.  A mathematical inconsistency in a theory tells you: try another path. Inconsistencies encountered on applying precursors of these ideas to a variety of physical phenomena suggested corrections and refinements.

To be emphatic: The part of this article following the Overview develops mathematical relationships of abstractions, including those relating to energy distribution system $S+\mathpzc{R}$ and Receipt  $\mathpzc{R}$. Those mathematical relationships reveal mathematical patterns. If the mathematical patterns correspond to measurements of physical systems, infer that the abstract mathematical relationships are analogs of physical relationships and that they provide an approximate description of those physical systems. (Similar considerations make computer simulations useful.) One then may hypothesize that the approximation is a general law, and subject that hypothesis to other tests. 

The utility of this approach arises if the chosen mathematical abstractions usefully idealize physical attributes of the universe, such as those considered in the following section. 

\subsection{Physical correspondences to the postulates}

The mathematical abstraction $S$ grows at the constant rate of length $L$ per radiation event in every direction. At any instant, how $\mathpzc{R}$ receives radiation is therefore undifferentiated as to place --- $\mathpzc{R}$ is homogeneous. 

From another perspective, radiation distribution at the same rate everywhere can be considered a radiation-space reference frame. Similar to $\mathpzc{R}$ as homogeneous radiation Receipt, our physical universe is isotropic to one part in 100,000 (Fixsen 1996) and homogeneous. 

The observed homogeneity and isotropy of our universe, analogous to an energy Receipt, imply a primeval atom (Lema\^itre 1950), a  `big bang' theory, similar to the postulated singularity \{0\} in section \ref{subsec Postulates}. The idea of a singularity \{0\} is analogous to the inference that the singularity \{0\} level for a circulatory system must be the source of energy --- food --- external to the circulatory system. 

The speed of light is the same in all reference frames, similar to the abstract $S+\mathpzc{R}$  radiating at a constant $L$ per radiation event in what may be considered the radiation-space reference frame. Our universe contains energy quanta. Similarly, there is some quantum attribute  in the almost featureless abstraction of $S$ leading to a volume in $S+\mathpzc{R}$ growing by a constant --- quantum --- $L$ per radiation event.

We perceive space to have three dimensions, like $\mathpzc{R}$. The cumulative number of  radiation  events generated from a radiation point of the abstraction  $S+\mathpzc{R}$  is counted by 1$L$, 2$L$, 3$L$, and so on. That resembles  counting  time and, in physical space, measuring distance. 

Postulate \ref{Postulate: S Transmits,R Receives100.100} connects $S$ and $\mathpzc{R}$.  $S+\mathpzc{R}$  is an idealization of the role of an  organism's circulatory system as an energy distribution system --- tubes scaling in number and size from the   heart, with each increment of energy supplied by a level of the circulatory system  entirely received  by a corresponding three dimensional mass (in $\mathpzc{R}$) that is part of the organism. Treating  $S+\mathpzc{R}$  and $\mathpzc{R}$ differently is required because they scale differently; also, one distributes energy, the other receives it. On the other hand, an organism's circulatory system is \textit{part} of the organism, and consists of tissues and tubes with volumes. So the circulatory system should be considered to be an analog of $S+ \mathpzc{R}$. Infer that the distribution of energy by the circulatory system to an organism it is part of is an instance of a principle pertaining to the distribution of energy in the universe because the same mathematical model appears to apply to the distribution of energy both within an organism and within the universe, namely scaling with a 4/3 exponent. 

$S+\mathpzc{R}$ as an idealization of what results in constant energy distribution through an organism's circulatory system arises from the circumstance that in April 2008 I found a 4/3 exponent for the scaling of energy supply via an organism's circulatory system tubes, and later that month noticed the similarity of that to an intermediate step in the proof of Stefan's law which involves black body radiation, as recounted above.

Physical attributes of our universe resemble the features of abstract $S$ and $\mathpzc{R}$, with radiation volumes manifested in a  conjoined  $S +\mathpzc{R}$. The abstractions are based on the postulates of section \ref{subsec Postulates}; mathematical inferences based on those postulates may model attributes of the physical universe. $S + \mathpzc{R}$ in this article denotes the union of $S$ and $\mathpzc{R}$, inhabited by volumes that radiate.

\subsection{Volumes, Lengths, and Scale Factors} \label{Sec-Vol-Length-SF}

The purpose of this section \ref{Sec-Vol-Length-SF} is mainly to develop the degrees of freedom notation used in this article, and  to comment on the notation and related concepts. 

\subsubsection{$S$ initiating $\mathpzc{R}$} \label{Sec-S-Initiating-R}

By Postulate \ref{Postulate: S Exists100.100}, \{0\} initiates homogeneous radiation --- at a constant rate of $L$ per radiation or pulse in  $S+\mathpzc{R}$. After (or concurrent with) this, $S$  comes into existence as the set of radiation events that have already occurred as of a point in time. 

The initial radiation also brings  $S +\mathpzc{R}$ into existence.  $S +\mathpzc{R}$ appears to grow in volume everywhere, based on the homogeneity of  $S+\mathpzc{R}$. It may be incorrect to say $S$ has parts. 

After the initial radiation event, $S$ is within a system  $S+\mathpzc{R}$ combined with the Receipt $\mathpzc{R}$ and yet maintains its own radiative character. $S$ has only one degree of freedom per pulse (or radiation event) relative to the preceding \textit{cumulative} radiation events;  radiation is constrained in  $S+\mathpzc{R}$ to travel along what we perceive as a straight line. 

Infer that, in part, $S$ causes space to grow everywhere  because  radiation volumes within $S+\mathpzc{R}$  scale by a 4/3 power, and then  scaled parts scale, and so on.  The proviso `in part' applies because The 4/3 RDFT plays a role in causing $\mathpzc{R}$ to grow \textit{relative} to $S + \mathpzc{R}$.

There is a push-pull question in regards to $S$ and $\mathpzc{R}$. Does $S$ compel $S +\mathpzc{R}$ to expand, because $\mathpzc{R}$,  with  only one degree of freedom per scaling, does not have enough `room' to accommodate $S +\mathpzc{R}$'s 4/3 degrees of freedom?    Or is  $S$ (or $S$ as it is manifested within   $S+\mathpzc{R}$) compelled to continue to scale by the expansion of $\mathpzc{R}$? Does the push-pull question make sense, and if it does, is it a question that physics can answer? 

Using the concept of degrees of freedom as applied to \textit{motion} tells us that radiation itself has one degree of freedom. This characterizes radiation, namely that in  $S +\mathpzc{R}$  radiation travels radially  in a straight line. The rate of speed of radiation in units of $L$ per radiation event has a one-to-one relationship with time in radiation's reference frame. Time is also linear, and each radiation event adds one unit of time and one radiation length $L$ to the existing cumulative radial radiation length of a radiation cone $G_k$ in  $S+\mathpzc{R}$. This one dimensional aspect of radiation appears to correspond, as noted above, to the one dimensional aspect of time in connection with space-time.

The Overview shows that hypothesizing how $S$ leads to $\mathpzc{R}$ is unnecessary to the mathematics used there.  Mathematics, however,  raises the question: what physical conditions lead to the mathematical model? This section considers what postulates might lay the foundation for the mathematics that ensues

Due to $\mathpzc{R}$'s three dimensions, an unconstrained point moving in $\mathpzc{R}$ has three independent degrees of freedom. Therefore, a spatial point cannot be within $S$ because at each point of  $S$  only one degree of freedom of  motion per radiation event is conferred. 

Looked at in another way, radiation occurs in time, is a dynamic event, involves motion only, not space. There is a one-to-one relationship between lengths $L$, added by $S$ in $S +\mathpzc{R}$, and time. Space is static. A radiation event moves (adds or joins) one generation of space to the preceding generations; or we might say, each next generation of space is embedded in the expanded space that comes into existence.  

A radiation volume that radiates at the rate of $L$ per radiation event must be in both $S$ (as radiation) and $\mathpzc{R}$ (being contained in a volume), here  denoted $S + \mathpzc{R}$.  A radiation volume has four degrees of freedom, three arising from the three dimensions of $\mathpzc{R}$ and one arising from $S$'s one degree of freedom with respect to radiation (or  with respect to radial motion in $S + \mathpzc{R}$). 

The first radiation generation creates new radiation or energy clusters, which in turn radiate or scale. When there have been two generations of radiation from \{0\} there has been only one generation of radiation from the new sub-clusters arising from the first generation radiation after \{0\}. Suppose this to be the case since scaling (radiation) in every generation originates from every part of $S$ in  $S+\mathpzc{R}$.

There are three systems to consider. One dimensional $S$ results in a constant stream of radiation events. In three dimensional $S + \mathpzc{R}$, radiation  has one degree of freedom; the cumulative radial length of a radiation cone increases linearly. $L$ plays a role in the concepts to follow   as a measuring stick for distances in $S + \mathpzc{R}$. $\mathpzc{R}$ is a three dimensional space that receives radiation, but itself does not radiate. $S$ and $\mathpzc{R}$ reside in $S + \mathpzc{R}$.  A radiation volume is not wholly in $\mathpzc{R}$ because $\mathpzc{R}$ itself has no radiation, and is not wholly in $S$ because $S$ is only radiation without volume.   

Some of these provisional ideas probably are not necessary for the mathematics leading to the $4/3$ ratio theorems developed below. The physics of how all this might occur plays no role in this section; only the mathematical machinery for deriving the degrees of freedom of a volume of radiation in $S + \mathpzc{R}$ relative to a corresponding volume in abstract three dimensional space $\mathpzc{R}$ per radiation event matters. 

A radial distance in $S + \mathpzc{R}$ is of necessity (it may be argued) a radial distance in $\mathpzc{R}$, since the spatial and dimensional characteristics of  $S + \mathpzc{R}$ are aspects of, or inherited by, $\mathpzc{R}$ at any given instant, that is, absent a radiation event. Length, area and volume are concepts that conceptually reside in three dimensional space; area and volume can be defined relative to length.  We also rely on the derivation of 4/3 scaling of radiation cone volume increments. Radiation cone volume increments are in $S + \mathpzc{R}$.  Radial space lengths that are of length $\ell$ are measuring sticks for $\mathpzc{R}$'s reference frame.

It  seems more logically sound to describe a radial radiation length $L$ as being in the $S + \mathpzc{R}$ reference frame, rather than in the $S$ reference frame. Perhaps it is best to think of $\ell$ as the space or $\mathpzc{R}$ measuring stick and $L$ as the radiation or $S+\mathpzc{R}$ measuring stick. 

What distinguishes $S + \mathpzc{R}$ from $\mathpzc{R}$ is that radial lengths are differently measured by $L$ and $\ell$; lengths are related by $\ell = (4/3) L$. Think of $\ell$ and $L$ as different lengths used to measure, indirectly, the same amount of energy. 

Designate  as $\ell$ the increment of length in $\mathpzc{R}$'s reference frame that corresponds  to $L$ in $S + \mathpzc{R}$'s reference frame. 

The fraction $\phi$ formed by the ratio of $\ell$ to $L$, to be determined, is:

\begin{equation} \label{Eq-defn-phi-ratio}
\phi = \frac{\ell_k}{L_k}.
\end{equation}

\noindent Equation (\ref{Eq-defn-phi-ratio}) is a consequence of \textit{the correspondence of quantum lengths} between $S + \mathpzc{R}$ and $\mathpzc{R}$, or the correspondence between the quantum lengths of $V_k$ and $\Theta_k$. From equation (\ref{Eq-defn-phi-ratio})

\begin{equation} \label{Eq-defn-phi-ratio-200}
L_k = \frac{\ell_k}{\phi}.
\end{equation}

\noindent \textbf{The dark energy question:} This article's modeling of dark energy reduces to: what is  $\phi$?

\subsubsection{Degrees of freedom of a scale factor}

The idea of degrees of freedom arises  in relation to three dimensional space. A particle has 3 degrees of freedom of motion in a 3 dimensional space. Implicit in this usage is the idea that the degrees of freedom are independent of each other, related to the orthogonality of the three (Cartesian) axes that can be used model three dimensional space. 

Also implicit in this usage is the idea that degrees of freedom of a system is \textit{relative} to degrees of freedom along a single axis or line: an area for example affords two degrees of freedom because points within it can be located by two orthogonal (linearly independent) axes. If there are two particles each with 3 degrees of freedom we say that the 2 particle system located in 3 dimensional space has $2 \times 3 = 6$ degrees of freedom, and so on. These observations are also considered in section \ref{subsubs DF-notation}.

Given the conventional deployment of the term `degrees of freedom', characterizing scaling as exhibiting `degrees of freedom of a scale factor' is a  generalization. Adopting this extended or generalized usage is helpful in modeling and solving various scaling problems. That does not require that degrees of freedom of a scale factor be a faithful description of reality, only that as a mathematical technique its mathematical relationships mirror physical relationships. This section explores degrees of freedom of a scale factor in more detail than does The Overview above. 

By postulate, Receipt (or space) $\mathpzc{R}$ has three dimensions. A radiation volume increment $V_k$ in $S + \mathpzc{R}$  corresponds to spatial sphere $\Theta_k$ in $\mathpzc{R}$. Spheres  scale in relation to each other isotropically along their three orthogonal axes. A sphere is used to model space in relation to scaling because a sphere is isotropic at all scales and space has been measured to be isotropic to one part in 100,000, very nearly isotropic (Fixsen 1996). A sphere is therefore used to model space in the context of scaling because it closely models observed attributes of space. 

With each radiation event that increases radial  radiation  distance (the radial length of a radiation cone $G_k$) by $L$, the radius  of a spatial sphere $\Theta_k$ corresponding to  $G_k$ increases by a radial increment $\ell = \phi L$. 

Consider $\ell$ in its setting in $\mathpzc{R}$. Let $\Theta_{k+1} = \theta_k \Theta_k > \Theta_k$. The scale factor $\theta_k > 1$ since radiation distance cumulatively growing by radial increments of $L$ causes the spatial volume corresponding to a radiation cone volume to grow. Denote a scale factor $\beta_k$ for $\ell$ such that $\ell_{k+1} = \beta_k \ell_k = \ell$ for all $k$. Here $\beta_k$ is the scale factor for the radial increment $\ell$ which is constant, and so  $\beta_k=1$. But $\beta_k$ \textit{relative} to the scale factor $\theta_k$ for the entire sphere (not just its incremental increase in volume) changes as $k$ changes; since the cumulative spatial volume $\Theta_k$ grows with each scaling, $\ell$ relative to distance $d(\Theta_k)$ from the center of $\Theta_k$ becomes a smaller proportion of $d(\Theta_k)$ with each scaling:

\begin{equation} \label{Eq-SF-as-DF100.100}
\begin{split}
\Theta_{k+1} &= \frac{4}{3} \pi (\ell_{k+1})^3 \\
&= \frac{4}{3} \pi (\beta_k \ell_k)^3 \\
&= \frac{4}{3} \pi (\beta_k)^3 (\ell_k)^3 \\
&= (\beta_k)^3 \left[\frac{4}{3} \pi (\ell_k)^3\right] \\
&= \theta_k \Theta_k. 
\end{split}
\end{equation}

\noindent Hence, $(\beta_k)^3 = \theta_k$ and $\beta_k= (\theta_k)^{1/3}$. The same relationship, $\beta_k= (\theta_k)^{1/3}$ that applies to the relationship between a length in a sphere, applies when the scaling of an increment of any spatial volume is compared to the scaling of an increment of spatial length $\ell$. Both reflect $\ell$ having one dimension and space having 3 dimensions relative to a spatial length.

Think of $\beta$ in the preceding example as the \textit{base} of a scale factor, terminology analogous to the idea of the base of a logarithm.   The exponent of the base of a scale factor is its degrees of freedom. For example, $\beta$ is the base of the scale factor  $\beta^k$. To measure scaling we take the exponent of the ratio of scale factors with a common base, as for example in 
\begin{equation}
Deg(\beta^m \ell / \beta^n \ell) =\log_\beta(\beta^{m-n})  \ deg = (m-n) \ deg.
\end{equation}
The dimensionality of $\mathpzc{R}$ can impute degrees of freedom to the base $\beta$ of the scale factor $\beta^m$ of $\ell$, for example, if we describe a volume $\Theta$ as $\beta^{3} \ell^3$. The degrees of freedom of the scale factor of a length in a three dimensional volume compared  to the scale factor of that volume is $1/3$, which we obtain by comparing $\beta_k$ to $(\beta_k)^3$. 

A space volume $\Theta_k$ incrementally adding a constant $\ell$ to its radius is a different circumstance than the situation described in the Overview, in which the side of a cube increases by a constant scale factor $\beta>1$; in that case the scaling of the cube side from one generation to the next is always $\beta$, and the scaling of $\Theta_k$ is always by $\theta = \beta^3$. Regardless of whether $\ell$ is constant or scales by $\beta>1$, the $Dim$ relationship between space volume and a length in space is the same: $Deg^*(\ell_k)= (1/3)Deg^*(\Theta_k)$.

\subsubsection{Degrees of freedom notation} \label{subsubs DF-notation}

By May 2012, it was apparent that degrees of freedom played an important role in mathematically modeling 3/4 metabolic scaling. With this in mind, a notation that compresses ideas relating to degrees of freedom of a scale factor makes it easier to isolate degrees of freedom concepts relevant to the 4/3 scaling of homogeneous energy distribution systems. With easier to manipulate degrees of freedom concepts, one can better focus on the relationship of those concepts.  An appropriate notation will free the observer and reader to focus on  salient principles, without having to translate similar words and expressions into understandable conceptual chunks repeatedly. 

The 4/3 Degrees of Freedom Theorem implies that a fractional  4/3 degrees of freedom is a sensible notion. Fractional degrees of freedom arises due to the comparing of scale factors. Degrees of freedom of a motion within a length, area and volume are described with whole numbers. A particle in a volume has three degrees of freedom \textit{relative} to its motion along a length. If  in space each gas molecule has 3 degrees of freedom of motion, then $n$ gas molecules have $3n$ degrees of freedom.  

We are not as interested in  degrees of freedom within one, two or three dimensions as  in the degrees of freedom of a scale factor \textit{relative} to another scale factor. As mentioned above, the \textit{relativity} of degrees of freedom is \textit{implicit} in the usual definition of dimension: a volume has three dimensions \textit{relative to} a length. 

Before v. 15 of this article, the $Deg$ notation went through various permutations. To devise an appropriate notation,  observe that to compare systems that scale, the \textit{ratio} of scale factors to a common base is effective when the length, area or volume belong to the same system. The comparison of energy densities involving dark energy led to the idea of a $deg$ unit, in order to be able to apply dimensional analysis to calculations. 

Let $Deg$ be a function that compares degrees of freedom of a scale factor for different things; the output is the exponent of the base of the scale factor. The unit $deg$ applies to the output, to track the appropriateness of the ratio.  There are  different contexts to which $Deg$ applies. 
\begin{itemize}
	\item Compare the $Deg$ relationships between length, area and volume; the context is  \textit{dimensionality}, an aspect of space or $\mathpzc{R}$.
	\item Compare how a part of a radiation distribution system scales when the number of radiation events grows; this is a radiation context. 
	\item Compare how a radiation cone volume increment scales to how space scales. The ratio thus obtained, unlike the preceding two cases, can be dimensionless since $\deg$ appears in both numerator and denominator. 
\end{itemize}

The $Deg$ function's use is in evaluating ratios. The unit $deg$ helps to illustrate that. So we have
\begin{equation} \label{Eq-Why-Deg-ofRatio-is-better}
\begin{split}
Deg(X_{k+1} / X_k) &= Deg( \beta^k X_{1} / X_1)  \\
& = \log_\beta(\beta^k) \ deg.
\end{split}
\end{equation}

On the other hand if we wrote
\begin{equation} \label{Eq-Ex-of-wrong-use-of-Deg}
\begin{split}
Deg(X_{k+1}) &= Deg( \beta^k X_{1})  \\
& = \log_\beta(\beta^k) \ deg
\end{split}
\end{equation}
we would have $X_1$ somehow vanishing from the argument of $Deg$ on the right side in equation (\ref{Eq-Ex-of-wrong-use-of-Deg}). The usage in equation (\ref{Eq-Why-Deg-ofRatio-is-better}) is superior to that of equation (\ref{Eq-Ex-of-wrong-use-of-Deg}). The notation $deg$ helps ensure that an equation obeys the rules of dimensional analysis. 

We can define a short form notation $Deg^*$ for comparing the $k^{th}$ to the $(k+1)^{st}$ generation of a scaled system:
\begin{equation}
\begin{split}
Deg^*(X_k) & \equiv Deg(X_{k+1} / X_k)  \\
& = Deg( \beta X_{k} / X_k)  \\
& = \log_\beta(\beta) \ deg \\
& = 1 \ deg.
\end{split}
\end{equation}

To evaluate the degrees of freedom of a system \textit{relative} to \textit{another} let the first scale while the other does not. Then take the logarithm, with the logarithm's base a common scale factor, of the fraction formed of the first over the second.

For example, find the effect of a cube $\Theta_k$ scaling by $\theta_k>1$ relative to a stationary version of itself.
 \begin{equation} \label{EqvVkHasDeg1-100.100}
\begin{split}
Deg\left( \frac{ \Theta_{k+1}}{ \Theta_k} \right) & = Deg \left( \frac{ \theta_k\Theta_k}{ \Theta_k} \right) \\
& =\log_{\theta_k} (\theta_k)  \ deg \\
& = 1 \ deg.
\end{split}
\end{equation}

Compare degrees of freedom  for a cube $\Theta_k$ scaling relative to the side $\ell_k$ which scales by $\beta_k$. As in equation (\ref{Eq-SF-as-DF100.100}), $Deg^*(\Theta_k) = 3Deg^*(\ell_k)$, and $Deg^*(\ell_k) = (1/3) Deg^*(\Theta_k)$.

Suppose that in $\mathpzc{R}$ the area $\mathpzc{A}_k$ of the side of a cube $\Theta_k$ scales by $a>1$. All relative to $\ell_k$, for $\Theta_k$ 

\begin{equation} \label{Eq-Deg-Theta-cf-Area100.100}
\begin{split}
Deg^*(\Theta_k) & = Deg^* (\mathpzc{A}_k) + Deg^* (\ell_k) \\ 
& = 3 Deg^* (\ell_k ) \\
&= 3 \ deg,
\end{split}
\end{equation} 
and $Deg^* (\mathpzc{A}_k) = (2/3) Deg^*(\Theta_k)$.

A system can have a fractional number of degrees of freedom \textit{relative} to another system. In a \textit{growth} context, a radiation cone scales differently relative to uniformly scaling volumes such as cubes and spheres, as section \ref{Sec-Scaling volumes and lengths in S} shows. 

Since we want to be able to compare the scaling of level 1 relative to level $k+1$, we use the abbreviation $Deg^{**}$, as in
\begin{equation}
\begin{split}
Deg^{**}(\ell_k) & = Deg(\beta^k \ell_{1}/ \ell_1) \\
& = \log_\beta(\beta^k)  \ deg \\
& = k \ deg
\end{split}
\end{equation} 
which makes explicit the implicit aspect of $Deg$ that degrees of freedom is determined relative to some other system attribute. It also has the result that the unit involved, such as $\Theta_k$ in the first line of equation (\ref{EqvVkHasDeg1-100.100}) naturally cancels out, rather than having to mandate that $Deg$ is the exponent of a scale factor. 

The notation distinguishes between $Deg( \ell_{k+1}/ \ell_k) \equiv Deg^*(\ell_k) = 1 \ deg$ and $ Deg( \ell_{k+1}/ \ell_1) = Deg(\beta^k \ell_{1}/ \ell_1) = k \ deg \equiv Deg^{**}(\ell_k) = k \ deg$.

The Overview in version 17 of this article adds the additional idea of subscripts for $Deg$ and $deg$, and also a $Dim$ function that is analogous to the $Deg$ function. This is a convenience the dispenses with the necessity of specifying verbally, for example, that equation (\ref{Eq-Deg-Theta-cf-Area100.100}) is all relative to $\ell_k$. The verbal specification can be compressed to a notational specification: $Deg_\ell^*(\Theta_k) = 3 \ deg_\ell$.

\subsubsection{Scaling lengths and volumes in $S+\mathpzc{R}$} \label{Sec-Scaling volumes and lengths in S}

In $S + \mathpzc{R}$ designate the $(k+1)^{st}$ radial increment of length for each pulse of radiation as $L_{k+1}= L_k =L$ for all $k \geq 1$.

In a growing sphere, a given length  scales from its center \textit{in the same way} along each of its three axes. That is not the case for radiation. 

In a radiation volume, a length that is radial from the source grows by  a  constant $L$ from one radiation  volume increment to the next while the diameter of the radiation cross-section orthogonal to the radial length \textit{increases} in size. A \textit{uniformly scaling} spherical volume cannot be used to model  \textit{relative} axial scaling (the scaling along the three axes) of radiation because \textit{scaling} is not uniform for each of  a radiation volume's three axes in the way that scaling is for a uniformly scaling spherical volume. A uniformly scaling spherical volume \textit{can} be used to model a homogeneously scaling isotropic space. 

Let $r_k$ be the \textit{average} radius of a radiation cone volume increment $V_k$ between two points on the radial axis that are at distances $kL$ and $(k+1)L$ from the source. Let the scale factor for $r_k$ be $\beta_k$ and for $V_k$ be $v_k$. $\beta_k>1$ since the cross-sectional area (and diameter, as above noted) of the radiation cone gets bigger as the radial distance from the source increases. The radial  radiation length increment $L$ is  constant. To discriminate between the scaling up of the radius of the average cross-sectional area of radiation as compared to the constancy of the radial increment $L$, the \textit{scaling} of radiation, in contrast to a uniformly scaling spherical space, must be modeled using a radiation cone or a volume that permits one to distinguish between the scaling of the cross-sectional area and the scaling of a length orthogonal to that area. 

The radial length of radiation cone $G_k$ is $kL$. Suppose the radius at the end of $G_1$ farthest from its apex is $r$. The radial length of  $G_k$ increases linearly with $k$. Since a radiation cone $G_k$ is in profile a triangle, the radius of the far end of a radiation cone volume increases in proportion: the radius at the far end of a radiation cone is $kr$. Here we ignore the curvature of the base of the far end of a radiation cone.

Since for $G_k$, radial length $kL$ and far end radius $kr$ increase \textit{linearly}, scaling would not at first instance appear capable of having a portion of a radiation cone volume grow by a power greater than one for volume increments, but it is  capable of that in the space reference frame relative to the radiation-space reference frame; that is the result of 4/3 power scaling for radiation cone volume increments. The expansion of space is made possible by the relative difference in scaling of the two reference frames. 

The scaling from one radiation cone volume increment to the next is $(k+1)/k$. Scaling of a radiation cone volume increment \textit{must} play a role, because we want to compare the scaling of a volume in $\mathpzc{R}$ to the way a radiation cone volume increment scales in $S + \mathpzc{R}$ \textit{per generation}. To derive The 4/3 RDFT it is necessary to compare scaling of radiation volume increment  $V_k$, which is $Deg^*(V_k)$, to the scaling of its corresponding space volume $\Theta_k$, which is $Deg^*(\Theta)$. Define a scale factor $s_k$ for an increment of radiation length $L_k=L$. This reveals that $s_k$ scales differently on a radiation event than in space comparing a length to a volume. The mathematical inconsistency that arises by applying both scalings to the same radiation cone volume increment is resolved by adopting radiation-space and space as distinct (though related) reference frames.

For the  scale factor $s_k$ of $L$ we have (dimensionally): $s_k=(v_k)^{1/3}$. This is an instance of the general case that a scale factor for a length has $1/3$ the degrees of freedom relative to the scale factor for a  volume. Since radiation radial length $L$ is constant, $s_k=(v_k)^{1/3}$ implies that as $k$ increases the radiation cone volume increment $V_k$ takes on a skinnier aspect; $L$'s proportion of the width of $V_k$'s cross-section --- its average diameter --- gets smaller. 

The volume of each radiation cone volume increment $V_k$ can be calculated in different reference frames, in space and in radiation-space). The same energy for $V_k$ in the radiation-space reference frame is in $V_k$ in the space reference frame  by Postulate \ref{Postulate: S Transmits,R Receives100.100}.

Compare the scale factor $\beta_k$ for $r_k$ to the scale factor $v_k$ for $V_k$ upon a radiation event in radiation-space.
\begin{equation} \label{Eq-beta-in v-100.100}
\begin{split}
V_{k+1} &= v_kV_k  \\
&= \pi (\beta_k r_k)^2 L_k \\
&= (\beta_k)^2 \pi (r_k)^2 L \\
&=  (\beta_k)^2 V_k \\
&= a_k V_k.
\end{split}
\end{equation}

\noindent Hence,  in radiation-space $(\beta_k)^2 = v_k$ and $\beta_k= (v_k)^{1/2}$. As well, if $a_k=(\beta_k)^2$ is the scale factor for the average cross-sectional area $A_k$,  $a_k=v_k$.

It follows that in radiation-space: 

\begin{equation} \label{Eq-Deg-Vk-cf-Area-k100.100}
Deg^*(A_k)=Deg^*(V_k), 
\end{equation}

\noindent where $A_k$ is the average cross-sectional area of $V_k$. That is so because  in radiation-space $Deg^*(L_k)=0 \ deg$  ($L$'s  radiation property) since $L_k$ is the same for all $k$.  

Consider the $Deg$ relationship of $V_k$  in radiation-space as follows:
\begin{equation} \label{Eq-Deg-Ratio-V-L}
\begin{split}
Deg^*(V_k) & = Deg(V_{k+1}/V_k) \\
& = Deg( A_{k+1}L/A_{k}L) \\
& = Deg( A_{k+1}/A_{k}) \\
& = Deg^*(A_k).
\end{split}
\end{equation}
Equation (\ref{Eq-Deg-Ratio-V-L}) omits $L$'s scale factor $s_k$ which only plays a role in comparing, dimensionally, how length scales to how volume scales. It follows that, for radiation, $Deg^*(A_k)=Deg^*(V_k)$.

The $Deg^*$ relationship  in radiation-space for a radiation volume and its average cross-sectional area contrasts with the dimensional $Deg^*$ relationship for a uniformly scaling spherical volume in space $\mathpzc{R}$ in equation  (\ref{Eq-Deg-Theta-cf-Area100.100}), where the scale factor along each radial axis is $\theta^{1/3}$. 

For a uniformly scaling spherical volume \textit{in space}, dimensionally 
\begin{equation}
Deg^*(\mathpzc{A}_k) = (2/3) Deg^*(\Theta_k),
\end{equation}
unlike the radiation volume increment relationship in (\ref{Eq-Deg-Vk-cf-Area-k100.100}). Radial length for radiation scales dimensionally by $1/3$ the power of the scaling of a radiation volume and, since radiation is at a steady $L$ per scaling, radial length \textit{relative to} the width of the average cross-section of a radiation cone volume increment is getting smaller.

\subsection{The average incremental scale factor of a cumulation}

Relative to the entire radiation cone $G_k$, a scale factor for \textit{cumulative} radial length $kL$ resulting in increased radial length $[(k+1)/k]L$ would be $(k+1)/k = \sigma$. Here distinguish  the scale factor $s$ for radial length $L$ from one radiation cone volume increment to the next from the scale factor $\sigma$ for the cumulative radial length of $G_k$. 

For an interval of length $kL$ for $G_k$ beginning from about $(k - 1/2)L$ to about $(k+ 1/2)L$, let $\sigma_{av}$ be $L$'s average scale factor.

To characterize the linearity of $G_k$'s cumulative radial length $kL$ as a scaling relationship seems odd. This odd approach is required because there is a dimensional scaling relationship per generation between radiation radial length $L_k$ and $V_k$ that is an outcome of the way $G_k$'s cumulative radial length $kL$ grows. 

$G_k$'s $L_k$ increases linearly  but $L$ has a dimensional \textit{scaling} relationship to a radiation cone volume increment $V_k$. The scaling relationship between $L_k$'s scale factor $s_k$ and $V_k$'s scale factor $v_k$, namely $s_k = (v_k)^{1/3}$ indirectly incorporates the linearity of $G_k$'s cumulating radial length $kL$, which is to say, reflects radial \textit{motion} in $G_k$'s fourth dimension. 

$Deg^*(L_k)=0 \ deg$ and $s_k = (v_k)^{1/3}$ compress information about radiation, respectively in radiation-space and in space reference frames.

The cross-sectional area of the  radiation cone volume increment $V_k$ scales by the square, $(\beta_k)^2 =[(k+1)/k]^2$,  of the base of the scale factor $\beta$ of its radius $r_k$. $V_k$'s scale factor $v_k$ scales in proportion to a cubed scale factor, $ [(k+1)/k]^3$, where the base of  the scale factor is the scale factor for a length.  $v_k$ changes from generation to generation of radiation events. Since the scale factor per generation changes with $k$, it is necessary to add the subscript $k$ to the scale factors.

Suppose that for some system a \textit{cumulative} length or amount of energy $D_k$ scales by a scale factor  $\sigma_{k}$ that might vary from one scaling to scaling next succeeding, such that  $D_{k+1} = \sigma_k D_k$. 

Take $D_{av}$ to be the \textit{mean} cumulative length (or energy) over the interval from $k - (1/2)k + 1$ to $k + (1/2)k$, that is, over an interval of $k$ scalings. Find the average scale factor $\sigma_{av}$ for the interval of $k$ scalings. An incremental scale factor, that is from the $i^{th}$ scaling to the $(i+1)^{st}$ scaling is $(i+1)/i$, and the average scale factor over the interval of $k$ scalings is

\begin{equation}
\sigma_{av} = \frac{ \sum (i+1)/i }{k}
\end{equation}

\noindent chosen so that the average scale factor $\sigma_{av}= (1+k)/k = 1 + 1/k$. Then perhaps

\begin{equation} \label{Eq-Av-SF&D-Nat-log100.100}
D_{k+1} = (\sigma_{av})^k D_{av} =  \left(1 + \frac{1}{k}\right)^k D_{av}
\end{equation}

\noindent over the interval of $k$ scalings. This observation connects the natural logarithm to scaling as discussed in section \ref{subsec-NatLog-LimitProof}. 

\subsubsection{Cumulative length and cumulative scaling} \label{subsubsec-Cum-Length}

In the $S + \mathpzc{R}$ reference frame the distance of the end of the cumulative radial length of the radiation cone $G_k$ from its source is, recalling that over an interval the average scale factor is $\sigma_{av}$, 

\begin{equation} \label{Eq-Dist(G_k)100.100}
\begin{split}
d(G_k) &= (\sigma_{av})^1L_1 + \ldots + (\sigma_{av})^{k}L_1= kL_1 \\
& = \log_{\sigma_{av}}((\sigma_{av})^k) L_1 \\
&=Deg(G_{k+1} /G_1)L_1 \\ 
& \equiv Deg^{**}(G_k)L_1  \\
&= kL_1
\end{split}
\end{equation}

\noindent using $d$ as a measure of the cumulation of any parameter (here suppressing $deg$); here the parameter is length. This measure of a cumulation works where the scale factor, $\sigma_{av}$ in this instance, is constant or where an average scale factor  over an interval is used. In other words, there has to be a relationship leading to $\log_\sigma(\sigma^k)=k$. 

For the cumulative radius of a corresponding spherical spatial volume $\Theta_1$  
\begin{equation} \label{Eq-Dist-Theta-100.100}
\begin{split}
d(\Theta_k) &=  (\beta_{av})^1 \ell_1 + \ldots + (\beta_{av})^{k}\ell_1 = k\ell_1 \\
&= \log_\beta((\beta_{av})^k) \ell_1 \\
&= Deg^{**}(\Theta_k) \ell_1. 
\end{split}
\end{equation}

Over an interval with $k$ scalings, the ratio of the cumulative distances in $\mathpzc{R}$ compared to the corresponding distance in $S + \mathpzc{R}$ is
\begin{equation} 
\frac{k\ell_1}{kL_1} = \frac{\ell_1}{L_1} =\phi
\end{equation}
or for constant $\ell$ and $L$
\begin{equation} \label{Eq-Cum-Dist-Ratio-100.100}
\frac{k\ell}{kL} = \frac{\ell}{L} =\phi.
\end{equation}

\noindent having regard for the definition of $\phi$ in equation (\ref{Eq-defn-phi-ratio}).

Thus $\phi$ measures the amount by which a distance in space $\mathpzc{R}$ scales radially (with hindsight, stretches) --- that is, in every direction --- \textit{relative to} radiation distance in $S + \mathpzc{R}$.

\subsubsection{Cumulative scaling and entropy} \label{subsubsec-CumScaling-Entropy}

In equation (\ref{Eq-Dist-Theta-100.100}), suppose that every length $\ell$ is proportional to an amount of energy, and that the equation expresses a relationship between the cumulative energy $d(\Theta_k)$ and the energy of its $k$ components each having energy proportional to $\ell$.

Take equation (\ref{Eq-Dist-Theta-100.100}) and divide each of the equivalent cumulative energies $d(\Theta_k)$ and $\log_{\beta_{av}}((\beta_{av})^k) \ell$ by $\ell$:

\begin{equation} \label{Eq-entropyAndscaling-100.100}
\frac{d(\Theta_k)}{\ell} =  \log_{\beta_{av}}((\beta_{av})^k).
\end{equation}

The right side of equation (\ref{Eq-entropyAndscaling-100.100}) is the log formula for entropy. The left side, continuing to treat the units as units of energy, is a cumulative energy divided by the unit of energy being scaled. Think of $d(\Theta_k)$ as the $dQ$ of thermodynamics, and think of $\ell$ as proportional to an amount of heat $T$ represented by one degree Kelvin in the particular system being considered. Then generalizing, based on these observations about equation (\ref{Eq-entropyAndscaling-100.100}), one can say entropy is the number of degrees of freedom of an amount of energy relative to a reference amount of energy that is being scaled. 

The observations in this section pertain to Stefan's Law discussed in section \ref{subsec-StefansLaw-Proof}.

\subsection{Terminology of $\Theta_k$}

In this article, $\Theta_k$ represents a spherical volume in $\mathpzc{R}$. In relation to the radiation volume increment $V_k$, $\Theta_k$ represents the spherical spatial volume corresponding to  $V_k$, or which receives $V_k$'s energy. When used in relation to the  entire radiation cone $G_k$, $\Theta_k$ represents the spherical spatial volume corresponding to $G_k$. When we compare $\Theta_{k+1}$ to $\Theta_k$, the context should make it obvious whether we are comparing $\Theta_k$ to a radiation cone $G_k$  or to a radiation cone volume increment $V_k$. 

$\Theta_k$ is also a way of distinguishing the dimensional nature of $V_k$ when in the radiation-space reference frame compared to the dimensional nature of $V_k$  in the space reference frame.

When each radiation cone volume increment $V_k$ corresponds to a spherical spatial volume $\Theta_k$ it is implicit that space grows everywhere; every radiation cone volume increment $V_k$ of every radiation cone creates a corresponding $\Theta_k$.

\subsection{Radiation-space and space, $Deg$ and length}

To compare the  degrees of freedom using the base of the scale factor for a radiation cone volume increment $V_k$ in  $S + \mathpzc{R}$ to the  degrees of freedom of   the base of the scale factor for a corresponding spherical spatial volume $\Theta_k$ in $\mathpzc{R}$ it is necessary to model a radiation volume as a radiation cone.

One might guess: To compare \textit{length} in $S + \mathpzc{R}$ to \textit{length} in $\mathpzc{R}$, as opposed to comparing scaling,  it is necessary to model a radiation volume as a sphere since the \textit{length} of the radius from a point is then the salient homogeneous attribute; per scaling, the distance from the center changes by the same (constant) length $L$. This guess is verified in section \ref{sec-prov-RLT-eq-Vol-Sph-Cone}. The reason why helps reveal some aspects of homogeneity in relation to scaling and degrees of freedom on one hand, and length on the other hand. 

First compare radiation-space and  space in relation to length. Radiation adds a constant unchanging radial length $L$ to $G_k$'s cumulative radial length. For a radiation cone, the radius of the average cross-sectional area of a radiation cone volume increment increases, unlike the unchanging radial length increments. 

The  constancy of radiation manifests itself as a constant radial length increment. In contrast, the homogeneity of space manifests itself as uniform radial scaling in all directions. 

To model a comparison between how radiation and space grow differently,  use a cone for one and a corresponding sphere for the other.  Two ways to proceed are as follows. Have a radiation cone with constant radial radiation length $L$ and compare the scaling of the uniformly scaling sphere to the   scaling of the radiation cone volume increments. Another way is to have a radiation sphere's radius growing by an unchanging length being transmitted into a spatial cone, where the length $\ell$ will be 4/3 that of $L$. 

Using the different shapes of a  cone and a sphere affords a way to model the different way a radiation volume changes, either from the point of view of scaling or of length, compared to space. Cones and spheres do not therefore necessarily provide an intrinsic difference between radiation-space and space; they just provide a way to model the difference in radiation radial length compared to a corresponding space radial length. One comparison is based on a ratio of degrees of freedom; the other is based on a ratio of lengths. 

\subsection{$Deg$ as a conceptual reference frame}

Observations about the exponent of the base of a scale factor in relation to cone and sphere volumes helped in deriving The 4/3 RDFT, as set out in The Overview. 

The cumulative length of a radiation cone is $kL$, and the cumulative length scales as $(k+1)/k$ at the $k^{th}$ level. This compels the use of subscripts $k$ for the scale factor $v_k$ for $V_k$ and the scale factor $a_k$ for $A_k$. It would be conceptually and notationally simpler to drop the subscripts for the scale factors. It may be possible to mathematically justify such an approach. Having used geometry and algebra as a scaffolding to arrive at The 4/3 RDFT, it may be possible to proceed as if the scaffolding is no longer necessary.

Simplified scaling might help in the proof of the natural logarithm theorem based on the limit of scaled intervals, as outlined in \ref{subsec-NatLog-LimitProof}.

\subsection{Using the algebra of degrees of freedom}

Ideas in this section will be deployed in the first proof of The 4/3 Ratio of Degrees of Freedom Theorem (The 4/3 RDFT) in the next section. The 4/3 Ratio of Degrees of Freedom Theorem leads to The 4/3 Ratio of Lengths Theorem, which in turn suggests an explanation for astronomical observations related to dark energy. The 4/3 RDFT itself suggests a possible explanation of vacuum pressure or dark energy.

\section{The 4/3 Ratio of Degrees of Freedom Theorem (The 4/3 RDFT)} \label{subsecThe4/3RDFTderivation100.100}

In section \ref{subsecThe4/3RDFTderivation100.100}, The 4/3 RDFT is first proved in section \ref{Sec-RDFT-proof-using-DF} using degrees of freedom observations. The proofs that follow it are treated as proofs or examples of particular instances of the general proposition that The 4/3 RDFT represents. In this portion of the article, the 4/3 scaling relationship is described as a theorem. In The Overview, it is described as a law which may be a better description of it.

\noindent \textbf{The 4/3 RDFT:}   A radiation volume in  $S +\mathpzc{R}$  scales homogeneously and isotropically transmitting degrees of freedom or energy into Receipt $\mathpzc{R}$. Relative to their respective scale factors, for radiation volume $G_k$ in $S + \mathpzc{R}$ and its corresponding spatial volume $\Theta_k$ in $\mathpzc{R}$, dimensionally $Deg^*(G_k) = (4/3) Deg^*(\Theta_k)$. It follows that $Deg^*(L) = (4/3) Deg^*(\ell)$.
The preceding sentence can be improved by using subscripts as described in The Overview. 

\subsection{Proof based on section \ref{secDegOfF100.100}} \label{Sec-RDFT-proof-using-DF}

$Deg^* (\Theta_k)  =3Deg^*(\ell_k)$ in $\mathpzc{R}$ (dimensionally) by Postulate \ref{Postulate: R Exists100.100}. 

Let $V_k$ be a radiation cone volume increment, part of a radiation cone $G_k$ in $S + \mathpzc{R}$ growing radially by $L$ per radiation event.  Per  radiation event,  when $G_k$ grows by $L$ in radiation-space, it grows radially by $\ell$ in space's reference frame.  

$V_k$ scales by $v_k$ without regard to how $L_k$ scales dimensionally relative to $V_k$. $V_k$'s average cross-sectional area $A_k$ scales by $a_k=(\beta_k)^2$ where $\beta_k$ is the base of the scale factor of the average radius $r_k$ of $V_k$. $\Theta_k$ scales by $\theta_k$ as in section \ref{secDegOfF100.100}. 

The relationship between $v_k$ and $a_k$  in radiation-space is 
\begin{equation}
\begin{split}
Deg\left(\frac{V_{k+1}}{V_k}\right)  &= v_k \\
&= Deg\left(\frac{A_{k+1}L}{A_kL}\right)  = Deg\left(\frac{a_kA_{k}LL}{A_kL}\right)  \\ 
&= a_k
\end{split}
\end{equation}
since in radiation-space $Deg^*(L)=0$.

To calculate how $V_{k+1}$ scales relative to $V_{k}$ in $S + \mathpzc{R}$ on a radiation event, one can attempt to compare scale factors relative to $v_k$ using $Deg$ ignoring the role of $Dim$, but that approach leads to $Deg_V^*(V_k) = (4/3) Deg_V^*(V_k)$.  While this incorrect way of proceeding still leads to a degrees of freedom ratio of $4/3 : 1$, it is necessary to put theory on a secure foundation. The problem arises because the context of $V_k$ scaling can be set in radiation-space in which $a_k=v_k$ and $Deg (L) = 0$ and at the same time in space in which $(a_k)^2/3 = v_k$ and $Deg (L) = (1/3) Deg (V_k)$. Calculation of $Deg$ must take place in a single reference frame. Scaling $V_k$ in a dimensional context repairs the problem. For $Dim$ the arguments of the function are not ratios. Then
 \begin{equation}\label{EqThe4-3RDFT-Dim200.100}
\begin{split}
	Dim_V([V_{k+1}]) & = Dim_V ([(a_kL)V_k] )  \\
	& =Dim_V ( [a_kA_kL][L]) \\
	& =Dim_V ( [V_k]) + Dim_V([L]) \\
	& =Dim_V ( [V_k]) + (1/3) Dim_V([V_k]) \\
&= (4/3) Dim_V([V_k])
	\end{split}
\end{equation}
which in a dimensional context is $Deg_V^*([LV_k]) = (4/3) Deg_V^*([V_k])$ on a radiation event.

\noindent On the other hand, 
 \begin{equation}\label{EqThe4-3RDFT-Theta-100.100}
\begin{split}
	Deg \left( \frac{\Theta_{k+1}}{\Theta_k} \right) &= Deg \left( \frac{\theta_k \Theta_k}{\Theta_k} \right) \\
	& = \log_{\theta_k} \left(\theta_k \right)  \ deg \\
		& = 1 \ deg
	\end{split}
\end{equation}

\noindent In effect, equation (\ref{EqThe4-3RDFT-Theta-100.100}) says that scaling a volume in space does not change its dimension. Relative to their respective scale factors, $Deg(V_k) = (4/3) Deg(\Theta_k)$. Put another way, dimensionally 
\begin{equation} \label{Eq-Ratio-DF-V-Theta100}
\frac{Deg(V_{k+1}/V_k)}{Deg(\Theta_{k+1}/ \Theta_k)} = \frac{(4/3) \ deg}{1 \ deg} = \frac{4}{3} : 1.
\end{equation}

Scaling by a $4/3$ power applies to the scale factor of each volume increment $V_k$ of a radiation cone relative to the corresponding $\Theta_k$ in the Receipt. As shown in The Overview, it is more precise to say that the ratio of degrees of freedom in equation (\ref{Eq-Ratio-DF-V-Theta100}) is a ratio of the degrees of freedom of $V_k$ in the radiation-space reference frame to its degrees of freedom in the space reference frame. In a sense, $\Theta_k$ is a conceptual stand in for $V_k$ in the space reference frame.  The 4/3 scaling relationship arises out of the different scaling of a quantum radial length in the radiation-space and space reference frames. The 4/3 scaling can be detected by considering volumes, but really is about quantum lengths in radiation-space and space reference frames. 

The 4/3 RDFT may also be described this way: the fractal dimension of $G_k$ in $S + \mathpzc{R}$ relative to  a corresponding $\Theta_k$ in $\mathpzc{R}$ is 4/3. This characterization of The 4/3 RDFT has a connection to Brownian motion discussed in section \ref{Subsec-Proof-4-3-RDFT-Brownian} below.

The $V_k$ as a distribution system  are the means by which the radiated quantity is distributed. They are not the energy source.  

The 4/3 RDFT does not apply to the whole radiation cone since the cumulative length $kL$ of a radiation cone $G_k$ scales by $(k+1)/k$  from one generation to the next. Cumulative length of a radiation cone  scales differently than the constancy of $L$ from one radiation cone volume increment $V_k$ to the next. To arrive at The 4/3 RDFT it is critical to recognize that $Deg^*(L_k)=0$, which is a consequence of  $L_k=L$ for all $k$. With constant $L$, a radiation cone volume increment scales differently in radiation-space than a sphere in space. 

Discriminating between the radiation and volume properties of $L$ is required. The radiation property of $L_k$ leads to $Deg^*(L_{k})=0$. The volume property leads to $Deg^*(L_k)=(1/3) Deg^* (V_k) \neq 0$. If the radiation property and the volume properties are not distinguished, then one would have both $V_{k+1}=v_kV_k$ on a radiation event and also $V_{k+1}= (v_k)^{4/3} V_k$  on a radiation event; both cannot be true in the same  reference frame. The mathematics requires two  reference frames.

By $L_k =L$ \textit{not scaling} --- being constant --- radially in the case of a radiation cone, compared to radius uniformly scaling in the case of a spherical volume in space $\mathpzc{R}$, might one expect $\Theta$'s degrees of freedom to be bigger than $V_k$? Section \ref{Section-4/3 RLT-InversenessOf-DF-Length} shows that it is the radial \textit{length} in $\Theta$  that is larger than the corresponding radial length in $V_k$, that is, $\ell_k = (4/3) L_k$. The lesser scaling in $S + \mathpzc{R}$ --- its truncated scaling in the radial direction compared to the scaling of the radius of its cross-sectional area --- compared to scaling in $\mathpzc{R}$ leads to space  distance stretching by $4/3$ \textit{relative to} radiation distance.

\subsection{Proof based on dimensions}

\begin{proof} A cumulative number of radiation events, like cumulative time, is linear, and so $Deg(S)=1 \ deg$. For space $\mathpzc{R}$, $Deg(\mathpzc{R})=3 \ deg$. Therefore 
\begin{equation}
\begin{split}
\frac{Deg(S + \mathpzc{R})} { Deg(\mathpzc{R} )} & = \frac{1 \deg +  3 \ deg}{ 3 \ deg}   \\
&= \frac{4\deg }{ 3 \ deg} \\
&= \frac{4/3 \deg }{ 1 \ deg}.
\end{split}
\end{equation} \end{proof}

The necessary explicit assumption implicit in the foregoing short proof is that distinct reference frames $S$ and $\mathpzc{R}$ exist, and that $S$ appears to us as part of $S + \mathpzc{R}$. 

From The 4/3 RDFT and the assumption that $Deg(\mathpzc{R})=3 \ deg$, $Deg (S + \mathpzc{R}) = (4/3) Deg(\mathpzc{R}) = (4/3) \times 3 \ deg= 4 \ deg$; if $\mathpzc{R}$ has 3 dimensions,  $S + \mathpzc{R}$ has 4 dimensions.  

The 4/3 RDFT is implicit in the perception that time is one dimensional and space  three dimensional, and thus in the four dimensional characterization of space-time.   

\subsection{Proof based on the 4/3 fractal dimension of Brownian motion} \label{Subsec-Proof-4-3-RDFT-Brownian}

Benoit Mandelbrot in his 2012 memoir (p. 247, Mandelbrot 2012) mentions ``My 4/3 conjecture about Brownian motion was chosen in 1998 [for the Mittag-Leffler Institute], when its difficulty had become obvious.'' 

In 2001, Lawler, Schramm and Werner (Lawler 2001) found that the fractal dimension of Brownian motion is 4/3. Their 2001 article sketches proofs that appeared in a sequence of four earlier articles which involved the mathematics of stochastic L\"owner evolution. The mathematics is lengthy, intricate and specialized --- A Guide to Stochastic L\"owner Evolution and its Applications by Kager and Wouter (Kager 2004) is over 75 pages long --- and entirely different than the algebra of degrees of freedom used to prove The 4/3 RDFT.

Lawler (p. 11, Lawler 2004) sets out the assumptions that apply to the stochastic  L\"owner (or Loewner) evolution process as including
\begin{itemize}
	\item Independent increments.
	\item Identically distributed increments. 
	\item Symmetry about the origin. 
\end{itemize}

\noindent These assumptions, if we consider average scale factors as corresponding to the scale factors involved in The 4/3 RDFT, match those leading to the proof of The 4/3 RDFT  using the relative degrees of freedom of scale factors. Lawler, Schramm and Werner's result is a particular instance of the general case. The 4/3 fractal dimension of Brownian motion taken as a specific instance of the general case provides another proof of The 4/3 RDFT.  

\subsection{Proof based on Stefan's Law} \label{subsec-StefansLaw-Proof}

Stefan's Law (also known as the Stefan-Boltzmann Law)  concerns black body energy radiation. In black body radiation both the radiation energy distribution system $S + \mathpzc{R}$ and the Receipt $\mathpzc{R}$ are contained within the same cavity, which reaches an equilibrium temperature. The cavity appears to represent one system, but in fact two distinct reference frames apply. 

An intermediate step in the proof of Stefan's Law, first derived by Boltzmann in 1884 (Boltzmann 1884), is described by Planck (p. 62, Planck 1914) as (using Planck's notation)

\begin{equation}\label{Eq PlanckOnStefansLaw}
\left( \frac{\partial S}{\partial V} \right)_T = \frac{4u}{3T}
\end{equation}

\noindent (and see  p. 742, Allen \& Maxwell 1939). In equation (\ref{Eq PlanckOnStefansLaw}), $\partial S$ represents change in entropy, $\partial V$ represents change in volume, and the subscript $T$ on the left side of the equation indicates that the equation applies at a given temperature.  In equation (\ref{Eq PlanckOnStefansLaw}), the expression $u/T$ gives the number of degrees of freedom in volume density of radiation $u$ (p. 61, Planck 1914) relative to a degree of absolute temperature.

A single degree $T$ of absolute temperature is proportional to the mean math length $\mu$ (not radiation volume density $u$ in the numerator in equation (\ref{Eq PlanckOnStefansLaw}) on the right hand side) as a scale factor of the total energy in a system. The cumulative temperature of a system is proportional to the degrees of freedom in the system, as reflected in the thermodynamic equation,
\begin{equation}
\frac{Q}{T} = S
\end{equation}
for $Q$ a quantity of heat and $S$ entropy. 

If the temperature is $nT= \mu^kT$, then $\log_\mu(\mu^k)=k$. The remarks in section \ref{subsubsec-CumScaling-Entropy} apply. Equation (\ref{Eq PlanckOnStefansLaw}) says that in a cavity, when a radiation volume $G_k$ grows, the change in its entropy --- degrees of freedom --- (the $\partial S$)   relative to the change in volume (the $\partial V$) scales by $4/3$ of $\mathpzc{R}$'s energy density $u$ relative to energy's scale factor $T$, another instance of The 4/3 RDFT. 

On average, within the cavity holding the black body radiation, there is homogeneous radiation or scaling. The setting is analogous to that described by the postulates in section \ref{subsec Postulates}. Proof of Stefan's Law is proof of a particular instance of The 4/3 RDFT.

\subsection{Proof by comparing volume scale factors for $S+\mathpzc{R}$ and $\mathpzc{R}$} \label{subsec-comparing-volume-SFs100.100}

All of the energy transmitted by a radiation cone $V_k$ is received by a corresponding space $\Theta_k$. Suppose energy is proportional to volume, and treat volume units as energy units. On a scaling of $V_k$ and $\Theta_k$, from section \ref{secDegOfF100.100} it follows that $(v_k)^{4/3}V_k = \theta_k \Theta_k$. Since $V_k = \Theta_k$ energy units, $v_k$ cannot be equal to $\theta_k$. This justifies choosing $V_k$'s scale factor to be represented by a symbol  different from $\Theta_k$'s in section \ref{secDegOfF100.100}. It also must be that for the respective scale factors $(v_k)^{4/3} =\theta_k$. That implies that dimensionally $Deg^*(V_k) = (4/3) Deg^*(\Theta_k)$.

\subsection{Proving The 4/3 RDFT with a result in cosmology} \label{subsub-Proving-RDFT-w-Cosmo}

This section relies on some cosmology. In Wang's notation (p. 17, Wang 2010;  p. 64, Ryden  2003), the cosmic scale factor stretching distance is $a(t)$, which in this article is denoted $\phi$. (Treatment of $a$ as a function of time is at odds with The 4/3 RDFT which assumes that the scale factor ratio for $S+\mathpzc{R}$ compared to $\mathpzc{R}$ is $4/3 \ deg : 1 \ deg$ at all scales. Let's pass over that difference for now.)

Wang writes (her notation) that cosmology reasons that for the energy density of matter, $\rho_m$,
\begin{equation} \label{Eq-Cosmo-Rho-m-a3-100.100-2}
\rho_m \propto \frac{1}{a^3}
\end{equation}

\noindent and for the energy density of radiation, $\rho_r$,
\begin{equation} \label{Eq-Cosmo-Rho-r-a4-100.100-2}
\rho_r \propto \frac{1}{a^4}.
\end{equation}

\noindent Identify radiation characterized in equation (\ref{Eq-Cosmo-Rho-r-a4-100.100-2}) with  scaling  in $S + \mathpzc{R}$ and matter as being included in space $\mathpzc{R}$. An energy density for radiation requires there be both a volume and radiation, that is, that there is a radiation volume, which is in  $S + \mathpzc{R}$. Suppose the same energy $E=\rho_r \times V_k = \rho_m \times\Theta_k$.

Matter energy density scales by distance by a fourth power in equation (\ref{Eq-Cosmo-Rho-r-a4-100.100-2})  and by  a third power in equation (\ref{Eq-Cosmo-Rho-m-a3-100.100-2}). The ratio of the powers of the base scale factors implies The 4/3 RDFT. The degrees of freedom of the distance scale factor for radiation in  $\mathpzc{R}$  compared to the degrees of freedom of the distance scale factor for $\mathpzc{R}$ is $4 \ deg:3 \ deg$. The ratio $4 \ deg:3 \ deg$ is equivalent to $4/3 \deg: 1 \deg$ which adds to the difficulty of deriving the appropriate mathematics. The two ratios  suggest each other only in hindsight.

\subsection{A possible proof based on a MIMO network}

In a multiple input multiple output (MIM0) antenna system (such as a cell phone network), if the number of antennas at each node are $M>1$, if `channel matrices are non-degenerate then the precise degrees of freedom $\eta^*_X = \frac{4}{3}M$' (Jafar and Shamai 2008). The set up Jafar and Shamai describe seems analogous to homogeneous radiation from an antenna system to recipient antennas. Their proof is entirely dissimilar to those above. If it is a special instance of The 4/3 RDFT, it is another proof of it. Is it a special instance?

\subsection{Network degrees of freedom and energy rate use} \label{subsec-DF-EnergyUseSF.100}

The Overview noted that for a mean path length $\mu$
\begin{equation} \label{Eq-log-Network-mu-unit201}
Q = \log_\mu(\mu^k) \mu.
\end{equation}
$Q$ supplies the energy for the network's output. Assume there is no loss of energy on transmission of energy to the output network. For time $t$ compare network output rates $Q/t$ to nodal or individual rates $\mu/t$ by dividing equation (\ref{Eq-log-Network-mu-unit201}) through by $t$:
\begin{equation} \label{Eq-log-Network-rate-per-mu202}
\frac{Q}{t} = \log_\mu(\mu^k)\frac{\mu}{t}= \frac{\log_\mu(\mu^k)}{t} \mu.
\end{equation}
$Q/t$ gives the network's rate of output for some parameter proportional to $Q$.

Let $\xi = r(\Theta_{k+1}) / r(\Theta_{k})$ be a ratio of rates $r(\Theta_k)$ of energy use by $\Theta_k$. We can consider $\xi$ to be the base of a scale factor $\xi^k$ such that $r(\Theta_{k+1}) = \xi r(\Theta_{k})$. Then 
\begin{equation} \label{Eq-Scaling-EnergyUse100.100}
r(\Theta_{k+1})  = \xi^k r(\Theta_{1})
\end{equation}

\noindent and $Deg^{*}(r(\Theta_{k}))=1 \ deg$ based on  $\xi$ as the base of scale factor for $ r(\Theta_{1})$.

Postulate \ref{Postulate: S Transmits,R Receives100.100} stipulates that  per scaling the amount of energy transmitted by a volume in $S+\mathpzc{R}$ equals the amount of energy received by a corresponding volume in $\mathpzc{R}$. That implies that 

\begin{equation} \label{Eq-f-Scaling-EnergyUse100.100}
V_k \times r(V_{k}) = \Theta_k \times r(\Theta_k)
\end{equation}

\noindent energy units. Are there phenomena where for equation (\ref{Eq-Scaling-EnergyUse100.100}) $\xi<1$? See section \ref {Section-3-4-Metabolic-Sc} on allometry.

\subsection{The NRT based on The 4/3 RDFT}

The base of the network's scale factors is the network's mean path length $\mu$.  For a network with $n=\mu^k$ nodes: 
\begin{itemize}
	\item There are $\log_\mu(\mu^k) = k$ generations or levels of scaled clusters based on the mean path length $\mu$. 
\end{itemize}
A  cluster of units (of energy or information for example) divides into $\mu$ clusters each with $1/ \mu$ as many units.  

In the particular case of a static network, this double role of $\mu$ presents quandaries: is $\mu$ a scale factor or is it a mean path length; how is it possible for a length such as the mean path length to also be a scale factor? how can it be that the average distance between network nodes is $\mu$ and yet, according to Jensen's inequality,   $\log_\mu(\mu^k) \mu= k \mu$ is the most efficient way of spanning the network?

The answers to these questions occur in preceding sections. 

For the first and second questions, $\mu$ is the base of the scale factor $\mu^k$. $\log_\mu(\mu^k) = k$ gives the number of lengths of size $\mu$ required to span a scaled path through a network's nested hierarchy of clusters scaled by $\mu$. Or, in other words, there are $k$ degrees of freedom associated with $\mu$ if we treat $\mu$ as the base of the scale factors for the network. Thus $\mu$ in this case is both the base of the scale factor and the characteristic length for a network. In this role, $\mu$ as the base of scale factors plays the same role as $L$ as a scale factor, and  as a length plays the same role as radiation length $L$.

For the third question concerning spanning the network, $\mu$ as the base of a scale factor applicable to the network, tells us, by reference to the exponent applied to it, the number of degrees of freedom in the network relative to $\mu$.

\subsection{A theory of emergence}

The 4/3 RDFT applied to homogeneous scaling may help explain the initiation and expansion of a system. The other concept needed is the linear relationship between a network rate and the network's degrees of freedom relative to a characteristic length.

The 4/3 RDFT causes the system to grow, perhaps initiates growth. As the number of scalings increases due to the pressure of  radiation distribution having 4/3 the degrees of freedom of the recipient space, the cumulative degrees of freedom in the recipient system increase. As the degrees of freedom of the recipient system increases, its collective rate increases, as The NRT indicates. 

These observations raise the push-pull question mentioned in section \ref{Sec-S-Initiating-R}.

For a system receiving energy, such as a system of organisms, as an environment ages the amount of stored energy resources (or accumulated store of solved problems) in the environment inhabited by the organisms increases. As the stored or accumulated energy or information resources increases,  available degrees of freedom increase, and the collective rate of change increases. The combined effect of The 4/3 RDFT and The NRT may predispose systems to emerge. 

These observations about emergence when they are applied to networks suggest why there are economies of scale. Larger and more developed societies, economies, languages and sciences have more degrees of freedom. Increasing opportunities to choose among beneficial  degrees of freedom might promote emergence and growth.

\section{Allometry and 3/4 metabolic scaling: A use of The 4/3 RDFT} \label{Section-3-4-Metabolic-Sc}

\subsection{Application of The 4/3 RDFT to metabolic scaling}

Some background to the problem of 3/4 metabolic scaling is discussed in \ref{Sec-Allom-3-4-Met-Sc} in this article. More is contained in Whitfield (2006). Now apply the idea of degrees of freedom of a scale factor in an attempt to account for 3/4 metabolic scaling. 

Treat the tubes of an organism's circulatory system as a uniformly scaled energy distribution system. 

Assume that the energy per unit volume of blood is constant. The same amount of energy flows in each level or generation of tubing. The average cross-sectional areas of a radiation cone's volume increments scale the same way as the cross-sectional areas of a circulatory system's tubes scale from one generation to the next and, likewise, have constant radiation per scaling or tube level. That is, the amount of blood flowing through the area of the cross-section at every level of an organism's circulatory system is the same; this fact is analogous to a constant rate of radiation.

Denote as $(Circ)_{k}$ the circulatory system of organism $k$. For $(Circ)_{k}$ level 1 is the heart as theory suggests. $(Circ)_{k}$ has $k$ scalings (or branchings) so that there are $k+1$ levels including all levels from level 1 to level $k+1$, the level of the capillaries.  Let $V_c$ be the volume of a capillary, and $\Theta_c$ be the spherical volume irrigated by a capillary volume $V_c$. 

Let $\mathfrak{r}_{k+1}$ represent the rate of metabolism of organism $k$. For the base of a scale factor $\xi$ let
\begin{equation} \label{Eq-MetabRate-scaling100}
\begin{split}
\mathfrak{r}_{k+1} & = \xi^{3/4} \mathfrak{r}_k \\
&= \xi^{(3/4)k} \mathfrak{r}_1.
	\end{split}
\end{equation}
\noindent Then
\begin{equation} \label{EqMetabDFScaling100.100}
\begin{split}
   Deg^*(\mathfrak{r}_k) &= Deg \left( \frac{\mathfrak{r}_{k+1}}{\mathfrak{r}_k}\right)  \\
	&= Deg \left( \frac{\xi^{3/4} \mathfrak{r}_k}{\mathfrak{r}_k} \right) \\
	& = \log_\xi(\xi^{3/4} )  \ deg \\
	& = (3/4) \ deg.
	\end{split}
\end{equation}

The energy supply capacity of the circulatory system for organism $k$ is increased by 4/3 power scaling. So $(Circ)_{k}$ with energy $E$ per time unit has capacity to supply energy to its capillaries of $v^{(4/3)k} E$ per time unit, by reason of The 4/3 RDFT. 

The foregoing would account for 3/4 metabolic scaling.

$Deg^*((Circ)_{k}) = (4/3) \ deg$. Then
\begin{equation} \label{Eq-3/4MetScaling100.200}
\begin{split}
(Deg^*((Circ)_{k}) \times  Deg^*(\mathfrak{r}_k) & = (4/3) \ deg \times (3/4) \ deg \\
& = 1 \ deg^2.
	\end{split}
\end{equation}

\noindent Equation (\ref{Eq-3/4MetScaling100.200}) implies that for organisms the product of the $Deg$ of energy supply capacity times the  $Deg$ of the rate of metabolism is invariant for all $k$, that is, for all sizes of organism.  

The likeliest explanation for this invariance involves intracellular chemistry. Cellular temperature affects rates of biochemical reactions (Gillooly 2001). If the biochemistry of cells is uniform or invariant for differently sized animals, which would be a conservative way for evolution to adjust for different body sizes, then the temperature of the cells \textit{in organisms with similar biochemistries} should lie in a narrow range. If the individual cell temperatures for similar organisms are the same, then larger organisms must have a reduced rate of metabolism, in light of equation (\ref{Eq-3/4MetScaling100.200}).

If cellular metabolism did not scale down by a 3/4 power when an animal's size scaled up, the extra 1/3 power (4/3 compared to 1) of energy supplied by the circulatory system compared to that of the cells receiving energy would  overheat the organism's cells. Adaptation in response to the 4/3 scaling of the energy distribution system, lowering the rate of cellular metabolism of organism $k+1$ compared to  organism $k$, avoids cellular damage caused by a body temperature too high and, as well, the extra effort required to obtain food for  unneeded energy. The 4/3 RDFT provides a new, plausible explanation of 3/4 metabolic scaling and the economies of organism scale.

The idea of the Brownian ratchet as a way to supply energy to cellular machinery seems to fit nicely with  The 4/3 RDFT. The exact mechanism and implementation of the Brownian ratchet in connection with The 4/3 RDFT may be subject to experimentation and further theoretical investigation. Is it the case that the 4/3 power increase in energy supply for a larger animal is constrained by a ratchet and pawl mechanism? Perhaps the cells in a larger organism can only take up so much of the $4/3$ power larger energy supply.

Suppose a larger  organism utilizes part of its extra $1/3$ degree of freedom of energy  for growth. Then the ratio of the degrees of freedom of the scale factors for metabolism will be less than $4/3 : 1$, and the exponent in $M^b$ will be higher than $3/4$ (since $b$ is inverse to 4/3), which has been found to be so (p. 871, Hulbert 2004). 

On the other hand, suppose that an organism cannot obtain the full benefit of the extra $1/3$ degrees of freedom (entropy) of its energy distribution system. Then the degrees of freedom  ratio may be somewhat greater than $4/3 : 1$, and the  degrees of freedom ratio for that organism's metabolism will be somewhat less than $3/4 : 1$, approaching $2/3 : 1$ (which might explain why some of the biological data misleadingly implies $2/3$ scaling). 

For an organism, the \textit{source} \{0\} of the energy supply --- food --- is external to the organism. Treating $S$ (instead of $S + \mathpzc{R}$) and $\mathpzc{R}$ as distinct reference frames still leads to a 4/3 ratio per generation implying The 4/3 RDFT but creates a logical gap: the circulatory system  (but not the energy supply source \{0\}) is part of the organism, not separate from it. Since the circulatory system as the energy distribution system \textit{is} both an energy distribution system and also has the attributes of volume and mass, treat the circulatory system as analogous to a radiation distribution system $S+\mathpzc{R}$. 

The metabolic scale factor 3/4 as the sum of an infinite series as supposed in WBE 1997 is incompatible with equation (\ref{Eq-3/4MetScaling100.200}) which is scale invariant. No summing or limiting process is required; the $4/3 \ deg: 1 \ deg$ ratio applies to each level or generation of the organism's circulatory system compared to its corresponding mass. 

WBE's comparing of scale factors combined with mean path length scaling, degrees of freedom of a scale factor, nested uniform scaling and the connection to Stefan's Law led to the derivation of The 4/3 RDFT. Metabolic scaling is one of the phenomena that, possibly, provides a clue to solving dark energy.  

Thus it may be that a feature of the circulatory system that resides in a human body, an object of biological study, models energy distribution of light in the universe in a way that provides a clue to dark energy, an object of astrophysical study.

\subsection{Lexical scaling and allometry issues}

Seeking a way to test the validity of  mean path length scaling as a model of lexical growth led to considering scaling in  biology (Whitfield 2006), metabolic scaling, the ideas in WBE 1997,  and investigating whether 3/4 scaling would emerge from algebraic manipulation of mean path length scaling. 

A first difficulty is that in 3/4 metabolic scaling the exponent of mass $M$ is \textit{fixed} as the mass increases, whereas in mean path length scaling for a system of size $n=\mu^k$ for some $k$, the exponent $k$ \textit{increases} as $n$ increases. Resolving this difference requires recognizing that a single arterial path from the aorta could be thought of as a single set of nested tubes. 

This is the approach taken in the first few versions of this arXiv article in 2008. In April 2008, the scaling approach adapted from lexical scaling gave 4/3 energy supply scaling instead of 3/4 metabolic scaling. 

The discrepancy between measured 3/4 metabolic scaling and 4/3 energy supply scaling is resolved by supposing that as energy supply capacity scales up by 4/3, the metabolic rate scales down by 3/4. The organism's circulatory system volume has a 4/3 entropy (and also energy capacity per unit time and fractal dimension) compared to the capacity of the organism's mass to use energy supplied to it. Since energy supply scales up by a 4/3 power, metabolism must slow by a 3/4 power to maintain the existing temperature of the organism's cells. 

For the mathematics to work, energy distribution and energy recipient systems and their respective scale factors must be distinct. 

An intermediate step of Stefan's Law also has a 4/3 entropy relationship. That suggests that the mathematics leading to the 4/3 entropy of metabolic scaling also applies to black body radiation, and to radiation distribution in general, and that 4/3 scaling for a radiated distribution compared to its recipient system reflects a general law. 

WBE 1997 (West 1997) supposed that capillaries distributed blood to a spherical volume. It was not clear why WBE 1997  used a spherical volume. The similarity of Stefan's law to 4/3 scaling and the involvement of radiated energy distribution suggests that spherical volumes work as distribution volumes in WBE 1997 because they scale homogeneously and isotropically, just as the cosmic background radiation is approximately homogeneous and isotropic. The connections between metabolic scaling and radiation suggest that 4/3 power scaling applies to cosmology. 

Since radiation can in principle increase continually, that implies that a model of metabolic scaling should not treat the initial generation as the capillaries. The reason is that energy that is scaling into smaller clusters would not result in clusters growing from small clusters to larger ones; and distance from the source would not increase by adding generations at the source, but by new generations being added at the points farthest from the source, just as counting adds to an existing cumulation. Thus the heart should be considered the first generation in the energy distribution system.

Several corrections and refinements occurred after these issues were addressed. So it is that this article that concerns dark energy began with problems seemingly remote from cosmology.

\section{The 4/3 Ratio of Lengths Theorem (The 4/3 RLT)} \label{Section-4/3 RLT-InversenessOf-DF-Length}

\noindent \textbf{The 4/3 RLT:}  Assume a radiation distribution system $S + \mathpzc{R}$  scales homogeneously and isotropically into a corresponding Receipt $\mathpzc{R}$. Then for radiation length $L$ relative to its corresponding Receipt (or spatial) length $\ell$, $\ell = (4/3) L$. (In equation (\ref{Eq-Cum-Dist-Ratio-100.100}), $\phi = 4/3$.)

\subsection{The 4/3 RLT based on The 4/3 RDFT}

Version 16 of this article compared, for homogeneous radiation, corresponding lengths $kL$ and $k\ell$ that  represent the same amount of energy use. This assumed that length $kL$ in $S + \mathpzc{R}$ becomes length $k \ell$ in $\mathpzc{R}$. A flaw in this approach is in using $Deg^{**}$ to obtain a length $k$. Length is a feature of radiation-space and of space; length is not appropriate to use for comparing scaling of energy which involves $Deg^{**}$ in radiation-space. The 4/3 ratio applicable to degrees of freedom is per radiation event. While $Deg_V^{*}(V_k)=(4/3)Deg_\Theta^{*}(\Theta_k)$, it does not follow that $Deg_V^{**}(G_k)=(4/3)Deg_\Theta^{**}(\Theta_k)$.

In order to obtain a 4/3 ratio of lengths, it is  best to proceed on the basis of a single radiation event. The ratio of lengths proof in version 16 is invalid. The 4/3 $Deg$ ratio is based on the relative dimensions of radiation-space and space, and $Deg^{**}$ has no role in obtaining that ratio. 

Proceed in a way similar to that set out in The Overview. 

From $Q/T = S$, where $Q=E$ is an amount of energy ($Q$ being heat), $T$ being a quantum amount of energy and $S$ being energy,
\begin{equation} 
E = S_L \times L
\end{equation}
and
\begin{equation} 
E = S_\ell \times \ell.
\end{equation}
Assume that $\ell=\phi L$, that is, that $\ell$ is the image in space of $L$ in radiation-space. 
For the same amount of energy $E$
\begin{equation} \label{Eq-equating-SL-Sell-re-E}
 S_L \times L = S_\ell \times \ell.
\end{equation}

The 4/3 RDFT shows that $Deg_V^{*}(V_k)=(4/3)Deg_\Theta^{*}(\Theta_k)$  and $S_L=(4/3)S_\ell$. Then from equation (\ref{Eq-equating-SL-Sell-re-E}) and substituting $(4/3)S_\ell$ for $S_L$
\begin{equation} 
(4/3)S_\ell \times L =  S_\ell \times \ell
\end{equation}
so $\ell= (4/3) L$. Hence $\phi= 4/3$ in $\ell= \phi) L$.

Consider $L$ and $\ell$ as measuring sticks for the same amount of energy. For the same amount of energy, $L$ turns over 4 times compared to $\ell$'s 3 times, because the ratio of degrees of freedom for a length $L$ in $S + \mathpzc{R}$ compared to a length $\ell$ in $\mathpzc{R}$ is $4/3: 1$. That degrees of freedom ratio requires that the corresponding length ratio is $L : \ell = 3: 4$, so $\ell = (4/3) L$.

Base another perspective on the volume property of $L$ applied to $Deg(V_{k+1} /V_k)=(4/3) \ deg$ (section \ref{subsec-DF-EnergyUseSF.100}). For the scale factor $s_k$, a scaling of $L$ from $k$ to $k+ 1$ is 
\begin{equation} \label{Eq-4-3RLT-fr-SFscompared100.100}
s_k = (v^{4/3})^{1/3}.
\end{equation}

\noindent Thus the radiation length traveled is 
\begin{equation} \label{Eq-4-3RLT-proof-vSF}
\log_v(v^{4/3 \times 1/3}) L = 4/3 \times 1/3 \times L.
\end{equation}
The scale factor for a scaling of $\ell$ from $k$ to $k+ 1$ is 
\begin{equation} \label{Eq-4-3RLT-fr-SFscompared200.200}
\beta_k = (\theta^{1})^{1/3}
\end{equation}

\noindent and for the space length scale factor $\beta_k$ the distance traveled is 
\begin{equation} \label{Eq-4-3RLT-proof-thetaSF}
\log_\theta(\theta^{1 \times 1/3}) \ell = 1/3 \times \ell. 
\end{equation}
\noindent Since the energy of a radiation length is fully transmitted into its spatial length, equating the right sides of equations (\ref{Eq-4-3RLT-proof-vSF}) and (\ref{Eq-4-3RLT-proof-thetaSF}), $(4/3)(1/3) L = (1/3)\ell$; and therefore $\phi = 4/3$. 

The 4/3 ratio applies only radial lengths, because the relationship of radius to area for cross-sectional areas is the same for radiation-space as for space. 

Space --- $\mathpzc{R}$ --- stretches radially by $4/3$ relative to a radial radiation distance in $S+\mathpzc{R}$.

Every increment $L$ of radiation in $S + \mathpzc{R}$ results in radial length in $\mathpzc{R}$ increasing by a constant $\ell$. The 4/3 stretching of space  represents radial stretching in $\mathpzc{R}$ \textit{relative} to radial radiation distance in $S+ \mathpzc{R}$. Characterizing the stretching of space requires the observer to recognize that radiation space distance in $L$ units has a distance image in space  in $\ell$ units; the stretching arises because $\ell = (4/3) L$. 

We can conceive of $L$ and $\ell$ as different measuring sticks for the same distance, but we \textit{perceive} a distance measured with measuring sticks $L$ as shorter than the same distance measured with measuring sticks $\ell$. If this difference between measuring sticks and $L$ and $\ell$ accounts for dark energy, among other observed phenomena, that implies the existence of different reference frames, in which the same light distance is perceived differently. 

The ratio for degrees of freedom of radiation length to a corresponding  space is inverse to the ratio of their lengths. For degrees of freedom of scale factors, for $S+\mathpzc{R}$ \textit{relative} to $\mathpzc{R}$, the scaling is $4/3 =\phi$. For \textit{length}, for $S+ \mathpzc{R}$ \textit{relative} to $\mathpzc{R}$, the scaling is $3/4 =1/ \phi$. If there is a lossless transmission of energy from $S + \mathpzc{R}$ to $\mathpzc{R}$, then the product of the degrees of freedom ratio 4/3 and the corresponding length ratio 3/4 equals 1, that is, is invariant. Does that invariance have some physical or other significance?

\subsection{Equivalence of The 4/3 RDFT and The 4/3 RLT}

Proving The 4/3 RDFT for scale factors implies  scaling for lengths in The 4/3 RLT is inverse to that for degrees of freedom. The 4/3 RDFT and The 4/3 RLT imply each other. Since The 4/3 RDFT and The NRT imply each other, evidence for any one of The 4/3 RDFT, The 4/3 RLT, and The NRT is also evidence for the other two. 

From another perspective, suppose the rate of energy use $\rho(E)=\rho(V_k)=\rho(\Theta_k)$. Then , 
\begin{equation}\label{EqClausius200.300}
	\begin{split}
	\rho(E) &= \rho(V_k) \\
	& = Deg^*(V_k) \times L \; \; \;\; \; \; \; \; \; \; \; \; \; \; \; \;\; \; \left(NRT \right)\\
	& = \left( \frac{4}{3} \right) Deg^*(\Theta_k) \times L \; \; \; \; \; \; \; \;     \left( \frac{4}{3}  RDFT \right)	\\
	& = Deg^*(\Theta_k) \times \left( \frac{4}{3} \right) \frac{\ell}{\phi}\\
	& = Deg^*(\Theta_k) \times \ell \; \; \; \; \; \; \; \; \; \; \; \; \;\; \; ( \rho(V_k) = \rho(\Theta_k))\\
	& = \rho(\Theta_k) \; \; \; \; \; \; \; \; \; \;\; \; \; \; \; \;\; \; \; \; \;\; \; \; \; \; \;\; \; \left(NRT \right).
	\end{split}
\end{equation}
The first and last lines of equation (\ref{EqClausius200.300}) are equal by assumption. $(4/3) \times (1/\phi) =1$, so $\phi = 4/3$. The relations in equation (\ref{EqClausius200.300} ) also work in the reverse direction.

\subsection{Proving The 4/3 RLT: maximum volume cone inscribed in a sphere} \label{subsec-MaxConeProof-RLT}

Fermat posed this problem (p. 238, Fermat Oevres): `pour trouver le c\^one de surface maxima qui peut \^etre inscrit dans une sph\`ere don\'ee', that is, to find the maximum surface area of a cone inscribed in a given sphere. (Cone surface excludes the area of the base of the cone). Lockhart in his recent book (p. 351--356, Lockhart 2012) poses and presents the solution of an equivalent problem: what inscribed cone has maximum volume for a given sphere? If the sphere has radius $r$, then the requisite cone has height (4/3)$r$, which can be proved by using trigonometry and by using calculus. 

Reorient the inscribed maximum volume cone so that its tip is at the center of the sphere. The radial length of the cone is $4/3$ the radius of the sphere according to Fermat's proof. 

Consider a sphere of homogeneous radiation from a point; its radius grows uniformly by $L$ per pulse. Characterize radiation by its homogeneous additions of length $L$ along every radial line from the point; this requires a \textit{radiation sphere}. Since we want to compare radial \textit{radiation} length to  radial spatial length, we must use a \textit{spatial cone} (a cone in the space reference frame) to reveal the difference between radiation length and spatial length.  

Suppose that all the energy from the portion of the sphere  that has the same volume as the inscribed cone is transmitted. Then a radiation length of 1 in the radiation sphere  corresponds to a spatial length in the spatial cone that is 4/3 as great, as in The 4/3 RLT. In other words, a \textit{ portion} of a radiation sphere having the same amount of energy as a maximal inscribed spatial cone gives an inscribed cone that has a height equal to 4/3 the radius of the radiation sphere. 

The mathematical relationship between the ratio of the height of a cone with maximal volume inscribed in a sphere to the sphere's radius can be considered to be another instance of The 4/3 RLT. 

\subsection{Proving The 4/3 RLT: Equal volume sphere and cone} \label{sec-prov-RLT-eq-Vol-Sph-Cone}

The ratio described by The 4/3 RLT also applies to a cone and sphere with equal volumes (as opposed to considering only part of radiation sphere volume, as in section \ref{subsec-MaxConeProof-RLT}). 

The volume of a sphere is
\begin{equation} \label{Eq-Vol-Sphere-formula}
\frac{4}{3} \pi r^3.
\end{equation}

\noindent The volume of a cylinder with height $h$ is
\begin{equation} \label{Eq-Vol-Cylinder-formula}
 \pi r^2 h.
\end{equation}

\noindent Equate equations (\ref{Eq-Vol-Sphere-formula}) and (\ref{Eq-Vol-Cylinder-formula}), letting $h=\phi r$. Then
\begin{equation} \label{Eq-Vol-Sphere-Eq-Vol-Cyl}
\frac{4}{3} \pi r^3 = \pi r^2 \phi r,
\end{equation}

\noindent which implies $\phi = 4/3$, and so $h= (4/3) r$. For the cylinder, now substitute a cone with average radius $r$ over its height. Then the space distance $h=\ell=(4/3) r = (4/3) L$, which is 4/3 of a radiation radial distance, as in The 4/3 RLT. 

This geometric relationship of cone and volume raises the question of how The 4/3 RDFT relates to geometry, and whether the geometry of space arises out of The 4/3 RDFT.

\subsection{Proving The 4/3 RLT using Clausius's mean path length ratio theorem} \label{subsecClausiusMPLTm}

Clausius in his 1858 paper (p. 140 in Brush's translation) remarks:
\begin{quotation}
The mean lengths of path for the two cases (1) where the remaining molecules move with the same velocity as the one watched, and (2) where they are at rest, bear the proportion of $\frac{3}{4}$ to $1$. It would not be difficult to prove the correctness of this relation; it is, however, unnecessary for us to devote our time to it.
\end{quotation}

Gas molecules all moving with the same average velocity correspond to a homogeneously radiating energy distribution in $S+\mathpzc{R}$. Stationary gas molecules correspond to the Receipt, $\mathpzc{R}$. Clausius's mean path length ratio theorem, generalized, is another proof of The 4/3 RLT.
 
Clausius's device of having  a configuration of gas molecules in which only one molecule moves  compared to all gas molecules moving randomly  is analogous to finding relative degrees of freedom by having one system scale (all parts in motion) while the other does not (only one moving gas molecule), as in section \ref{subsubs DF-notation}. I adapted Clausius's  device for section \ref{subsubs DF-notation}'s definition of the relative degrees of freedom of scale factors. The adaptation is achieved by creating a ratio of a system scaled by the base of a scale factor compared to the  system before it has been so scaled, as in $vV/V$.

The system of gas molecules in which only one is moving may be better characterized as a system where the number of equally sized cubic sub-volumes equals the number of moving gas molecules. Then the stationary system has a mean path length based on the average distance between pairs of centers of the cubic sub-volumes. The $n$ moving molecules correspond to a radiation distribution system. The $n$ stationary sub-volumes correspond to static space. 

Clausius's 1860 proof (p. 434, Clausius 1860) is not widely reproduced because early on his result was considered incomplete; it is not included in Brush's compilation of readings  on Kinetic Theory  (1948). To obtain the result Clausius started by finding that for two particles $\mu$ and $m$ and the angle $\vartheta$ between them ``the relative velocity between $\mu$ and $m$'' is

\begin{equation} \label{Eq-Clausius-MPL-Thm100.100}
\sqrt{u^2 + v^2 - 2uv cos\vartheta}
\end{equation}

\noindent and then calculating an integral formed of the product in equation (\ref{Eq-Clausius-MPL-Thm100.100}) multiplied by the number of particles $N$, with the result divided by $N$ to obtain a mean velocity.

The editor of Maxwell's works adds a note (p. 387, Vol. 1, Maxwell, 1890) that Clausius had assumed that the mean relative velocity for the two cases was of the same form, whereas Maxwell assumed the moving particles obeyed a statistical distribution. Longair (p. 7, Longair 2013) observes that Clausius's calculation did not agree with the ratio of the specific heat capacities for a fixed pressure ($C_p$) and constant volume ($C_v$). Clausius's 1857 paper predicted $C_p/C_v$ would be 1.67 whereas experiments revealed it was about 1.4.  Maxwell's innovative statistical approach led to an accurate estimate $C_p/C_v= \sqrt{2}$ and laid the foundation for statistical mechanics. 

In order to model the ratio $C_p/C_v$, the amount of energy in each specific heat capacity is required. It is necessary to take into account the distribution of velocities. Clausius was aware that actual velocities varied: ``we must assume that the velocities of the several molecules deviate within wide limits on both sides of this mean value'' (p. 113, Clausius 1860).  Clausius's result in relation to gas molecules gives a ratio of lengths which does not accord with the measured values for the ratio $C_p/C_v$. It does, however, prove a particular instance of The 4/3 RLT, and, if The 4/3 RLT is valid and connects to dark energy, is on that account  remarkable, a noteworthy addition to his contributions to the science of thermodynamics.

\section{A proposed explanation of dark energy} \label{SecMainDE100.100}

The foregoing suggests that the 4/3 degrees of freedom of $S + \mathpzc{R}$, for volumes, areas and lengths, relative to $\mathpzc{R}$, is a  vacuum pressure --- the pressure of empty space. Cosmologists designate vacuum pressure as $\Lambda$ (capital lambda in the Greek alphabet), the cosmological constant, and as dark energy. They note that pressure, as in vacuum pressure, and energy  have the same units. Since $4/3 \ deg_L: 1 \ deg_L$ arises from radiation having more degrees of freedom than space, the situation can be described this way: for a radiation event, vacuum pressure equals radiation pressure equals the ratio of degrees of freedom of $S + \mathpzc{R}$ relative to $\mathpzc{R}$ for a volume.

This point of view also suggests a relationship between $(4/3) \deg$ and energy. Since The 4/3 RDFT applies at all scales, that suggests that the relationship between $(4/3) \deg$ and energy also applies at all scales.

The following explanation for dark energy is based mainly on The 4/3 RLT.

\subsection{Theory} \label{subsecDE-Theory100.100}

Suppose $\mathpzc{R}$ corresponds to what we conventionally think of as the space part of space-time (or radiation-space) $S + \mathpzc{R}$. Matter resides in $\mathpzc{R}$. Suppose $S$ corresponds to what we perceive as radiation in the radiation part of radiation-space and that a radiation volume in $S + \mathpzc{R}$ creates a vacuum, or \textit{degrees of freedom}, pressure \textit{relative to} a spatial volume in $\mathpzc{R}$.

Cosmology assumes that the energy density of space is the sum $\rho_\Lambda +\rho_\gamma +\rho_M = \rho_c$, where $\rho_c$ is the critical density. Cosmology defines 
\begin{itemize}
	\item $\Omega_\Lambda \equiv \rho_\Lambda /\rho_c$,
	\item $\Omega_{\gamma} \equiv \rho_\gamma /\rho_c$, and
	\item $\Omega_M \equiv \rho_M /\rho_c$.
\end{itemize}
Then $\Omega_{\Lambda} + \Omega_M + \Omega_{\gamma} =1$. $\Omega_{\Lambda}$ is the energy density of dark energy $\rho_\Lambda$ relative to $\rho_c$.

The  currently measured  $\Omega_{\gamma,0}$ (the  0 in $\Omega$'s subscript $\gamma,0$ denotes the present) is only $8.4 \times 10^{-5}$ (p. 67 Ryden 2003), much smaller than the energy densities of matter and dark energy. Accordingly, $\Omega_\Lambda + \Omega_M \approx 1$, disregarding $\Omega_\gamma$ since it is so small compared to measured values for $\Omega_\Lambda$ and  $\Omega_M$.

Then, 
 \begin{equation} \label{Eq-ratioEnDensities200.200}
\begin{split}
\frac{\Omega_\Lambda}{\Omega_M } & = \frac{\rho_\Lambda / \rho_c}{\rho_M / \rho_c}= \frac{\rho_\Lambda}{\rho_M }.
\end{split}
\end{equation}

Construct a spatial cube in $\mathpzc{R}$ (in which matter resides) with side $\ell$ and measure the same cube in $S + \mathpzc{R}$ but using as a measuring stick $L = (3/4) \ell$. Regardless of whether we use $L$ or $\ell$ to measure the side of the cube, the amount of energy in the cube is the same. 

The vacuum's energy density $\rho_\Lambda$ is the \textit{ratio} of the energy density $E / L^3= E / [(3/4) \ell]^3= \rho_\gamma$ of the cube in measured in $S + \mathpzc{R}$ using quantum length $L$ compared to the energy density  $E/\ell^3 = \rho_M$ of the same cube measured in $\mathpzc{R}$  using  quantum  length $\ell$ which is:
 \begin{equation} \label{Eq-ratioEnDensities100.100-2}
\begin{split}
\rho_\Lambda & = \frac{\rho_\gamma}{\rho_M} = \frac{\rho_\gamma/ \rho_c}{\rho_M / \rho_c} = \frac{\Omega_\gamma}{\Omega_M} \\
& = \frac{E /L^3}{E /\ell^3} =\frac{E/[(3/4) \ell]^3}{E /\ell^3} = \frac{4^3/3^3}{1} \\
& = \frac{64}{27} : 1  \approx \frac{0.7033}{0.2967} :1 \\
&= \frac{0.7033}{0.2967}. 
\end{split}
\end{equation}

The mathematics predicts that the ratio of $\Omega_\Lambda$ to $\Omega_M$ should be  about  $0.7033 : 0.2967$.

The cube containing energy $E$ is measured in two ways, one way using the characteristic measuring stick for radiation $L$ and the second way using the characteristic measuring stick for space $\ell$.  This may take some getting used to. 

$\mathpzc{R}$ does \textit{not} stretch relative to itself in terms of the number of characteristic lengths $k \ell$ but only relative to the \textit{number} of corresponding radiation characteristic lengths  $kL$ in $S + \mathpzc{R}$. The distance is the same, but the measurement of it changes depending whether we measure the distance in a radiation-space reference frame using $L$ or in a space reference frame, using $\ell=(4/3)L$. 

The vacuum pressure arises from the dimensional ratio $Deg^*(V_k)/Deg^*(\Theta_k) = (4/3) \ deg / 1 \ deg$. Similarly, vacuum energy density  $\rho_\Lambda$  is the \textit{ratio of relative energy densities} described in equation (\ref{Eq-ratioEnDensities100.100-2}), namely $\rho_\gamma / \rho_M$.  A hurdle raised by the nature of vacuum pressure is the difficulty in distinguishing between the ratios $(4/3) \ deg : 1 \ deg$  and $4 \ deg : 3 \ deg $. It is the \textit{ratio} $(4/3) \ deg : 1 \ deg$ that characterizes dark energy. Dark energy is a consequence of the different degrees of freedom that radiation has compared to space. Clausius may have been the first physicist to observe  this ratio   when he compared the mean path lengths of gas molecules all moving to gas molecules that were stationary. 

The 4/3 RLT makes another prediction. 

Suppose light from a kind of standard candle $SC$ with intrinsic brightness $B$ travels from object $SC_1$ distance $d_1=1$ distance benchmark to Earth. Suppose light from another such object $SC_2$ with the same intrinsic brightness $B$ travels distance $2d_1$ to Earth. Distance $d_1$ is the light distance in $S+\mathpzc{R}$ in terms of radiation-space quantum length units $L$, but radial distance in $\mathpzc{R}$ stretches by 4/3 when evaluated in terms of space quantum length $\ell$. The light from $SC_2$ takes 4/3 as long to reach Earth because of the 4/3 radial stretching of space, so $SC_2$ appears 
\begin{equation}
\frac{B}{(4/3)d_1} = (3/4)B, 
\end{equation}
\noindent 3/4 as bright, or 25\% fainter, relative to $SC_1$ compared to what would be expected if the ratio of corresponding distances between $\mathpzc{R}$ compared to $S+\mathpzc{R}$ was $1:1$ instead of $4/3: 1$. 

A third prediction. Let $\dot{R}(t)$ be a speed such as 50 miles per hour (mph), $t$ an elapsed time such as 2 hours and $R(t)$ as a distance. Then the distance $R(t)$ traveled in 2 hours would be $t \times \dot{R}(t) = 2$ hours $\times 50$ mph $= 100$ miles. If we know the speed and distance, the time required to travel a distance is distance divided by speed equals trip duration,
\begin{equation} \label{EqHubbleReason100.100}
	\frac{R(t)}{\dot{R}(t)} = t,
\end{equation}

\noindent or, in other words, to travel 100 miles at 50 miles per hour takes $(100/50)$ (miles/miles per hour) $=2$ hours.

Apply the same approach to degrees of freedom per time unit. If $R(t)  kL =\phi kL = (4/3)kL$ then $\dot{R}(t) =4/3$ would be the rate of scaling. Since the ticking of each of the $k$ scalings is constant, we can think of $k$ time units as proportional to the radiation distance $kL$. Inverting the left side of  (\ref{EqHubbleReason100.100})
\begin{equation} \label{EqHubbleReason100.200}
	\frac{\dot{R}(t)}{R(t)} = \frac{1}{t}.
\end{equation}

The 4/3 RLT therefore implies
\begin{equation} \label{EqHubbleReason200.300-2}
	\frac{\dot{R}(t)}{R(t)} = \frac{4/3}{(4/3)t} = \frac{1}{t} =H(t),
\end{equation}
where $H(t)$ is Hubble's constant. Equation (\ref{EqHubbleReason200.300-2}) conforms to the usual relationship between speed, time and distance, as described by equation (\ref{EqHubbleReason100.100}).

In the next section we compare the predictions in this section to observations.

\subsection{Observation and the Benchmark Model} \label{subsecDE-Obsvn100.100}

If vacuum pressure results from the 4/3 degrees of freedom of radiation relative to space then vacuum pressure is, for a corresponding volumes, 
\begin{equation} \label{Eq-Vac-Press-ratio}
\frac{Deg(S + \mathpzc{R})}{  Deg (\mathpzc{R})} = \frac{(4/3) \deg}{1 \ deg}, 
\end{equation}
where $\mathpzc{R}$ is space and $S+\mathpzc{R}$ is radiation plus space, that is, radiation-space. 

The Seven Year Wilkinson Microwave Anisotropy Probe (WMAP) (Jarosik et al.) measured (p. 2)  dark energy density $\Omega_{\Lambda}=0.728^{+0.015}_{-0.016}$, with a $68\%$ confidence limit (also  Komatsu 2011). The Planck satellite March 2013  (p. 11, Planck Collaboration 2013) measured dark energy density $\Omega_{\Lambda}=0.686 \pm{0.020}$, with a $68\%$ confidence limit. 

The 4/3 RLT predicts $\Omega_\Lambda = 0.7033$ compared to $\Omega_M$ which lies close to (only $0.0037$ --- one half per cent --- different than) the midpoint, 0.707, of  the Planck and WMAP measurements. The difference in the Planck and WMAP measurements is $0.042$, eleven times greater than $0.0037$. Since the measured isotropy of the cosmic microwave background radiation CMB is good to one part in 100,000, the match between theory and observation should be close. Current measurements of $\Omega_{\Lambda}$ can not tell us that The 4/3 RLT explanation of dark energy is wrong; the midpoint of the WMAP and Planck measurements, too close to be a mere coincidence, suggests that, on the contrary, The 4/3 RLT is valid. If The 4/3 RLT is valid, then so is the equivalent The 4/3 RDFT.

Observations provide a range of estimates for $\Omega_M$ (Sec. 1.6, Weinberg 2008), but the realistic case based on the data is $\Omega_M = 0.3$ ( p. 50, 51, Weinberg 2008). Ryden (p. 67, Ryden 2003) calls the theory based on data  from observations the Benchmark Model. (In her book, Ryden uses $\epsilon$ for energy density instead of $\rho$ and $m$ for matter instead of $M$.) 

The Benchmark Model has  $\Omega_M,0 \approx 0.3$. Hence (in the notation of this article), and following Ryden 
\begin{equation}
	\frac{\rho_{\Lambda,0}}{\rho_{M,0}} = \frac{\Omega_{\Lambda,0}}{\Omega_{M,0}} \approx \frac{0.7}{0.3} \approx 2.3. 
\end{equation}

\noindent In the past (Ryden notes), mass in the universe occupied a smaller volume scaling by a scale factor $a$; mass energy density changed with $\rho_{M,0}/a^3$. 
\begin{equation}  \label{Eq-DE-Obs-100.100}
	\frac{\rho_\Lambda(a)}{\rho_M(a)} = \frac{\rho_{\Lambda,0}}{\rho_{M,0}/a^3}=\frac{\rho_{\Lambda,0}}{\rho_{M,0}}a^3.
\end{equation}
\noindent The implicit assumption  in equation (\ref{Eq-DE-Obs-100.100}) is that $a$ changes with time, consistent with $a$ itself being scaled by a quantum scale factor. (The $a$ used by Ryden corresponds to $R(t)$, the scale factor for the R-W metric.) Using the idea of two reference frames enables us to dispense with the inference that $a(t)$ changes over time. Stretching occurs on every radiation event, for $\mathpzc{R}$ relative to $S+ \mathpzc{R}$ due to their different quantum lengths (their different characteristic measuring sticks). 

(We may suspect therefore that in the two slit experiment the appearance of waves in the case of the two slits is the stretching of space; it is the background  in motion, a background that moves at all scales, space expanding at all scales, that results in what is observed.)

Ryden writes (p. 67), having regard to equation (\ref{Eq-DE-Obs-100.100}): ``If the universe has been expanding from an initial very dense state, at some moment in the past, the energy density of matter and $\Lambda$ must have been equal.'' This implies that the left side of equation (\ref{Eq-DE-Obs-100.100}) equals 1. Then solving for $a =a_{M,\Lambda}$ in that (p. 68, Ryden 2003),
\begin{equation}  \label{Eq-DE-Obs-100.500}
	a_{M,\Lambda} = \left (\frac{\Omega_{M,0}}{\Omega_{\Lambda,0}} \right)^{1/3} \approx \left( \frac{0.3}{0.7}  \right)^{1/3} \approx 0.75 = \frac{3}{4}.
\end{equation}

\noindent Equation (\ref{Eq-DE-Obs-100.500}) matches the result in section \ref{subsecDE-Theory100.100}. Equation (\ref{Eq-DE-Obs-100.500}), however, differs from The 4/3 RLT, which says that the ratio of energy densities has been the same at all epochs. Since equation (\ref{Eq-DE-Obs-100.500}) is true for all epochs, it was also true at the beginning. The reasoning that Ryden describes differs in that respect from The 4/3 RLT.

Second, the 1998 observations of Type Ia supernovae (Fig. 13, Riess 1998) used as standard candles found that these supernovae appeared ``about 25\% fainter, that is, farther away than expected'' (Dark Energy Survey 2013); and Cheng (p. 259, Cheng 2010) as predicted by The 4/3 RDFT.

Third, Hubble's constant $H(t)$ is defined as
\begin{equation}
	\frac{\dot{R}(t)}{R(t)} = H(t)
\end{equation}

\noindent (p. 49, Ryden 2003; p. 42, Bernstein 1995) for the scale factor $R$ in the R-W metric. `Classic' cosmological solutions (p. 37, Liddle 2003) which apply a fluid equation suggest that the elapsed age of the universe $t_0$ for the early radiation dominated universe was  
\begin{equation} \label{EqAgeUni200.100}
t_0=1/2 \times 1/H_0
\end{equation}
\noindent and for the current matter dominated universe is 
\begin{equation} \label{EqAgeUni200.200}
t_0  =2/3 \times 1/H_0 
\end{equation}

\noindent (p. 37, Coles 2002; p. 40, Weinberg 2008). Equation (\ref{EqAgeUni200.200}) implies an age 
\begin{equation} \label{EqAgeUni200.300}
t_0 = (6.5 - 10) \times 10^9 h^{-1} \; \; \mathrm{years}
\end{equation}
\noindent  (p. 84, Coles 2002), $h\equiv H_0/100 \mathrm{km} \mathrm{s}^{-1} \mathrm{Mpc}^{-1}$. $h=0.67$ (p. 11, Planck Collaboration 2013), which, using equation (\ref{EqAgeUni200.300}), leads to an underestimate compared to the currently estimated age of the universe, about 13.7 billion years (p. 84, Coles 2002). Equation (\ref{EqHubbleReason100.200}) gives $H(t) = 1/t$, a result consistent with cosmological observation, and with the usual relationship, described by equation (\ref{EqHubbleReason100.200}), of speed, time and distance. 

The 4/3 RLT thus provides three cosmological explanations or predictions consistent with corresponding observations.

\section{Other possible connections to cosmology}

\subsection{Time, time's arrow, and the universe's expansion}

In $S + \mathpzc{R}$'s reference frame there is a constant increase in the cumulative radiation distance which increases at an invariant rate of $L$ per scaling. The increase in the cumulative radiation distance is unidirectional, that is, it only increases and does not decrease, which via The 4/3 RDFT induces a unidirectional expansion of $\mathpzc{R}$ relative to $S+\mathpzc{R}$. 

Both the expansion of space, induced in $\mathpzc{R}$ by the different scaling of $S + \mathpzc{R}$ compared to  $\mathpzc{R}$, and time are unidirectional,  possibly explaining time's arrow.  ``This seems to suggest that this expansion was produced by the radiation itself'' (Lema\^itre 1927), as  Lema\^itre suggested many years ago. The 4/3 RDFT is consistent with Lema\^itre's suggestion.  

The constancy of the speed of light is mirrored by the constancy of the progress of time in $S$'s and $S + \mathpzc{R}$'s reference frames. Switching from absolute Newtonian time to an absolute speed of light in $S$'s and $S + \mathpzc{R}$'s reference frame illuminates the correspondence between light speed and time. Quantum time must be stretched in space $\mathpzc{R}$ the same way quantum length in stretched in space $\mathpzc{R}$.

Adapt the approach of section \ref{subsecThe4/3RDFTderivation100.100}. Compare how $V_{k}$ scales relative to itself, to how $V_k$ scales relative to  $L_k$.

 \begin{equation}\label{EqTime=Expansion100.100} 
\begin{split}
	Deg \left( \frac{V_{k+1}}{V_{k}} \middle/ \frac{V_{k+1}}{L_{k+1}} \right) &= Deg \left( \frac{(v_k)^{4/3}}{(v_k)^{1/3}} \right) \\
	& = 4 \ deg. 
	\end{split}
\end{equation}

Equation (\ref{EqTime=Expansion100.100}) is another way of showing that $S+\mathpzc{R}$ has four degrees of freedom relative to a radiation length in $\mathpzc{R}$. $\mathpzc{R}$ --- space --- itself scales dimensionally with three degrees of freedom. Equation (\ref{EqTime=Expansion100.100}) provides further justification for looking at radiation-space as four dimensional. 

Constant radiation at the rate of $L$ per scaling in $S +\mathpzc{R}$ (in proportion to the travel time for radiation) results in 4/3 degrees of freedom for a radiation volume in $S+\mathpzc{R}$ relative to a corresponding spatial volume in $\mathpzc{R}$, which leads to the acceleration of the universe. In other words, constant radiation causes the universe to expand and accelerate; constant radiation is also proportional to time in $S + \mathpzc{R}$'s reference frame since $L$ is added at a constant rate in $S + \mathpzc{R}$'s reference frame. 

Years ago Thomas Gold speculated that ``The large scale motion of the universe thus appears to be responsible for time's arrow'' which he thought might be deduced ``from an observation of the small scale effects only'' (Gold 1962). If The 4/3 DFT applies at all scales, then it would apply at the quantum level. At the same time, The 4/3 DFT would explain the expanding universe. 

\subsection{Inflation and the horizon problem}

Examine the horizon problem in cosmology --- the problem of how remote parts of the universe connect --- in $S + \mathpzc{R}$'s reference frame. 

The horizon problem arises because space expands, with the result that over time light speed has a reducing ability to span space. In $S + \mathpzc{R}$'s reference frame however, things are not too far apart for light to reach them. In $S + \mathpzc{R}$'s reference frame, the largest radial distance accrued to date arises from the distance light has travelled since it began its journey. 

Similarly, in quantum mechanics,  entangled particles might be connected in $S + \mathpzc{R}$'s reference frame and only appear to be disconnected when measured in $\mathpzc{R}$'s reference frame. 

The relationship $\ell=(4/3) L$ is scale invariant: $k \ell=k[(4/3) L] = (4/3) k L$. It applies from quantum to cosmological scales. 

The theory of inflation in cosmology attempts to account for the horizon problem and the flatness problem. The factor required for inflation ranges between $4 \times 10^7$ to  $5 \times 10^{29}$ (p. 204, Weinberg 2008). If quantum length $L$ is proportional to Planck time or Planck length, the number of iterations of $L$ in the first few seconds of the universe might be sufficient to account for inflation. 

If dark energy, the (4/3) $deg$ of $S+\mathpzc{R}$ compared to the 1  $deg$ of $\mathpzc{R}$, is the same at all places and times, then the dark energy accelerating the universe today could have inflated the universe at the universe's beginning. The same principles that explain dark energy may also explain the universe's early inflation. It is simpler to suppose one mechanism accounts for inflation and dark energy than to suppose  distinct mechanisms. It is simpler to suppose a constant relationship $\ell = (4/3)L$ at all times than to posit an initial inflation followed by deceleration in turn succeeded by currently observed acceleration. 

\subsection{Special relativity and Bondi's $k$ calculus}

The physicist Hermann Bondi's \textit{Relativity and Common Sense} was  published in 1962 (Bondi 1980). In it he explains special relativity using a $k$ calculus. He takes $k$ to be the ratio of the time interval for light transmission to the time interval for light reception (p. 72). Using $k$, he derives the Lorentz transformations that characterize special relativity (Ch. X). Bohm (Bohm 1996), Freund (Freund 2008), and d'Inverno (sec. 2.7-2.12, d'Inverno 1992) discuss the $k$ calculus in their books on relativity. 

In this section, I will use $m$ instead of $k$ as an exponent of the base $s$ of scale factors $s^m$, to avoid confusion with Bondi's $k$ calculus notation.

Bondi gives an example of two observers Alfred and David at rest relative to each other. Bondi shows that  two inertial observers have the same scale factor $k$, so that a round trip of light flashes starting with Alfred takes $k \times k T = k^2 T$ time units, for $T$ an interval of time. The exponent of $k$ in the $k$ calculus resembles degrees of freedom of the base of a scale factor. If  the  algebra of degrees of freedom is analogous to Bondi's $k$ calculus the algebra of degrees of freedom may apply to special relativity.

A difference between the  algebra of degrees of freedom and Bondi's $k$ calculus is that $s$ is the base of  a scale factor for radiation length $L$ in the algebra of degrees of freedom, whereas $k$ is a ratio of the time intervals for transmission relative to reception. 

Suppose there is a length of time proportional to the base  $s$ of scale factors for $L$, and that degrees of freedom apply to $s$ as in the algebra of degrees of freedom.

For observers $A$ and $B$ at rest relative to one another, each would be scaling by the same power of $s$, say $s^m$. A round trip of light flashes would scale as $(s^m) \times (s^m) = (s^m)^2$. If we set $k \equiv s^m$, this is the same result that Bondi obtains for a round trip light flash in these circumstances. If the two inertial observers are not at rest relative to each other, then the exponents of the base of the scale factor $s$ for each of them would differ.

By then following Bondi's derivation of special relativity, one may infer that the algebra of degrees of freedom leads to a result equivalent to that obtained by Bondi. 

\section{Natural logarithm theorems} \label{secNatLogThms}

Since mathematics is often based on the idealization of natural phenomena, the following derivations of the natural logarithm suggest that there is a physical basis for the natural logarithm. 

Most of these derivations of the natural logarithm seem to use some aspect of the idea of degrees of freedom of the base of a scale factor.

\subsection{Scaling proof}

\textit{Natural logarithm theorem 1:} A homogeneously and isotropically radiating energy distribution system  $S + \mathpzc{R}$  has the natural logarithm as the base of its scale factors.

\begin{proof} Lengths in $S + \mathpzc{R}$  scale with one degree of freedom at any given time: over an interval, for (an average) base $s$   (or $s_{av}$) of a scale factor,  $s^k$ itself scales by its base, $s$. The \textit{rate of change of the scale factor over the interval} $s^k$ is $s$ for each $k$. What applies for $k=1$ applies for all intervals of size 1, i.e. from $k$ to $k+1$:
\begin{equation}
	\frac{ds}{dt} =s,
\end{equation}

\noindent which implies that over that interval $s \approx e$, the natural logarithm. For a large number of scalings,  $s \approx e$ very closely. \end{proof} 

An observation similar to this in connection with lexical scaling led to this scaling proof which arrived unsought. This proof suggested looking for a second proof that might support the validity of the scaling proof. That led to the lexical  benefit proof of the natural logarithm theorem.  

\subsection{Lexical benefit proof}

The idea of the lexical  benefit proof is that the per node benefit of networking must be equal to the per node cost of networking, in terms of energy or information per node. 

\textit{Natural logarithm theorem 2a:} For a finite isotropic network $\mathpzc{R}$, the base of the logarithmic function  describing how information scales when per node network benefit equals network cost is the natural logarithm.\\
\noindent \textbf{Proof:} The contribution of networking to the multiplication of capacity per transmitting node of $\mathpzc{R}$'s $n=\sigma^\eta$ nodes as a proportion of the network's entropy $\eta$ is
\begin{equation}\label{Eq SM-Ec 500.20}
\begin{split}
	 \frac{d \eta}{dn}&=\frac{ d \left[\log_{\sigma}\sigma^{\eta}\right]} {d(\sigma^{\eta})}\\
&=\frac{1}{\ln\left(\sigma\right){\sigma^{\eta}}}.
\end{split}
\end{equation}
The per node reception of the increase in capacity $\eta$ due to networking as a proportion of $\eta$ is $1 / n = 1 / \sigma^\eta$. For an isotropic network, which is an idealized network of maximum efficiency having regard for Jensen's inequality, the contribution to the increase in capacity per transmitting node, as described in Equation (\ref{Eq SM-Ec 500.20}), equals the increase in capacity per receiving node, so 
\begin{equation}\label{Eq SM-Ec 600.20}
\frac{1}{\sigma^{\eta}} =\frac{1}{\ln\left(\sigma\right){\sigma^{\eta}}}\Rightarrow \ln(\sigma)=1\Rightarrow \sigma = e.
\end{equation}

\subsection{Average degrees of freedom proof}

This average degrees of freedom proof is  an updated perspective on the lexical benefit proof, employing the concepts of degrees of freedom and nestedness  instead of the idea of benefit and cost per node.  The NRT implies that the number of degrees of freedom for a whole system is the same as the number of degrees of freedom per node that is nested within the system. That leads to:

\textit{Natural logarithm theorem 2b:} A network homogeneously scaled and its individual nodes have the same number of degrees of freedom, and  are thus scaled by the natural logarithm measured in steps.

\begin{proof}  For a system, for each single scaling, the average number of scalings per node is
\begin{equation} \label{NatLogThm2-100.100}
	\frac{d \log_s(n)}{dn} = \frac{1}{\ln (s) \times n}
\end{equation}
for average scale factor $s$ over an interval.

The average number of degrees of freedom --- scalings --- per node per single scaling for a system of $n$ nodes is also 
\begin{equation} \label{NatLogThm2-100.200}
	\frac{1}{n}.
\end{equation}

Equating equations (\ref{NatLogThm2-100.100}) and (\ref{NatLogThm2-100.200}) implies $s=e$. \end{proof}

\subsection{Nestedness proof}

\textit{Natural logarithm theorem 3:} For a network, the mean path length $\mu =s$ as the base of a scale factor  spans $S$ in one generation $d\eta$, which is equivalent to $\eta$ segments each $\mu$ units; that implies $s = e$.

\begin{proof} The observed equivalence reduces to

\begin{equation} \label{Eq-NatLogThm-using-ds-deta}
	\frac{ds}{d \eta} = \eta \times s
\end{equation}

\noindent which implies $s=e^\eta$, which for $\eta=1$ implies $s=e$. \end{proof}

Here $ds$ is the distance proportional to the scale factor $\mu$ and $d\eta$ is the number of generations to be spanned, so the left side of equation (\ref{Eq-NatLogThm-using-ds-deta}) tells us the value of the mean path length per generation and the right side of the same equation gives us the distance measured in units of $\mu$. This proof takes advantage of the apparent paradox that the left side and the right side of equation (\ref{Eq-NatLogThm-using-ds-deta}) are different ways of spanning the network.

\subsection{Proposed limit proof} \label{subsec-NatLog-LimitProof}

\textit{Natural logarithm theorem 4:}  In equation (\ref{Eq-Av-SF&D-Nat-log100.100})
\begin{equation} \label{Eq-Nat-log-limit.100.100}
D_{k+1} = (\sigma_{av})^k D_{av} =  \left(1 + \frac{1}{k}\right)^k D_{av}.
\end{equation}

\noindent Let $k$ increase. Then
\begin{equation} \label{Eq-Nat-log-limit-incr-n.100.100}
lim_{n \rightarrow \infty}  \left(1 + \frac{1}{k}\right)^k = e.
\end{equation}

If for example $k = 10^6$, the scale factor of $D_{av}$ will be very close to the value of the natural logarithm. The scale factor for a time comparable to Planck time --- about $5.39 \times 10^{-44}$ second --- would accommodate $10^6$ scalings in a small fraction of a second. This may explain why the natural logarithm so commonly is the average scale factor for natural phenomena. The characterization and attributes of the average scale factor used in equation (\ref{Eq-Nat-log-limit.100.100}) requires mathematical justification and perhaps revision.

\subsection{Measured mean path lengths}

The natural logarithm theorems imply that the measured mean path lengths  for isotropic information radiating networks, measured as the number of steps for nodes to connect on average, should be close to the value of $e \approx 2.71828$. That observation combined with The 4/3 RLT might explain some measurements of mean path lengths for actual networks. 

The mean path length for English words has been measured as 2.67 (Ferrer i Cancho 2001), for neurons in \textit{C. elegans} as 2.65 (Watts \& Strogatz 1998) and for the neurons of the human brain as 2.49 (Achard et al. 2006), all close to the value of $e$, consistent with these networks being isotropic energy (or information) distribution systems. 

A human network receiving information (a Receipt) should have a mean path length equal to 
\begin{equation}
	\frac{4}{3} \times e = \frac{4}{3} \times 2.71828 \approx 3.624.
\end{equation}

\noindent The mean path length for a network of 225,226 actors (Watts \& Strogatz 1998) was measured as 3.65, pretty close to 3.624, less than one per cent different. These measurements are consistent with The 4/3 RLT, where the mean path length $\ell$ of the Receipt was 4/3 the mean path length of the information distribution system, so that $\ell = (4/3) L$. Concepts that relate to expansion of the universe thus appear to connect to the way that information is distributed within societies.

\subsection{Efficiency and the number of sub-systems}

According to The NRT and Jensen's Inequality, for maximum efficiency the natural logarithm should scale systems;  the number of subsystems per generation should equal the natural logarithm. But an organism might have a circulatory system, digestive system, and respiratory system, for example. Three systems departs from the ideal scaling afforded by scaling by the natural logarithm, approximately 2.71828. Does that mean that organisms, for example, cannot achieve maximal efficiency, because a whole number of their sub-systems can not equal the natural logarithm? 

Natural logarithm scaling can be achieved if the functions of the sub-systems that are discretely named overlap. One should find that for sub-systems such as those comprising an organism there is frequently overlap in their different functions; part of the reason may be that the most efficient scaling is by the natural logarithm. 

\section{Entropy}

\subsection{Transmitting degrees of freedom} \label{subsection-DF-matching-principle}

All of the foregoing suggests that energy transmission can be modeled by the transmission of degrees of freedom. Per radiation event, 4/3 degrees of freedom of a scaling radiation distribution system cannot fit into a space that scales with only 1 degree of freedom, unless  space distance grows radially by 4/3 relative to a corresponding radiation length. For one degree of freedom of some system to transmit exactly, the other system must be of the same kind -- spatial or radiating -- and must also have  one degree of freedom, a matching principle. 

If one system is a homogeneously radiating system and the other an isotropic recipient system, then the product of the ratio of their relative degrees of freedom and the ratio of their relative lengths is one. 

\subsection{Definitions of entropy}

The historical development of entropy began with Sadi Carnot's 1824 monograph (Carnot 1960) that described an ideal heat engine modelled on a steam engine. In 1848, William Thomson, Lord Kelvin (Kelvin 1848) devised an absolute temperature scale; a degree Kelvin varies in proportion to the volume of an ideal gas. 

Equipped with the concept of absolute temperature, for an ideal heat engine's reversible heat cycle Clausius in 1865 (p. 331, Clausius 1865) derived, in effect, that

\begin{equation} \label{Eq-entropy-equality-dS-t}
\frac{dQ_1}{T_1} = dS_1 = dS_2=\frac{dQ_2}{T_2},
\end{equation}

\noindent where $S$ represents entropy, $dS$ represents change in entropy, $Q$ represents heat, $dQ$ represents change in heat, $T$ is temperature in degrees Kelvin, subscript 1 is for the portion of the ideal heat cycle during which heat is removed, and subscript 2 is for the portion of the ideal heat cycle during which heat is added. Clausius coined the term entropy  (p. 111-135, 357, Clausius 1865). 

Boltzmann's \textit{H} Theory (Boltzmann 1872, 1898) led to a probability and logarithm-based description of entropy.  Planck gives a derivation of the logarithm-based characterization of entropy  (Part III, Ch. I, Planck 1914).

The physical nature of entropy (Vol. 1, Ch. 44, Feynman 1963)  is obscured by the use of the degree of absolute temperature  in equation (\ref{Eq-entropy-equality-dS-t}). 

The degree Kelvin is derived from the Centigrade scale that assigns one degree between the freezing and boiling of water, based on the invented concept of temperature. 

Entropy as the number of scalings of a scale factor of a designated quantum amount is simpler. A simpler characterization of a concept is easier to understand and to deploy as a problem solving tool, and, if it better and more fundamentally describes physical reality, improves the likelihood that the concept will help  reveal an underlying  physical process. 

\subsection{Entropy as degrees of freedom}

The matching principle described in section \ref{subsection-DF-matching-principle} coupled with perfect efficiency (an assumption in Carnot's ideal heat engine) requires that  
\begin{equation} \label{Eq-entropy-equality-principle-appld}
Deg\left(\frac{dQ_1}{T_1}\right) = Deg\left(\frac{dQ_2}{T_2}\right).
\end{equation}

Consider equation (\ref{Eq-entropy-equality-dS-t}).  For a given ideal heat engine with temperature $T_1$ higher than temperature $T_2$, with $\epsilon$ a unit of energy and $e$ the base of the scale factors, let
\begin{equation} \label{EQ-dQ1-in-E1}
dQ_1= \log_e(e^{x+m+n})\epsilon \text{ and } T_1= \log_e(e^{m+n})\epsilon \text{ and} 
\end{equation}
\begin{equation} \label{EQ-dQ2-in-E1}
dQ_2= \log_e(e^{x+m})\epsilon \text{ and } T_2= \log_e(e^{m})\epsilon. 
\end{equation}
Then
\begin{equation} \label{EQ-dS1-dS2-in-E1}
Deg\left(\frac{dQ_1}{T_1}\right) = Deg\left(\frac{dQ_2}{T_2}\right)=x  \ deg 
\end{equation}
\noindent relative to the base  $e$ of the scale factors. When temperature $T_2=\log_e(e^{m})\epsilon$ increases to $T_1= \log_e(e^{m+n})\epsilon$, equality of the left hand side and the middle expression in equation (\ref{EQ-dS1-dS2-in-E1}) must result since the heat engine's working substance matches the temperature of the heat reservoir it is brought into contact with. 

This suggests that implicit in  the Clausius characterization of entropy for an ideal heat engine is the concept of degrees of freedom.

\section{Discussion}

\subsection{How to describe The 4/3 RDFT}

Which is the best characterization of The 4/3 RDFT: 
\begin{itemize}
	\item $Deg ( S ) = (4/3) Deg ( \mathpzc{R} )$, 
	\item  $Deg_S ( S + \mathpzc{R} )  = (4/ 3) Deg_\mathpzc{R}( S + \mathpzc{R} )$,
\item $Deg( S + \mathpzc{R} ) =  (4/3) Deg ( \mathpzc{R} )$  or
\item $Dim( S + \mathpzc{R} ) =  (4/3) Dim ( \mathpzc{R} )$?
\end{itemize}

The first choice is likely wrong. The 4/3 RDFT involves the relative degrees of freedom of a radiation volume compared to a spatial volume, and the system $S$ cannot be characterized as a volume, because from the dimensional point of view it has (like time) only one degree of freedom. A radiation volume can only be in $S + \mathpzc{R}$ since it has a volume, and so is not in $S$ and has radiation, and so is not in $\mathpzc{R}$. Radiation, space and time occur together in our perception, inhabited by stars and space and planets. This observation favors using $S + \mathpzc{R}$ instead of $S$ to describe where radiation volumes reside. The model of dark energy requires a comparison of the degrees of freedom of a radiation volume \textit{relative to} a corresponding space. 

Astronomical observation also suggests the first choice is wrong. If the first choice were correct, then it would follow that $\rho_\gamma / \rho_M$ would be $64:27$, but $\rho_\gamma$ has been measured to be much smaller than $\rho_M$. 

The second choice, using a subscript for $Deg$ emphasizes that The 4/3 RDFT compares a radiation volume in $S + \mathpzc{R}$ to a spatial volume in $\mathpzc{R}$. $Deg_\mathpzc{R}(S + \mathpzc{R}) = Deg(\mathpzc{R})$ so a subscript $\mathpzc{R}$ for $S + \mathpzc{R}$ unnecessary; just use $Deg(\mathpzc{R})$. $S + \mathpzc{R}$ when compared to $\mathpzc{R}$ necessarily connotes a radiation volume; so a subscript $S$ is unnecessary. 

The advantage of the third choice over the second is succinctness.

Based on the distinctions between $Deg$ and $Dim$, the last choice is probably most accurate. Radiation-space has four dimensions. Space has three dimensions. The ratio of their dimensions, relative to a volume in space, is $4/3: 1$. The third choice is acceptable if one bears in mind that $Deg$ is being used to characterize $Dim$ applied to equivalence classes of lengths, areas and volumes. 

\subsection{Dark energy's difficulty}

If dark energy can be solved using existing theories of cosmology, it must be a hard problem because it has so far defied solution.

If dark energy can not be solved using the existing theories of cosmology, then trying to solve it within cosmology is currently impossible. 
 
It is impossible to know whether or not there is a missing theory that will solve dark energy without having a theory to test. Suppose it could somehow be inferred that there is a missing theory.  Since there are for all practical purposes a limitless supply of phenomena that might provide a vital analogical clue to the missing theory, the task of identifying such a phenomenon for a theory physics does not know about is impossible.

Hence, solving dark energy within existing  theories of cosmology is likely either very difficult or impossible.

Even after finding a 4/3 scaling ratio,  it is difficult to find the means by which such scaling is achieved. 4/3 scaling per generation may be mathematically impossible if there is only one system that has scaling; likely one system scaling requires  scaling over time. 4/3 scaling per generation probably could only occur if there are two systems, but to all appearances we live in one universe.  To accept as a novel notion that there are two real and distinct reference frames is a formidable conceptual and psychological challenge.  

We  can not  detect $S$ and $\mathpzc{R}$  as separate frames of reference through direct perception, but not only through observation and inferences based on those observations. 

Suppose that there does exist some hitherto unidentified principle that can solve dark energy. Then the only way for it to be located is for that principle to be mathematically inferred in some other context. Then by seeking to  test its validity in a variety of other contexts, eventually it might be tested against dark energy. 

The hypothesis of this article is that The 4/3 RDFT is the missing principle and appears to be a law of nature.

The question of whether average IQs improve because language improves led to the question of a scale factor for lexical growth, led to 3/4 metabolic scaling, led to a connection to Stefan's Law and black body radiation, led to a radiation proof of The 4/3 RDFT, and led eventually to a possible explanation of dark energy. The chance of following such an inferential path is probably  slight. 

Likely, other paths also lead to  The 4/3 RDFT. The development of the 4/3 scaling of degrees of freedom began with observations about 3/4 metabolic scaling, hardly an obvious starting point for deciphering dark energy. Many degrees of freedom in reasoning might have to be explored to find an explanation for dark energy.  

\subsection{The fluid equation}

The fluid equation used in cosmology suggests that the formula for Hubble's parameter transitions over time, as described in section \ref{subsecDE-Obsvn100.100}. The results in this article suggests that $H(t)=1/t$. If The 4/3 RLT is valid, and if the fluid equation cannot be reconciled with it, then the fluid equation may be inapplicable to the determination of Hubble time.  

\subsection{Is The 4/3 RLT proved?}

The 4/3 RDFT and The 4/3 RLT are logically equivalent. The 4/3 RDFT and The NRT can be derived from each other. I have not in this article set out all the various results from The NRT, but there are several that suggest The NRT is valid, including the natural logarithm theorems.  Considering the number of proofs for The 4/3 RDFT and The 4/3 RLT alone, there are strong grounds for thinking The 4/3 RLT is valid. The NRT bolsters the case, in its applications including the natural logarithm theorems. If The 4/3 RDFT is a general law of nature, then Brownian motion, Clausius's ratio of mean path lengths, Stefan's Law, the cosmological scaling of energy densities, metabolic scaling, and dark energy are all instances of the general law.

Scaling and related concepts has been considered in connection with a variety of phenomena in three other articles on arXiv (Shour, Lexical growth; Isotropy; Intelligence) as well as in earlier versions of this article. Scaling and degrees of freedom suggest connections to, and possible explanations for, various problems, such as the ergodic hypothesis and the many worlds thesis (Everett 1957), lexical growth, increasing average IQs, economic growth rates, special relativity, quantum mechanics. 

That complexity arises from the manifestation of degrees of freedom available to a variety of systems seems not unlikely. The algebra of scaling, though unfamiliar in light of its novelty, can be simply described. 

The manner in which these ideas developed plays a role in my view of their plausibility.  The  question in September 2007 involving scaling was whether increasing information compression of language could explain increasing average IQs. Solutions raised questions which in turn led to other solutions and more questions. The solutions belong to nature; observed instances of nature's solutions support the generality of the 4/3 ratio theorems as laws of nature. 

\subsection{Some questions}

Do the following questions make sense, and if so, what are their answers? 
\begin{itemize}
	\item Are there cosmological observations other than those described in this article that are consistent with The 4/3 RDFT and The 4/3 RLT?
	\item Is it possible to measure radiation pressure  to test The 4/3 RDFT?
	\item What is $S$?
	\item Why is $Deg(\mathpzc{R})=3$? Is that implicit in the nature of $S$ or in the emergence of a universe or peculiar to our universe? Could there exist a universe with $Deg(\mathpzc{R}) \neq 3$? If so, what would be its attributes? 
	\item  Does The 4/3 RDFT determine the geometry of our universe? How?
	\item  Are radiation cones and space spheres, and space cones and radiation spheres dual relationships? 
	\item Is  the cosmic scale factor $a(t)$ is the same at all epochs?
	\item  Is 4/3 in the formula for the volume of a sphere related to The 4/3 RDFT?
	\item Is the 4/3 ratio of degrees of freedom theory in MIMO an instance of The 4/3 RDFT?	  
	\item If the measured mean path length for neurons in the human brain is accurate, why is it less than $e$?
	\item Can general relativity be described in terms of The 4/3 RDFT?  
	\item  Is there a way to relate tensors to The 4/3 RLT?
	\item Is energy  related to The 4/3 RDFT? \\
	\item Does the concept of degrees of freedom explain Everett's (1957) many worlds hypothesis?
	\item Does the concept of degrees of freedom resolve the problems raised by the ergodic hypothesis?
	\item Is $s \times i$ a scale factor in Schr\"odinger's equation analogous to the role of the scale factor $s$ relative to $L$?
	\item Do other phenomena instance The 4/3 RDFT and The 4/3 RLT?
	\item Are dark energy and dark matter related?
\end{itemize}

If the ideas sketched in this article are valid, these questions may be only a beginning.\\

\end{document}